\documentclass[%
 reprint,
 amsmath,amssymb,
 aps,
]{revtex4-2}

\usepackage{graphicx}
\usepackage{dcolumn}
\usepackage{bm}

\begin{document}

\preprint{APS/123-QED}

\title{Optimal Cell Shape for Accurate Chemical Gradient Sensing\\in Eukaryote Chemotaxis}

\author{Daiqiu Mou}

\author{Yuansheng Cao}
 \email{yscao@tsinghua.edu.cn}
\affiliation{
 Department of Physics, Tsinghua University, Beijing, 100084, China
}

\date{\today}

\begin{abstract}
Accurate gradient sensing is crucial for efficient chemotaxis in noisy environments, but the relationship between cell shape deformations and sensing accuracy is not well understood. Using a theoretical framework based on maximum likelihood estimation, we show that the receptor dispersion, quantified by cell shape convex hull, fundamentally limits gradient sensing accuracy. Cells with a concave shape and isotropic error space achieve optimal performance in gradient detection. This concave shape, resulting from active protrusions or contractions, can significantly improve gradient sensing accuracy at the cost of increased energy expenditure. By balancing sensing accuracy and deformation cost, we predict that a concave, three-branched shape as optimal for cells in shallow gradients. To achieve efficient chemotaxis, our theory suggests that a cell should adopt a repeating ``run-and-expansion" cycle. Our theoretical predictions align well with experimental observations, implying that the fast amoeboid cell motion is optimized near the physical limit for chemotaxis. This study highlights the crucial role of active cell shape deformation in facilitating accurate chemotaxis.
\end{abstract}

\maketitle

\section{Introduction}
Chemotaxis, the directed motion of cells in response to chemical gradients, plays a crucial role in diverse biological phenomena, including immune response, neural development, and cancer metastasis\cite{van2004chemotaxis}. Unlike bacteria, which use temporal integration during swimming for gradient sensing \cite{segall1986temporal,sourjik2002receptor}, eukaryotic cells exploit their larger size to directly sense spatial gradients across their bodies. This occurs through ligand binding to membrane receptors, triggering downstream signaling pathways that establish cellular polarity and motility, ultimately leading to directed cell movement\cite{swaney2010eukaryotic}.

Eukaryotic chemotaxis shows remarkable sensitivity, detecting gradients as shallow as $1\sim 2\%$  concentration difference across the cell, corresponding to a mere 10-molecule disparity between front and rear\cite{song2006dictyostelium,van2007biased,ueda2007stochastic,fuller2010external}. This high accuracy requires a fundamental understanding of the physical limits of gradient sensing in these systems. Existing models primarily address receptor dynamics in simplified geometries\cite{endres2008accuracy,endres2009accuracy,mora2010limits,lalanne2015chemodetection}, similar to the Berg-Purcell limit \cite{berg1977physics} for concentration sensing. Alternatively, some models explore specific kinetic features like receptor cooperativity\cite{hu2010physical}, extracellular ligand degradation\cite{segota2017extracellular}, and spatiotemporal integration\cite{iijima2002temporal,rappel2008receptor,iglesias2008navigating}. However, during chemotaxis, fast-moving cells like \textit{Dictyostelium}, neutrophils, and T cells exhibit highly dynamic, irregular shapes, characterized by amoeboid motion\cite{andrew2007chemotaxis,skoge2010gradient,miao2017altering}. This dynamic cell deformation alters the spatial distribution of receptors, potentially impacting the cell's ability to explore its chemical environment. The exact way in which cell shape limits gradient sensing in eukaryotes remains an open question.

 Here, we explore the physical limits and optimization principles of cell shape for accurate chemical gradient sensing, with minimal assumptions. We apply a maximum likelihood estimation approach that captures the maximum information a cell can extract from an instantaneous measurement. This provides a unified framework to discuss diverse mechanisms that can improve gradient sensing accuracy. We prove that the gradient inference noise has a universal lower bound dictated by the cell shape's convex hull. Accuracy of gradient direction detection, measured by the chemotactic index, is (near) optimal for cells with an isotropic error space, and the optimized instantaneous gradient measurement depends solely on the signal-to-noise ratio.
 Despite potential modulation of receptor dynamics, the most effective strategy for enhancing gradient sensing accuracy is  deforming the cell to a concave shape. Balancing sensing accuracy and deformation energy cost, our theory predicts a three-branched structure as the optimal shape for shallow gradients, consistent with observations in fast-moving amoeboid cells. This suggests amoeboid cells operate at the physical limit for chemotaxis, potentially using additional energy for active shape tuning to further enhance chemotaxis accuracy.

\section{Model}
 
\begin{figure*}[t]
    \centering
    \includegraphics[width=\textwidth]{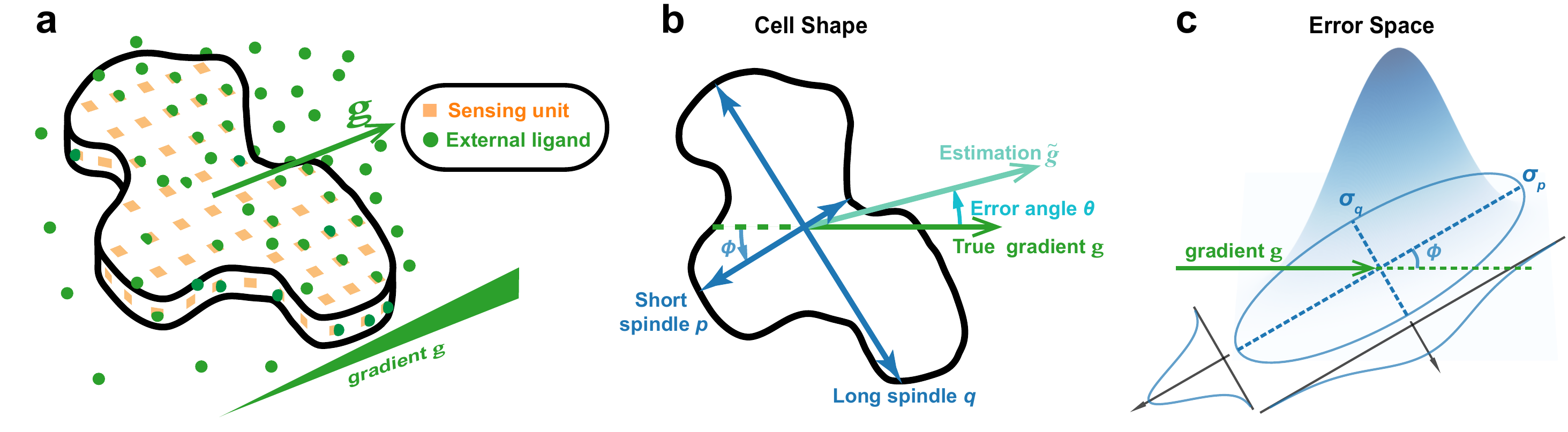}
    \caption{Illustration of cell sensing chemical gradient and error space. (a) a 3D cell in chemical gradient with receptors on membrane. (b) a polarized 2D cell (polarity described by angle $\phi$) and the estimated gradient $\Tilde{\bm{g}}$ compared to true gradient $\bm{g}$, with an error angle $\theta$. (c) the error space of the cell in b (spindles aligned with the cell in (b) ) can be represented by a Gaussian distribution.}
    \label{fig:1}
\end{figure*}

We explore the physical limits and optimization principles governing cell shape's impact on chemotaxis accuracy, employing minimal assumptions. We consider a cell in a shallow chemical profile $c(\bm{r})\approx c_0+\bm{g}\cdot\bm{r}$, where $\bm{g}=\nabla c$ represents the chemical gradient. As depicted in Fig.~\ref{fig:1}a, the cell senses the gradient by integrating local concentration information, $c_i$, from individual receptor units positioned at $\bm{r}_i$. A receptor unit can represent a single receptor, a cluster of cooperative receptors \cite{hu2010physical}, or even a single cell in collective sensing scenarios \cite{camley2016emergent,ellison2016cell}. Each unit estimates the local concentration $c_i$ with an estimation value $\hat{c}_i$ and variance $\sigma_i^2$ independently (we ignore the estimation correlation between nearby receptors due to ligand rebinding and diffusion \cite{bialek2005physical,endres2009accuracy}). Since sensing accuracy becomes problematic only at shallow gradient ($\delta c\ll c_0$)\cite{endres2008accuracy}, we assume only $c_0$ can affect $\sigma_i$. Each concentration measurement can be viewed as a sample drawn from a distribution $\rho_i(\hat{c}_i,\sigma_i^2)=\frac{1}{\sqrt{2\pi}\sigma_i}\exp{[-(\hat{c}_i-c_0-\bm{g}\cdot\bm{r}_i)^2/2\sigma_i^2]}$. Here, we assume $\sigma_i$ only depends on the average concentration, $c_0$, and the intrinsic variations of receptor properties (e.g., different binding constants \cite{hopkins2020chemotaxis}), but it does not contain gradient information. 

The cell needs to extract gradient information, i.e., an estimate (denoted as $\Tilde{\bm{g}}$) of the gradient vector $\bm{g}$ from these concentration estimates. Here, we employ maximum likelihood estimation (MLE) based on the measurement set $\mathcal{L}=\log(\prod_i\rho_i)$\cite{endres2009maximum,camley2017cell,ipina2022collective}. Note that $\mathcal{L}$ does not incorporate historical measurement, making this an instantaneous estimator. Maximizing $\mathcal{L}$ is equivalent to minimizing the following loss function
\begin{equation}
    \ell(\Tilde{\bm{g}},\Tilde{c}_0)=\sum_i\frac{1}{\sigma_i^2}(\hat{c}_i-\Tilde{c}_0-\Tilde{\bm{g}}\cdot 
    \bm{r}_i)^2,
    \label{MLE}
\end{equation}
where $\Tilde{c}_0$ and $\Tilde{\bm{g}}$ are the unknown parameters to be inferred. (see Appendix.~\ref{sec:app:GLS}).

Minimizing Eq.~\ref{MLE} yields the inferred gradient direction
\begin{equation}
    \Tilde{\bm{g}} = \bm C^{-1} \sum_i \alpha_i\hat{c}_i (\bm{r}_i-\bm{r}_0) ,
    \label{MLE_infer}
\end{equation}
and the Cramér-Rao inequality provides the associated inference uncertainty
\begin{equation}
    \mathrm{Cov}(\Tilde{\bm{g}}) = \sigma_c^2 \bm C^{-1}.
    \label{MLE_error}
\end{equation}
Here, $\sigma_c^2=1/(\sum_i1/\sigma_i^2)$ is the total concentration sensing error. $\alpha_i=\sigma_i^{-2}/(\sum_i1/\sigma_i^2)$ reflects the error contribution at position $\bm{r}_i$. $\bm{r}_0=\sum_i\alpha_i\bm{r}_i$ denotes the center of receptor units weighted by the error contribution. The symmetric covariance matrix $\bm C$ describes the positional correlation with elements $\bm C_{uv}=\sum_i\alpha_i(u-u_0)(v-v_0)$ for $u=x,y,z$ and $v=x,y,z$. For identical and uniformly distributed (i.u.d.) receptors on cell membrane, $\sigma_i=\sigma$, thus $\bm{r}_0$ becomes the geometric center of the cell shell. Here, we assume the cell has precise information of the receptor positions\cite{ipina2022collective}. The estimation for $\Tilde{c}_0$ is given in Appendix.~\ref{sec:app:cO_est}.

The MLE approach and Eq.~\ref{MLE_error} provide a unified framework for exploring various strategies to improve gradient sensing accuracy. 
The uncertainty in the inferred gradient direction arises from two key factors: (1), $\sigma_c^2$, which represents the error associated with estimating the average concentration $c_0$, using multiple receptors in shallow gradient. Many mechanisms, including the classic Berg-Purcell limit \cite{berg1977physics} and its refinements \cite{bialek2005physical,endres2009maximum,govern2012fundamental}, receptor cooperativity and adaptation \cite{tu2013quantitative,owen2023size,hu2010physical}, and nonequilibrium sensing \cite{lang2014thermodynamics,govern2014energy,owen2020universal}, have been proposed to enhance concentration sensing accuracy.
(2), the positional uncertainty matrix $\bm C$, which is determined by cell shape and the spatial arrangement of receptor units. Here, we focus on the relatively unexplored role of cell shape, as represented by the matrix $\bm C$, in gradient sensing. To isolate the effect of cell shape, we keep the total number of receptors fixed throughout this work. Consequently, receptor density varies as the cell surface area  (or contour for 2D cells) changes. 

\section{Receptor dispersion bounds the sensing accuracy}
\begin{figure}
    \centering
    \includegraphics[width=\linewidth]{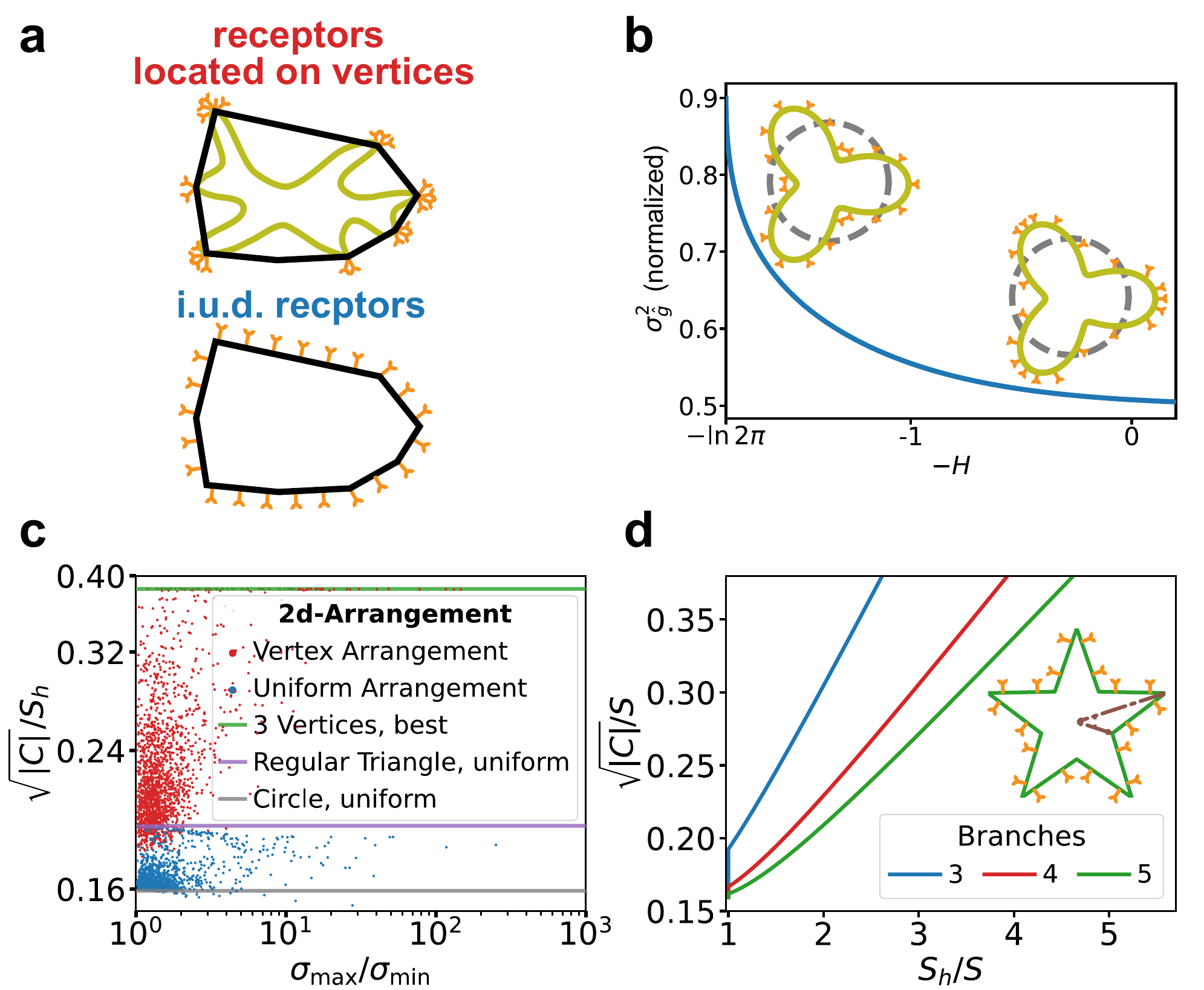}
    \caption{The total noise in gradient sensing is determined by receptor distributions and cell shape. (a) Top, receptors distributed on vertices of a polygon or a concave cell with the same vertices. Bottom, identical and uniformly distributed (i.u.d.) receptors on a convex cell. (b) Sensing noise $\sigma_g^2$ decreases in a concave cell as the non-uniformity (-$H$) of the receptors distribution increases. The shape (green lines) is parameterized in polar coordinates by $r=1+0.5\cos3\theta$, and the sensing uncertainty is normalized by a circular cell with the same area (gray line). (c) Sampling $\sqrt{|\bm C|}/S$ for different shapes and receptor distributions. (d) $\sqrt{|\bm C|}/S$ for different convex hull in a 3-, 4-, and 5-point regular star shapes with fixed size and i.u.d. receptors. 
    }
    \label{fig:sampling}
\end{figure}

The symmetric matrix $\bm C$ can be diagonalized to have three principle axes of polarity $I_p,I_q$ and $I_w$. Consequently, the covariance in Eq.~\ref{MLE_error} decomposes into three components $\sigma_p^2,\sigma_q^2$ and $\sigma_w^2$, with the relationship $\sigma_{p,q,w}=\sigma_c/\sqrt{I_{p,q,w}}$. These components satisfy the equation
\begin{equation}
    \sigma_p\sigma_q\sigma_w=\frac{\sigma_c^3}{\sqrt{|\bm C|}}.
    \label{noise relation}
\end{equation}
Eq.~\ref{MLE_infer} and \ref{MLE_error} spans the error space of gradient inferred by the cell, which can be described by a 3D Gaussian distribution, as shown in Fig.~\ref{fig:1}c. Notice that the isotropy of cell shape and corresponding error space are not the same: a cell with evenly distributed protrusions has an isotropic error space. The characteristic value $(\sigma_p\sigma_q\sigma_w)^2$ represents the generalized variance, which we define here as the total noise.

The dimensionless quantity $\mathcal{S}=|\bm{g}|^3/(\sigma_p\sigma_q\sigma_w)$ denotes signal-to-noise ratio (SNR) throughout this paper (For a spherical cell, the SNR returns to the often used form that depends on cell diameter as in \cite{endres2009accuracy,endres2008accuracy}). For i.u.d. receptors, the distortion of the error space reflects the geometric polarity of the cell shell, as shown in Fig.~\ref{fig:1}b and c.

The total noise in Eq.~\ref{noise relation} is determined by receptor localization on cell surface and cell shape. Using negative Shannon entropy $-H$ as a measure of receptor dispersion (see details in supplemental material \cite{supp}), Fig.~\ref{fig:sampling}b shows that the sensing accuracy for a given shape decreases as receptor distribution becomes more uniform. This suggests that receptor clustering and relocalization at the cellular scale enhance sensing accuracy. This large-scale receptor clustering is well-documented in bacteria such as \textit{E. coli}. In eukaryotic cells, although receptor localization on microvillli has been observed\cite{cai2022t,yang2022membrane}, the overall receptor distribution remains nearly uniform. In \textit{Dictyostelium}, receptors are evenly distributed, even in the presence of dynamic protrusions\cite{xiao1997dynamic}. Based on these observations, we assume receptors are i.u.d., unless specified in the following.

$|\bm C|$ reaches its maximum when receptors are located at the vertices of convex polygons (Fig.~\ref{fig:sampling}c). This also applies to concave shapes where all vertices coincide with those of the convex shape in Fig.~\ref{fig:sampling}a. 
In Appendix.~\ref{sec:app:Bound}, we demonstrate that for any arbitrary shape, $|\bm C|$ is upper-bounded by the cell's 3D convex hull $V_h$: $|\bm C|< 9V_h^2/4$. For 2D cells, the bound is the 2D convex hull $|\bm C|< S_h^2/4$. 
Through numerical sampling, we found that the minimal noise is achieved by distributing the receptors only on the vertices of a 3-point shape (Fig.~\ref{fig:sampling}c), a result also reported in \cite{alonso2024receptors}.
Further results for 3D cells are provided in  Table.~S1 from the supplemental material (SM)\cite{supp}. Importantly, the bound on $|\bm C|$ leads to a universal lower bound on the total noise of gradient sensing:
\begin{equation}
    \sigma_p\sigma_q\sigma_w> \frac{2\sigma_c^3}{3V_h}.
    \label{noise lower bound}
\end{equation}
For 2D cells, the total noise has a lower bound of $\sigma_p\sigma_q> 2\sigma_c^2/S_h$. For convex cells, the convex hull size is equal to the cell size ($V_h=V$). However, concave cells have a larger convex hull $V_h>V$ due to their inward curvature. As a result, for cells with the same volume, concave shapes can achieve a potentially lower total noise level due to the advantage of their larger convex hulls.

Unlike the case with vertex-localized receptors (Eq.~\ref{noise lower bound}), there is no simple bound for i.u.d. receptors. Nevertheless, the convex hull remains a good predictor for $|\bm C|$, as shown in Fig.~\ref{fig:sampling}d. With fixed cell size and i.u.d. receptors, $\sqrt{|\bm C|}/S$ is proportional to the convex hull $S_h$ for regular star shapes (see Fig.~S4 in the SM\cite{supp} for model details). Thus, convex hull serves as a simple measure for receptor dispersion and the total noise level of gradient sensing for certain cell shape groups. Below we will explore the optimal cell shape for accurate gradient sensing under realistic constraints, with fixed cell volume/size and uniformly distributed receptors.

%%%%%%%%%%%%%%%%%%%%%%%%%%%%%%%%%%%%%%%%%%%%%%%%%%%%%%%%%%%%

\section{Cell with isotropic error space is optimal in chemotaxis accuracy}
For biological relevance, we focus on the accuracy of the inferred gradient \textit{direction} by the cell. Consistent with previous studies\cite{endres2008accuracy}, we define a chemotactic index (CI) as the average cosine of the angle between the true gradient $\bm{g}$, and the inferred direction, $\hat{\bm{g}}$, within the error space:
\begin{equation}
    \text{CI}=\langle\cos\theta\rangle=\left\langle\frac{\bm{g}\cdot\hat{\bm{g}}}{|\bm{g}||\hat{\bm{g}}|}\right\rangle.
    \label{CI_def}
\end{equation}
This index, ranging from 0 (completely random inference) to 1 (no bias), quantifies the average tendency of a cell to infer the chemical gradient using the MLE method. Here, $\Hat{\bm{g}}$ represents the inferred gradient vector, with its mean and covariance provided by Eq.~\ref{MLE_infer} and \ref{MLE_error}, respectively.

Most eukaryotic cells need to adhere to a substrate or extracellular matrix for movement and chemotaxis. In this work, we consider a 3D cell migrating on a substrate located on the $xy$ plane, reflecting the typical setup in chemotaxis experiments and the natural environment for many chemotactic cells (For a swimming cell in 3D matrix see discussion and SM). Consequently, only the in-plane gradient component guides the cell's movement, and we assume $\bm{g}$ points along the x-direction. We assume i.u.d. receptors on the membrane and an irregular disk-shape cell. This reduces our problem to finding the optimal shape in 2D space, with a 2D error space with axes $\sigma_p$ and $\sigma_q$ represented by a Gaussian distribution. The total noise bound (analogous to Eq.~\ref{noise lower bound}) becomes $\sigma_p\sigma_q\geq 2\sigma_c^2/S_h$, where $S_h$ is the convex hull of the 2D shape. 

\begin{figure}
    \centering
    \includegraphics[width=\linewidth]{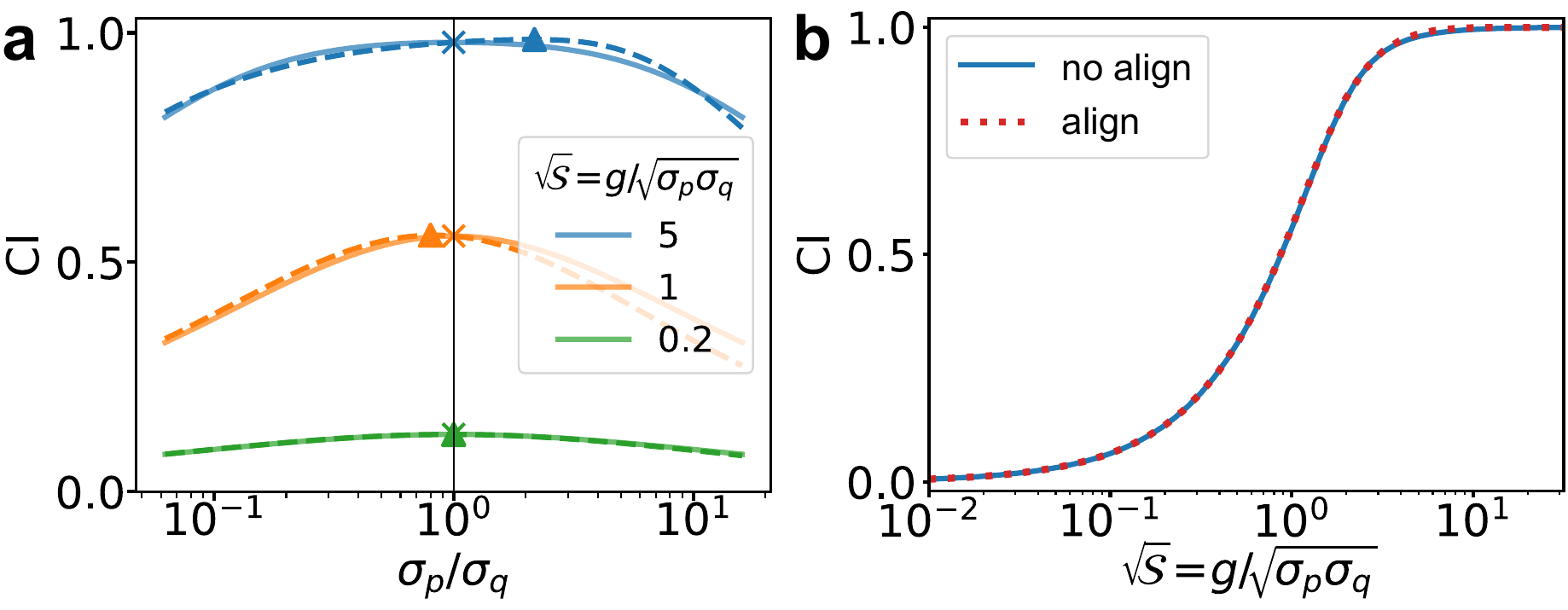}
    \caption{Cell with isotropic error space is (near) optimal for accurate gradient sensing. (a) the optimal CI when a cell can align its error space orientation to any direction (solid line and cross marks) or align with previous measurement (dashed line and triangle marks). Marks denote the optimal CI. (b) the CI dependence on signal-to-noise ratio.}
    \label{fig:2}
\end{figure}

CI is fully determined by three independent parameters: error space aspect ratio $k=\sigma_p/\sigma_q$ that measures the space anisotropy (here we use a generalized aspect ratio $0<k<\infty$), error space orientation angle $\phi$ (the angle between p-axis and $\bm{g}$), and the SNR (defined as $|\bm{g}|^2/(\sigma_p\sigma_q)$ in 2D), as show in Fig.~\ref{fig:1}c. During chemotaxis, a cell can actively tune its shape to maximize CI. Controlling the total noise and error space aspect ratio, $k$, only requires precise information about the relative positions of receptors $(\bm{r}_i-\bm{r}_0)$, which is feasible\cite{ipina2022collective}.
However, controlling the space orientation requires information about gradient direction, as $\phi$ is the relative angle to the true gradient.

In practical scenarios, a cell can only adjust the angle $\phi$ with finite accuracy. Given a prior probability distribution, $p_{\phi}(\phi)$ for $\phi$, the expected CI from one independent measurement (or instantaneous CI) should be averaged over the ensemble of possible angles
\[
\text{CI}=\int \cos\theta\, p(\hat{\bm{g}}|\phi,k) p_{\phi}(\phi) \,d\hat{\bm{g}}d\phi.
\]

The orientation dynamics during chemotaxis may depends on the measurement history \cite{wang2013rho,wu2014three,aquino2014memory,karmakar2021cellular,skoge2014cellular,hu2011geometry,hiraiwa2014relevance,nakamura2024gradient}.
If a cell has no prior information about the direction of the true gradient, it can only choose a random direction to align its spindle, resulting in $p_{\phi}(\phi)=1/2\pi$. The optimal CI is always achieved when the error space is isotropic, i.e. $k=1$ (Fig.~\ref{fig:2}a). However, if a cell can align $\phi$ based on previous measurements, a non-monotonic dependence on aspect ratio $k$ is observed, as shown in Fig.~\ref{fig:2}b. The peak of the optimal aspect ratio is found near $k=1$, with slight deviations. Even in an unrealistic scenario where a cell already knows the precise gradient direction, the improvement of anisotropic error space is marginal compared to isotropic error space (see SM and Fig.~S2~S3). 

These analysis suggests that the potential improvement in gradient sensing by fine-tuning error space anisotropy is minimal.
Therefore, we conclude that an isotropic error space is (nearly) optimal for gradient inference. As a result, the optimal CI depends solely on the SNR:
\begin{equation}
    \text{CI}= \frac{\sqrt{2 \mathrm{\pi}\mathcal{S}} \cdot\exp (- \mathcal{S}/4)}{4}
    \left[I_0\left(\frac{\mathcal{S}}{4}\right)+I_1\left(\frac{\mathcal{S}}{4}\right)\right].
    \label{CI_circle}
\end{equation}
Here, $I_{0,1}$ is the first (second)-order modified Bessel function of the first kind (see derivation in the SM\cite{supp}).
Similar equation has been obtained in \cite{endres2008accuracy} for a spherical cell with absorbing receptors. The sole dependence of CI on SNR has also been postulated in \cite{ueda2007stochastic} and tested in \cite{amselem2012control}, and here we show that it is the theoretically optimized CI under realistic conditions.

\section{Concave cells can improve accuracy significantly}
\begin{figure*}[t]
    \centering
    \includegraphics[width=\textwidth]{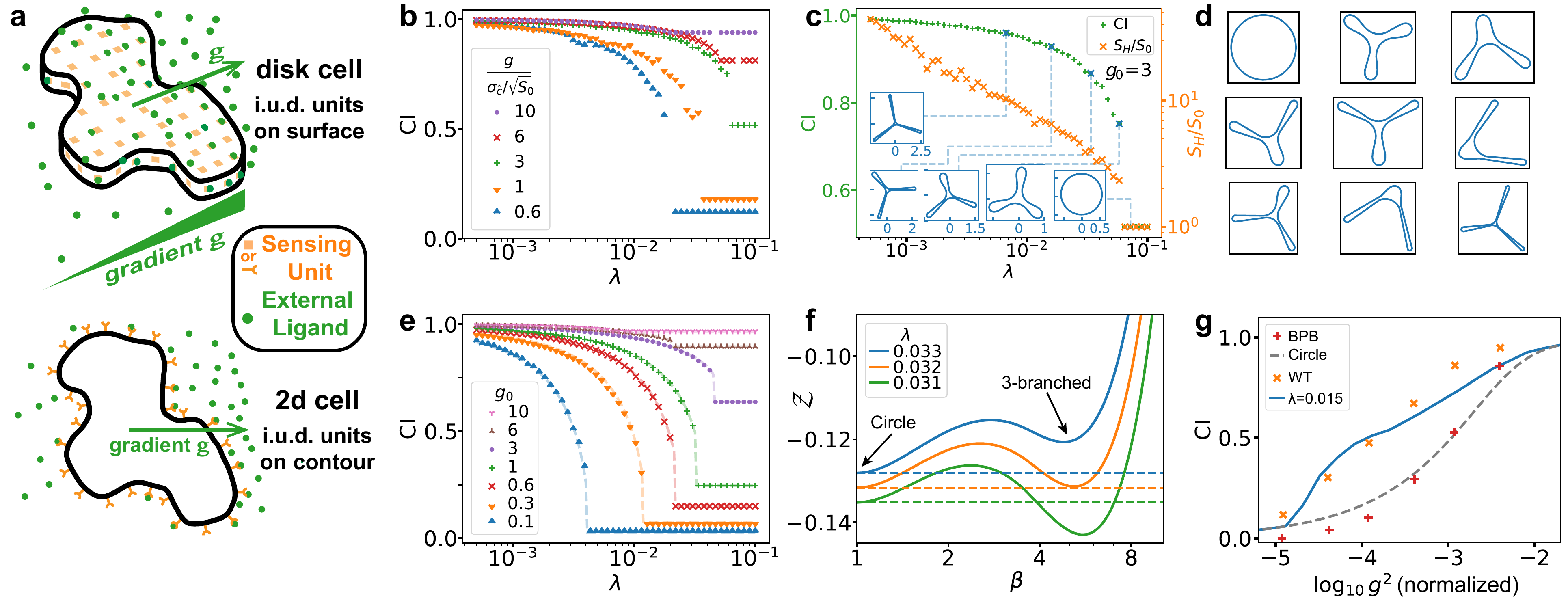}
    \caption{Optimal cell shape for accurate gradient sensing under deformation energy constraint. 
    (a) Illustration of a 3D disk shape cell (top) and a 2D cell (bottom) with i.u.d. receptors on surface and contour, respectively. 
    (b) optimal CI for different (normalized) gradient $g_0$ and marginal energy cost $\lambda$. (c) CI for optimal shape and its corresponding convex hull. (d) Possible optimal cell shapes from gradient descent searching. 
    (e) Optimized $\mathrm{CI}$ for 2D cells, where all receptors are evenly distributed on the contour. The dots denote the simulation results, and the dashed lines are calculated on the $\beta$-parameterized 3-tentacle-shape.
    (f) 2D Optimization landscape over a subspace of 2D shapes. 
    (g) CI under different deformation ability. 
    Experiment data (dots) is taken from \cite{tweedy2013distinct},
    where BPB treated cells are regarded as no deformation ability.
    Blue line corresponds to $\lambda=0.015$ and gray dashed line for a circular disk cell ($\lambda=\infty$). CI is averaged between the run and expansion states at low SNR\cite{tweedy2013distinct}.
    }
    \label{fig:3}
\end{figure*}

Because a cell with isotropic error space approaches optimal CI under the constraint of total noise, and from Eq.~\ref{CI_circle}, CI is a monotonically increasing function of SNR (Fig.~\ref{fig:2}d), the only effective strategy to enhance CI is to reduce the total noise itself or to increase the SNR. The effect of cell shape on SNR can be decomposed into two dimensionless contributions,
\[
\mathcal{S}=\frac{|\bm{g}|^2}{\sigma_p\sigma_q}=\frac{|\bm{g}|^2S}{\sigma_c^2}\cdot \frac{\sqrt{|\bm C|}}{S},
\]
corresponding to two strategies. The first term, $|\bm{g}|^2S/\sigma_c^2$, represents one straightforward (yet crucial) approach that increases the cell's surface area. Cells can achieve this (while conserving volume) by increasing cell-substrate adhesion, reducing contractility, or increasing internal pressure in confined spaces. Beyond that, cells can further improve CI by adopting a concave shape through protrusions and contractions that enlarge convex hull within a given size, as indicated by $\sqrt{|\bm C|}$/S. To isolate this effect, we maintain a constant cell thickness, thereby fixing the the top and bottom surface areas (and below we use the dimensionless gradient $g_0=|\bm{g}|\sqrt{S}/\sigma_c$). This is equivalent to the scenario that the side membrane can sustain bending until a maximum value. In Fig.~\ref{fig:sampling}d and Fig.~S4 in the SM\cite{supp}, we show that the term $\sqrt{|\bm C|}/S$ can theoretically approach infinity, causing CI to saturate at 1. However, this scenario comes at the cost of a diverging cell surface.

 In eukaryotic cells, the active force responsible for maintaining a concave shape can arise from the accumulation of force-generating molecules, including septin, actin or myosin at localized regions. They consume free energy (e.g., from ATP hydrolysis) and perform work against cell periphery deformation forces (such as tension, bending and cytoskeleton elasticity). For simplicity, we only consider the energy expenditure due to membrane/cortex tension. This is equivalent to minimize a free-energy-like objective function
\[
\mathcal{Z}=-\mathrm{CI}+\lambda(L-L_0)+\epsilon|S-S_0|^2.
\]
Here, $\lambda$ is the marginal energy cost of improving CI determined by the cell's deformability. A higher $\lambda$ represents increased membrane tension or decreased active force. Finally, $S_0$ is the fixed area size,  $L_0=2\sqrt{\pi S_0}$ is the circle circumference (with size $S_0$) and $\epsilon$ is a Lagrange multiplier (note $\lambda$ is not a Lagrange multiplier).

For various gradient and energy cost $\lambda$, we search the global minima of $\mathcal{Z}$ using gradient descent method in the space of all cell shapes (CI is calculated using the no-alignment scenario, and methods see the SM\cite{supp}). Interestingly, only two distinct shapes emerged as optimal: a circle and a three-branched structure, as illustrated in Fig.~\ref{fig:3}b-d. Due to fluctuations in optimization algorithm, some two-branched shape also emerges as near optimal, as in Fig.~\ref{fig:3}d (also Fig.~S5 in the SM\cite{supp}). For a given gradient, CI exhibits a discontinuous transition at a critical value $\lambda_c$. When $\lambda>\lambda_c$, the optimal shape is a circle. Conversely, when $\lambda<\lambda_c$, the three-branched shape becomes favorable. In steep gradient, the three-branched shape offers only marginal improvement in CI. However, in shallow gradient $\mathcal{S}\leq 1$, the transition from a circular to a three-branched shape leads to a significant enhancement in CI, with the latter achieving values more than two times greater than the former (Fig.~\ref{fig:3}b and c). A cell can always attain a CI close to 1 for a wide range of SNR, provided $\lambda$ is sufficiently small. At shallow gradient, $\lambda_c$ decreases as $g_0$ decreases. The optimal cell shape's convex hull $S_h$ (yellow dots in Fig.~\ref{fig:3}c) increases monotonically with CI. This indicates that the convex hull serves as a good predictor for the a cell's ability to accurately detect gradients.

For a better understanding of how optimal shapes vary under different conditions, we applied the optimization algorithm to 2D cells. In this scenario, the sensing units are uniformly distributed along the contour of the shape (Fig.~\ref{fig:3}a bottom), unlike being located on the whole surface of a 3D cell (Fig.~\ref{fig:3}a top). Interestingly, the results for 2D cells show consistency with those obtained for 3D disk shapes, as shown in Fig.~\ref{fig:3}e. To understand the transition near $\lambda_c$, we introduce a simplified representation using a single parameter $\beta$ (Fig.~S6). As $\beta$ changes from $1$ to $\infty$, the shape smoothly transitions from a perfect circle ($\beta=1$) to infinitely long, three equally spaced tentacles. The optimal CI in space defined by $\beta$ closely resembles the results from the optimization algorithm (dashed line in Fig.~\ref{fig:3}e, and the optimization results are shown in Fig.~S6 in the SM\cite{supp}). Fig.~\ref{fig:3}f visualizes the landscape of $\mathcal{Z}$ as a function of $\beta$. The circular shape consistently represents a local minimum at $\beta=1$. However, the three-branched shape only becomes the global minimum when $\lambda<\lambda_c$. Thus, the phase transition near $\lambda_c$ arises from the double-well landscape of $\mathcal{Z}$ in cell shape space. The substantial difference in cell perimeter between the two local minima translates to a significant gain in CI. This implies that close to the critical point, a minor enhancement in deformability can lead to a dramatic increase in CI.

Our theory suggests that for optimal CI with minima energy cost, cells should adopt only a few distinct shapes (Fig.~\ref{fig:3}d): a circular shape in steep gradient and a three-branched shape in shallow gradient (two-tentacle-shape is near optimal, see discussions in SM\cite{supp}). This finding aligns with observations of \textit{Dictyostelium} cells performing chemotaxis in shallow gradients. These cells utilize a pseudopod-splitting mechanism that results in a two-branched (or multiple-branched) leading edge \cite{insall2010understanding,skoge2010gradient,bodor2020cell,andrew2007chemotaxis,alonso2024persistent}, resembling the concave and three-branched shape in our simulations. Notably, it was demonstrated in \cite{tweedy2013distinct} that 85$\%$ of the shape variability in chemotacting \textit{Dictyostelium} cells can be captured by just two principal components (PCs): PC1, corresponding to elongation, and PC2, associated with pseudopod splitting. It was further observed that PC1 dominates at high SNR, while PC2 becomes dominant at low SNR. These experimental findings are consistent with our predictions. 
Additionally, for a given $\lambda$, the optimal CI decreases with SNR, with a sharp drop corresponding to the transition from the three-branched cell shape to a circle. This transition shifts to lower SNR as $\lambda$ decreases, as shown in fig.~\ref{fig:3}g. In \textit{Dictyostelium}, treatment with p-bromophenacyl bromide (BPB) that inhibits pseudopod-splitting (corresponding to increase in $\lambda$) leads to sharp decline in CI. The optimized CI versus SNR curves for strong and weak deformability conditions agree with theoretical prediction (Fig.~\ref{fig:3}g). This consistency between our theoretical predictions and experimental data suggests that stereotypical cell shapes represent optimized strategies for achieving accurate chemotaxis at the fundamental physical limit \cite{tweedy2013distinct}. These findings highlight the crucial role of active forces in promoting eukaryotic cell chemotaxis.

\begin{figure*}[t]
    \centering
    \includegraphics[width=\textwidth]{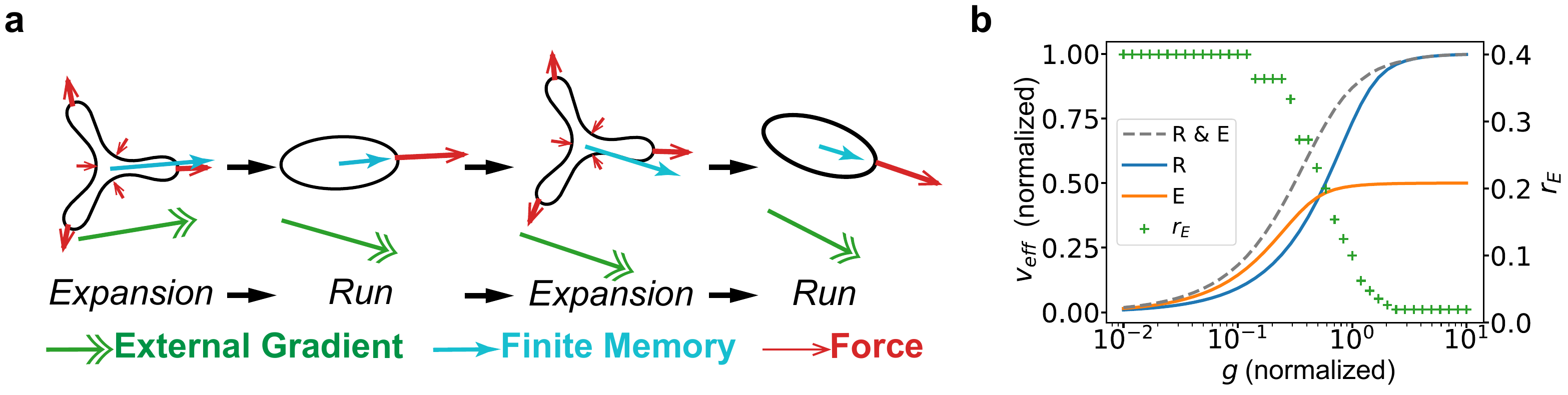}
    \caption{
    The ``run-and-expansion" strategy for efficient chemtoaxis.
    (a) Cells with finite memory cyclically switch between expansion phase and run phase.
    (b) The hybrid strategy (R\&E) improves chemotaxis efficiency, measured by the effective speed towards gradient direction $v_\mathrm{eff}$, compared to the run-only (R) or expansion-only (E) strategy under different gradients. 
    The green crosses indicate the proportion of expansion mode in the optimal strategy.
    }
    \label{fig:RE}
\end{figure*}

The three-branched shape, although optimal for gradient sensing accuracy, suffers in cell mobility due the protrusions not being consistently aligned. The elongation mode, featuring a single, broad pseudopod, is optimized for unidirectional motion. In real scenarios, when a cell measures the gradient direction with high accuracy, it should establish a persistent polarity and travel towards that direction as quickly as possible. Experimental studies have shown that amoeboid cells like \textit{Dictyostelium} and T cells alternate between the elongated and multi-branched modes during chemotaxis in shallow gradients \cite{tweedy2013distinct,cavanagh2022t}. We propose a ``run-and-expansion" cycle for eukaryotic cell chemotaxis\cite{tweedy2013distinct} in shallow gradients, as illustrated in Fig.~\ref{fig:RE}a and detailed in SM and fig. S8. During the expansion phase, the cell moves at a slower speed but exhibits a more accurate gradient sensing. It establishes a dynamic memory that aligns with the measured direction\cite{wang2013rho,nakamura2024gradient} and decays over time (see SM). This memory guides the run phase, where the cell adopts a single protrusion and moves at a faster speed but with less accurate sensing. We use a dynamic model to search for the optimal switching time between these two phases. Compared to a strategy relying solely on run or expansion, this hybrid strategy achieves both high speed and accurate sensing, as measured by the effective speed $v_\mathrm{eff}$ along the gradient direction (Fig.~\ref{fig:RE}b). The proportion of the expansion phase increases as SNR decreases, because noise effect becomes dominant and the three-branched shape guarantees the accurate sensing. These results are consistent with the experimental results\cite{tweedy2013distinct}.

\section{Discussion}

Our study reveals a fundamental role for cell shape in accurate gradient sensing, complementing the established importance of receptor dynamics. We introduce a general theoretical framework that demonstrates how cell shape inherently limits the ability to sense gradients. We show that receptor dispersion, influenced by cell shape as quantified by its convex hull, significantly impacts sensing accuracy. When considering the cost of deformation, the optimal cell shape in shallow gradients emerges as a concave form with isotropic error space.  These findings suggest that cells can actively tune their shape to approach the physical limit of gradient sensing, providing a new target for understanding and potentially manipulating cellular navigation.

How cells implement this MLE mechanism remains elusive. One potential scenario involves localized initiation of downstream signaling pathways by receptors. This could lead to a spatial correlation between these signaling molecules \cite{singh2022sensing} that mimics the spatial correlation matrix, $\bm C$. However, it is important to recognize that the MLE method represents a fundamental physical limit on the accuracy of gradient information obtainable from a single measurement. The agreement between theoretical predictions and experimental observations implies that efficient chemotacting cells operate close to this limit. 

Cell memory is critical for efficient chemotaxis\cite{novak2021bayesian}, as shown in Fig. 5. CI in eq.\ref{CI_circle} is an instantaneous measure of gradient direction. In a steady concentration gradient, memory can improve sensing accuracy in a manner similar to the Berg-Purcell limit: the variance is inversely related to the number of independent measurements. Although it is clear how bacteria implement this strategy through adaptation circuits\cite{tu2013quantitative}, it remains unclear how eukaryotic cells memorize gradient. One potential candidate is the slow global inhibitors in the local-excitation-global-inhibition (LEGI) model\cite{levchenko2002models,xiong2010cells}. In bacterial chemotaxis, memory is essential as cell measures the gradient direction by comparing temporal concentration signals. In contrast, eukaryotic cells can directly infer the gradient from a single measurement. However, in eukaryotic chemotaxis, memory is not always beneficial. For example, many cells undergo chemotaxis in environments with constantly changing gradient directions, such as neutrophils chasing pathogens or \textit{Dictyostelium} aggregation through cAMP waves. In these scenarios, long memory can actually hinder the accuracy of gradient inference. In the micropipette experiments, \textit{Dictyostelium} or neutrophils can switch polarity within a minute\cite{janetopoulos2008directional,hadjitheodorou2021directional}, which is a typical activation timescale of the chemotaxis signaling pathways. This suggests that these cells are unlikely to have a long memory of the previous gradient direction, and thus, an instantaneous, accurate measurement is crucial for their function.

The intuition behind our theory is straightforward: for a cell with limited size, the most effective strategy for exploring its surroundings is to extend protrusions radially. The three-branched shape represents the first configuration in n-branched shapes that preserves the isotropic error space, making it optimal in CI while minimizing the deformation energy cost. Our analysis on how isotropy and concavity influence gradient sensing can be generalized to 3D, free-moving (swimming) cells as well. Following similar principles, we predict that the optimal strategy of detecting shallow gradient in a 3D environment for those cells would be a four-branched structure. This is verified by numerical simulations, with similar cell shape transitions on deformability observed (see SM\cite{supp} for details). This shape is similar to what have been observed in T cells\cite{cavanagh2022t}. Furthermore, since the optimal cell shape resembles an isotropic error space, only the gradient magnitude, rather than its direction, will influence the sensing accuracy. Thus, the predicted shape is optimal for detecting gradients in multiple directions.

Many migratory cells (or cell clusters) possesses long, thin finger-like structures such as filopodia (length comparable to cell diameter) that are used to sense mechanical cues and adheres to the extracellular matrix. When a cell encounters multiple signals, they may not follow that predicted shape above. It would be fascinating to investigate whether the principles of governing the growth of finger-like structure align with the rules established by our theory for free space exploration, and how cells switch between different strategies based on their environment and biological context.

Our theory has the potential to be extended to collective gradient sensing, where a group of cells collaborate to detect spatial gradient\cite{camley2016emergent,ellison2016cell,fancher2017fundamental}. Cell colonies exhibit varying degrees of fluidity and deformability depending on cell-cell adhesion strength and motility\cite{bi2015density,bi2016motility}. How cells use these properties to influence their collective chemotaxis capabilities is an exciting direction for future research.

\begin{acknowledgments}
We thank Dr. Robert Endres for valuable discussion and comments. D. M. and Y. C. are supported by National Natural Science Foundation of China under grant 12374213.
\end{acknowledgments}

\appendix
\section{Maximum Likelihood Estimation \& Generalized Least Squares}\label{sec:app:GLS}

In order to sense spatial concentration differences,
    a cell (or a group of cells in collective sensing) must have the ability to perform local concentration sensing.
Our model achieve this by the use of local sensing units that operate independently.
Each unit would give an estimate (i.e., a signal) of 
    local concentration $\hat{c}_i$. The estimation has an expectation of the true concentration $c_i=c_0 + \bm g\cdot\bm r_i$ 
    and a variance $\sigma_i^2$.
Here, $\bm r_i$ is the position of local unit,
    $c_0$ is the concentration at $\bm r=\bm 0$,
    and $\bm g = \nabla c$ is the gradient direction.
We define $W=\sum\nolimits_j\sigma_j^{-2}$ and $\alpha_i = \sigma_i^{-2}/W$.
Clearly, $\sum\nolimits_i \alpha_i=1$ and $\alpha_i\ge 0$.
For convenience,
    we set the original point of coordinate system $\bm r=\bm 0$ to be the weighted average position of all local sensing units, calculated as
    $\sum\nolimits_i \alpha_i \bm r_i$.

By viewing each local sensing unit as a cluster of 
    independent, identical sub-units,
    the central limit theorem dictates that the local concentration signal, $\hat{c}_i$,
        follows a Gaussian distribution $\mathcal{N}(c_i, \sigma_i^2)$.
Mathematically, the joint probability distribution function (PDF) is (note the independent assumption among different local units),
\begin{equation*}
    \begin{split}
        &\mathrm{PDF}\left(\{\hat{c}_i\}\,\big|\, c_0, \bm g, \{\sigma_i^2\}\right)\\
        = 
        &\prod\nolimits_i \frac{1}{\sqrt{2\mathrm{\pi}}\sigma_i} 
        \exp\left[-\frac{(c_0 + \bm g\cdot\bm r_i - \hat{c}_i)^2}{2\sigma_i^2}\right].
    \end{split}
\end{equation*}
Using maximum likelihood criterion, 
    the best estimation $\tilde{c}_0, \tilde{\bm g}$ 
    is given by minimizing objective
\begin{equation}
    \ell = 
        \sum\nolimits_i 
        \frac{\alpha_i}{2}
        \left(c_0 + \bm g\cdot\bm r_i - \hat{c}_i\right)^2,
        \quad 
        \alpha_i \propto \sigma_i^{-2}
    \label{eq:ap:loss}
\end{equation}
At the minimum of Eq.~\ref{eq:ap:loss}, it gives
\begin{equation}
\begin{split}
    \frac{\partial \ell}{\partial g_x} = 
    &\sum\nolimits_i \alpha_i 
    \left(c_0 + \tilde{\bm g}\cdot\bm r_i - \hat{c}_i\right)
    \cdot x_i
    = 0\\
    \text{ i.e.}\quad
    &\sum\nolimits_i \alpha_i (\tilde{\bm g}\cdot\bm r_i) x_i
    = \sum\nolimits_i \alpha_i \hat{c}_i x_i
    \label{eq:min_condition} 
\end{split}
\end{equation}
where $g_x = \partial_x c,\, x_i=\bm r_i \cdot\hat{\bm x}$ are the component in the $x$-direction.
Note that the origin point $\sum\nolimits_i \alpha_i \bm r_i=\bm 0$.
Expand the left-hand side of Eq.~\ref{eq:min_condition} of $x$-direction gives
\begin{equation}
    \begin{split}
        &\sum\nolimits_i \alpha_i (x_i \tilde{g}_x + y_i \tilde{g}_y + z_i \tilde{g}_z) x_i\\
        =\, &\tilde{g}_x \sum\nolimits \alpha_i x_i^2
            + \tilde{g}_y \sum\nolimits \alpha_i x_i y_i
            + \tilde{g}_z \sum\nolimits \alpha_i x_i z_i\\
        = &\begin{bmatrix}
            C_{xx} & C_{xy} & C_{xz}
            \end{bmatrix}
            \cdot
            \begin{bmatrix}
                \tilde{g}_x\\\tilde{g}_y\\\tilde{g}_z
            \end{bmatrix}
    \end{split}
\end{equation}
Here, we defined $C_{uv} = \sum_i \alpha_i (u_i - u_0) (v_i - v_0)$
where $u, v$ stand for any directions of $x, y, z$, 
    and $u_0 = \sum\nolimits_i \alpha_i u_i,\,v_0 = \sum\nolimits_i \alpha_i v_i$.

Recalling that $\alpha_i = \sigma_i^{-2} / \sum\nolimits_i \sigma_i^{-2} \ge 0$ and $\sum\nolimits_i \alpha_i = 1$,
we can interpret $\alpha_i$ as a normalized probability distribution across the space,
    which reflects the weight assigned to the detection results of each local concentration signal, $\hat{c}_i$.
In this framework, $C_{uv}= \langle uv \rangle - \langle u \rangle \langle v \rangle$ 
    can be regarded as the covariance under the probability distribution $\alpha_i$. 
    Here, $\langle\cdot\rangle$ denotes the expectation of the variable $\cdot$ under this corresponding distribution.
Building on this concept, we define the covariance matrix as
\begin{equation}
    \bm C = \begin{bmatrix}
        C_{xx} & C_{xy} & C_{xz}\\
        C_{xy} & C_{yy} & C_{yz}\\
        C_{xz} & C_{yz} & C_{zz}\\
    \end{bmatrix}
\end{equation}
representing the extent to which the cell unfolds in space.
In this way, Eq.~\ref{eq:min_condition} can be rewritten 
    (in all directions) as
    $ \bm C \bm g = \sum\nolimits_i \alpha_i \hat{c}_i \bm r_i$.
Therefore
\begin{equation}
    \tilde{\bm g} = \bm C^{-1} \sum\nolimits_i \alpha_i \hat{c}_i \bm r_i.
    \label{eq:ap:est_res}
\end{equation}

To analyze the uncertainty of Eq.~\ref{eq:ap:est_res}, 
    we rewrite $\hat{c}_i = c_i + \delta_i$,
    where $\mathbb{E}(\delta_i) = 0$, $\mathrm{Var}(\delta_i)=\sigma_i^2$.
Then the right-hand side of Eq.~\ref{eq:min_condition} can be written as
\begin{equation*}
\begin{split}
    \sum\nolimits_i  \alpha_i \hat{c}_i \bm r_i
    &= \sum\nolimits_i  \alpha_i (c_{0} + \bm r_i\cdot \tilde{\bm g} + \delta_i) \bm r_i\\
    &= \bm 0 + \bm C \tilde{\bm g} + \sum\nolimits_i \alpha_i \bm r_i \delta_i.
\end{split}
\end{equation*}
With independence hypothesis on the estimation unit $\hat{c}_i\perp\!\!\!\perp \hat{c}_j$, i.e. $\delta_i\perp\!\!\!\perp\delta_j$,
we get the covariance of $\sum\nolimits_i  \alpha_i \hat{c}_i \bm r_i$ over direction $u, v$ as
\begin{equation*}
\begin{split}
    &\mathrm{Cov}\left(
        \sum\nolimits_i \alpha_i u_i \delta_i,
        \sum\nolimits_j \alpha_j v_j \delta_j
    \right)\\
    = &\sum\nolimits_{ij} \alpha_i\alpha_j u_i v_j \mathrm{Cov}(\delta_i, \delta_j)\\
    = &\sum\nolimits_i \alpha_i^2 u_i v_i \cdot \sigma_i^2\\
    = &\sum\nolimits_i \alpha_i u_i v_i \cdot W^{-1}\\
    = &W^{-1} C_{uv}.
\end{split}
\end{equation*}
So the covariance matrix of $\sum\nolimits_i  \alpha_i \hat{c}_i \bm r_i$ should be $W^{-1}\bm C$.
The covariance matrix $\bm C$ is real and symmetric, 
    so its inverse matrix $\bm C^{-1}$ is also symmetric.
Thus
\begin{equation}
\begin{split}
    \mathrm{Cov}\left[\tilde{\bm g}\right] &= 
    \mathrm{Cov}\left[\bm C^{-1}\cdot \sum\nolimits_i \alpha_i \hat{c}_i \bm r_i\right] \\
    &= 
    \bm C^{-1} \cdot W^{-1} \bm C \cdot \left(\bm C^{-1}\right)^\mathrm{T} \\
    &=
    W^{-1} \bm C^{-1}.
\end{split}
    \label{eq:ap:est_cov}
\end{equation}

Alternatively, we can also calculate the Fisher information matrix 
\begin{equation*}
\begin{split}
    I_{uv} &= -\mathbb{E}\left[
        \frac{\partial^2}{\partial g_u\partial g_v} 
        \log \mathrm{PDF}\left(\{\hat{c}_i\}\,|\, c_0, \bm g, \{\sigma_i^2\}\right)
        \right]\\
        &= \mathbb{E}\left[W \sum\nolimits_i \alpha_i u_i v_i\right]\\
        &= W C_{uv}\\
        &\hspace*{6.5em}
            \bm I = W \bm C,
\end{split}
\end{equation*}
and the Cramér-Rao bound of $\tilde{\bm g}$ holds 
    $\mathrm{Cov}(\tilde{\bm g}) \ge \bm I^{-1} = W^{-1} \bm C^{-1}$. This also
    shows that our method gives the best uncertainty.

In fact, there is a much more concise way to get Eq.~\ref{eq:ap:loss} 
    without the introduction of Gaussian distribution assumption.
Define
\begin{equation*}
\begin{split}
    \bm R = \begin{bmatrix}
        1   &x_1    &y_1    &z_1\\
        1   &x_2    &y_2    &z_2\\
        1   &x_3    &y_3    &z_3\\
        \vdots  &\vdots &\vdots &\vdots\\
    \end{bmatrix},\quad
    \bm \beta = \begin{bmatrix}
        c_0 \\ g_x \\ g_y \\ g_z
    \end{bmatrix},\quad 
    \bm c = \begin{bmatrix}
        \hat{c}_1 \\ \hat{c}_2 \\ \hat{c}_3 \\ \vdots
    \end{bmatrix},\quad 
    \\
    \bm \Sigma = \begin{bmatrix}
        \sigma_1^2 &  0 &0  &\dots\\
        0   & \sigma_2^2 &0  &\dots\\
        0   &0  &\sigma_3^2 &\dots\\
        \vdots  & \vdots &\vdots &\ddots\\
    \end{bmatrix}
        = W^{-1}\begin{bmatrix}
            \alpha_1 &  0 &0  &\dots\\
            0   & \alpha_2 &0  &\dots\\
            0   &0  &\alpha_3 &\dots\\
            \vdots  & \vdots &\vdots &\ddots\\
    \end{bmatrix}^{-1}
\end{split}
\end{equation*}
This question can be expressed as 
    finding best unbiased linear estimator of $\tilde{\bm \beta}$
    for $\bm c = \bm R \bm \beta + \bm\delta$,
    where $\mathbb{E}[\bm \delta] = 0,\, 
    \mathrm{Cov}[\bm \delta]=\bm \Sigma$.
And the solution is given by\cite{aitken1936iv}
\begin{equation}
\begin{split}
    \tilde{\bm \beta} 
    &=\mathrm{argmin}_{\bm b}(\bm c-\bm R\bm b)^\mathrm{T} \bm\Sigma^{-1} (\bm c-\bm R\bm b)\\
    &=(\bm R^\mathrm{T} \bm \Sigma^{-1} \bm R)^{-1} \bm R^\mathrm{T} \bm \Sigma^{-1} \bm c
    ,\quad \\
    \mathrm{Cov}(\tilde{\bm \beta}) 
    &= (\bm R^\mathrm{T} \bm \Sigma^{-1} \bm R)^{-1}
    = W^{-1}\begin{bmatrix}
        1 & \bm0 \\
        \bm0 & \bm C
    \end{bmatrix}^{-1}.
\end{split}
    \label{eq:GLS_solution}
\end{equation}
More importantly, it can be proven that this estimator is
 the best linear unbiased estimator (BLUE), 
    holding the least uncertainty for gradient $\tilde{\bm g}$
    among all linear unbiased methods,
    even without Gaussian distribution assumption.
    
Eq.~\ref{eq:GLS_solution} also provides the concentration estimator 
    $\tilde{c}_0 = \sum\nolimits_i \alpha_i \hat{c}_i$ 
    under Eq.~\ref{eq:min_condition},
    while its uncertainty $\mathrm{var}(\tilde{c}_0)=W^{-1}$.
And it can be regarded as 
    not only the concentration uncertainty of origin point $\bm r=\bm 0$,
    but also the best achievable uncertainty for estimating the average concentration across the space,
    as shown in next section.

\section{The estimation for $\tilde{c}_0$}\label{sec:app:cO_est}

Consider the case if the cell only need to 
    estimate the average concentration $c_0$ (without any care of the gradient)
    through an unbiased linear mapping $\hat{c} = \sum\nolimits_i a_i \hat{c}_i$
    ($\sum\nolimits_i a_i=1$).
The uncertainty of $\hat{c}$ should be
\begin{equation*}
    \sigma_{c}^2
    =\mathrm{Cov}\left(\sum\nolimits_i a_i \hat{c}_i, \sum\nolimits_j a_j \hat{c}_j\right)
    = W^{-1} \sum\nolimits a_i^2 / \alpha_i.
\end{equation*}
Again, we use the independence hypothesis 
    over the estimation unit $\hat{c}_i\perp\!\!\!\perp \hat{c}_j$.
Using Lagrange multiplier method, 
    we define $L = W^{-1} \sum\nolimits a_i^2 / \alpha_i - \lambda \left(\sum\nolimits_i a_i-1\right)$. 
Therefore
\begin{equation*}
    \frac{\partial L}{\partial a_i} = 2 W^{-1} \frac{a_i}{\alpha_i} - \lambda = 0, 
\end{equation*}
which gives $a_i/\alpha_i = \mathrm{const}$,
    i.e. $a_i = \alpha_i$, and
    $\tilde{c}_0$ just gives the least uncertain estimation of the average concentration.

Note here, 
    the best measurement noise $\sigma_{c}^2 = W^{-1}$.
In other words, 
    factor $W^{-1}$ is just the uncertainty 
    for the concentration estimation.

\section{Bound from cell shape convex hull}\label{sec:app:Bound}

The total noise in the gradient estimation can be expressed as
    $\sigma_p^2 \sigma_q^2 \sigma_w^2 = \sigma_c^6 / |\bm C|$
    (or $\sigma_p^2 \sigma_q^2 = \sigma_c^4 / |\bm C|$
        for 2D cases).
Matrix $\bm C$ encodes information about the covariance between the positions of the local sensing units distribution,
    i.e. $\alpha_i = \sigma_i^{-2} / \sum\nolimits_j \sigma_j^{-2}$.
In other words, 
    covariance matrix $\bm C$ reflects how dispersed 
    the sensing units are in space.
Intuitively, the wider the sensing units are spread out (larger dispersion in the $\{\alpha_i, \bm r_i\}$ distribution), the larger the determinant of the covariance matrix, $|\bm C|$, is expected to be. This connection between the spreading and the determinant motivates the concept of convex hulls. The convex hull, denoted by $V_h$, represents the smallest convex set that encloses all the data points (positions of sensing units) in a space.
    Specifically, we show below that
    $|\bm C| < S_h^2 / 4$ for 2D cases 
    and $|\bm C|< V_h^2 \cdot 9/4$ for 3D cases.

\begin{figure}
    \centering
    \includegraphics[width=\linewidth]{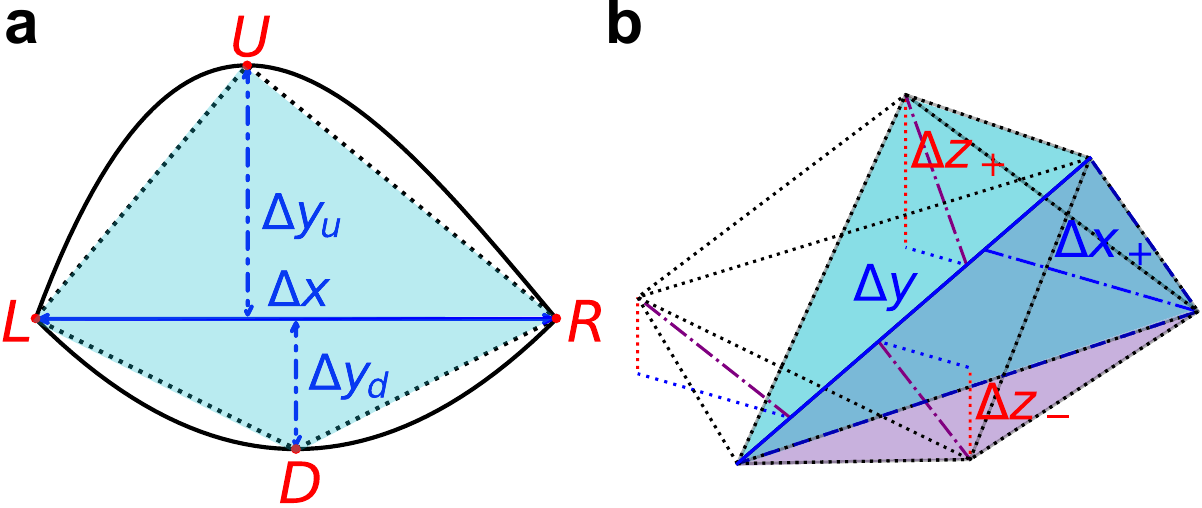}
    \caption{Convex Hull where all the sensing units located within.
    The black dotted lines give a convex polyhedral sub-region of the total shape, and the colored surfaces show the minimum volume(/area) used in our proof.
    (a) Convex hull in 2D space.
        The solid black lines denote an arbitrary 2D convex hull.
        $L, R, D, U$ are the points 
            with the maximum displacements 
            in $x_-, x_+, y_-, y_+$ directions, respectively.
    (b) Convex hull in 3D space.
        We use color blue to mark the lines in $xy$-plane,
            and red for lines in $z$-direction.}
    \label{fig:Bound}
\end{figure}

We first consider the 2D shape,
    which can be viewed as the projection of 3D cells onto a plane.
This projection reduces $\bm C$ to be a $2\times 2$ matrix.
As shown in Fig.~\ref{fig:Bound}a, 
    the solid black lines represents a 2D convex hull of the projected cell,
    with total area $S_h$ (i.e. volume in 2D space).
The distribution of weights, $\{\alpha_i\}$, is restricted to 
    lie within and on the surface of this area. This limitation plays a crucial role in bounding
    the determinant of the covariance matrix $\bm C$.

Considering the longest segment $LR$ as the direction of the $x$-axis. 
   For any point, $P$, within the convex hull, its $x$-coordinate must fall within the range $x_L \le x \le x_R$.
Therefore, $\mathrm{var}(x) \le \Delta x^2 / 4$, 
   where $\Delta x$ is the length of $LR$.
Similarly, let points $U$ and $D$ represent the maximum displacements  
   in the positive and negative y-directions, respectively. 
Following the same logic, we have $\mathrm{var}(y) \le \Delta y^2 / 4$,
   where $\Delta y = \Delta y_u + \Delta y_d$ and 
   $\Delta y_u, \Delta y_d$ is $U, D$'s displacements on $y$-axis.
Therefore, the determinant $\det \vert\bm C\vert = \mathrm{var}(x) \mathrm{var}(y) - \mathrm{cov}(x,y)^2$ is constrained to
\begin{equation*}
   \det \vert\bm C\vert 
   \le \mathrm{var}(x) \mathrm{var}(y)
   \le \Delta x^2 \Delta y^2 / 16.
\end{equation*}
The area of the convex hull, $S_h$, satisfies the following relation
\[
S_h \ge \Delta x (\Delta y_u + \Delta y_d) / 2 = 
   \Delta x \Delta y  / 2.
\]
Combining these two inequalities, we arrive at the final bound
\[
|\bm C| \le S_h^2/4.
\]    

For 3D scenario, as shown in Fig.~\ref{fig:Bound}b,
 we take the longest segment of the convex hull as $y$-axis, 
    which leads to a variance bound for the y-coordinate: $\mathrm{var}(y) < \Delta y^2/4$, where $\Delta y$ is the length of this segment.

For the $x$ and $z$-directions, we utilize polar coordinates on the projected $xz$-plane.   
 Here, we identify the line segment with 
        the largest radial displacement, denoted by $\Delta x_+$, and use it to define the positive $x$-axis orientation. 
Consequently, we obtain variance bounds for the $x$ and $z$-coordinates:
\[
\mathrm{var}(x) < \Delta x_+^2,\;\;\mathrm{var}(z) < (\Delta z_+ + \Delta z_-)^2/4.
\]
Combining these two variance bounds, we arrive at an inequality for the determinant
    \[|\bm C| \le \mathrm{var}(x)\mathrm{var}(y)\mathrm{var}(z)
    \le \Delta x_+^2 \Delta y^2 (\Delta z_+ + \Delta z_-)^2 / 16.\]   
Starting from the triangle formed by $\Delta y$ and $\Delta x_+$,
    we have \[V_h \ge \Delta x_+ \Delta y / 2 \cdot (\Delta z_+ + \Delta z_-)/3.\]   
Therefore, we get 
\[|\bm C| 
\le V_h^2\cdot 9 / 4\] 
for 3D cases.

\bibliography{ref}

\begin{thebibliography}{63}%
\makeatletter
\providecommand \@ifxundefined [1]{%
 \@ifx{#1\undefined}
}%
\providecommand \@ifnum [1]{%
 \ifnum #1\expandafter \@firstoftwo
 \else \expandafter \@secondoftwo
 \fi
}%
\providecommand \@ifx [1]{%
 \ifx #1\expandafter \@firstoftwo
 \else \expandafter \@secondoftwo
 \fi
}%
\providecommand \natexlab [1]{#1}%
\providecommand \enquote  [1]{``#1''}%
\providecommand \bibnamefont  [1]{#1}%
\providecommand \bibfnamefont [1]{#1}%
\providecommand \citenamefont [1]{#1}%
\providecommand \href@noop [0]{\@secondoftwo}%
\providecommand \href [0]{\begingroup \@sanitize@url \@href}%
\providecommand \@href[1]{\@@startlink{#1}\@@href}%
\providecommand \@@href[1]{\endgroup#1\@@endlink}%
\providecommand \@sanitize@url [0]{\catcode `\\12\catcode `\$12\catcode `\&12\catcode `\#12\catcode `\^12\catcode `\_12\catcode `\%12\relax}%
\providecommand \@@startlink[1]{}%
\providecommand \@@endlink[0]{}%
\providecommand \url  [0]{\begingroup\@sanitize@url \@url }%
\providecommand \@url [1]{\endgroup\@href {#1}{\urlprefix }}%
\providecommand \urlprefix  [0]{URL }%
\providecommand \Eprint [0]{\href }%
\providecommand \doibase [0]{https://doi.org/}%
\providecommand \selectlanguage [0]{\@gobble}%
\providecommand \bibinfo  [0]{\@secondoftwo}%
\providecommand \bibfield  [0]{\@secondoftwo}%
\providecommand \translation [1]{[#1]}%
\providecommand \BibitemOpen [0]{}%
\providecommand \bibitemStop [0]{}%
\providecommand \bibitemNoStop [0]{.\EOS\space}%
\providecommand \EOS [0]{\spacefactor3000\relax}%
\providecommand \BibitemShut  [1]{\csname bibitem#1\endcsname}%
\let\auto@bib@innerbib\@empty
%</preamble>
\bibitem [{\citenamefont {Van~Haastert}\ and\ \citenamefont {Devreotes}(2004)}]{van2004chemotaxis}%
  \BibitemOpen
  \bibfield  {author} {\bibinfo {author} {\bibfnamefont {P.~J.}\ \bibnamefont {Van~Haastert}}\ and\ \bibinfo {author} {\bibfnamefont {P.~N.}\ \bibnamefont {Devreotes}},\ }\bibfield  {title} {\bibinfo {title} {Chemotaxis: signalling the way forward},\ }\href@noop {} {\bibfield  {journal} {\bibinfo  {journal} {Nature reviews Molecular cell biology}\ }\textbf {\bibinfo {volume} {5}},\ \bibinfo {pages} {626} (\bibinfo {year} {2004})}\BibitemShut {NoStop}%
\bibitem [{\citenamefont {Segall}\ \emph {et~al.}(1986)\citenamefont {Segall}, \citenamefont {Block},\ and\ \citenamefont {Berg}}]{segall1986temporal}%
  \BibitemOpen
  \bibfield  {author} {\bibinfo {author} {\bibfnamefont {J.~E.}\ \bibnamefont {Segall}}, \bibinfo {author} {\bibfnamefont {S.~M.}\ \bibnamefont {Block}},\ and\ \bibinfo {author} {\bibfnamefont {H.~C.}\ \bibnamefont {Berg}},\ }\bibfield  {title} {\bibinfo {title} {Temporal comparisons in bacterial chemotaxis.},\ }\href@noop {} {\bibfield  {journal} {\bibinfo  {journal} {Proceedings of the National Academy of Sciences}\ }\textbf {\bibinfo {volume} {83}},\ \bibinfo {pages} {8987} (\bibinfo {year} {1986})}\BibitemShut {NoStop}%
\bibitem [{\citenamefont {Sourjik}\ and\ \citenamefont {Berg}(2002)}]{sourjik2002receptor}%
  \BibitemOpen
  \bibfield  {author} {\bibinfo {author} {\bibfnamefont {V.}~\bibnamefont {Sourjik}}\ and\ \bibinfo {author} {\bibfnamefont {H.~C.}\ \bibnamefont {Berg}},\ }\bibfield  {title} {\bibinfo {title} {Receptor sensitivity in bacterial chemotaxis},\ }\href@noop {} {\bibfield  {journal} {\bibinfo  {journal} {Proceedings of the National Academy of Sciences}\ }\textbf {\bibinfo {volume} {99}},\ \bibinfo {pages} {123} (\bibinfo {year} {2002})}\BibitemShut {NoStop}%
\bibitem [{\citenamefont {Swaney}\ \emph {et~al.}(2010)\citenamefont {Swaney}, \citenamefont {Huang},\ and\ \citenamefont {Devreotes}}]{swaney2010eukaryotic}%
  \BibitemOpen
  \bibfield  {author} {\bibinfo {author} {\bibfnamefont {K.~F.}\ \bibnamefont {Swaney}}, \bibinfo {author} {\bibfnamefont {C.-H.}\ \bibnamefont {Huang}},\ and\ \bibinfo {author} {\bibfnamefont {P.~N.}\ \bibnamefont {Devreotes}},\ }\bibfield  {title} {\bibinfo {title} {Eukaryotic chemotaxis: a network of signaling pathways controls motility, directional sensing, and polarity},\ }\href@noop {} {\bibfield  {journal} {\bibinfo  {journal} {Annual review of biophysics}\ }\textbf {\bibinfo {volume} {39}},\ \bibinfo {pages} {265} (\bibinfo {year} {2010})}\BibitemShut {NoStop}%
\bibitem [{\citenamefont {Song}\ \emph {et~al.}(2006)\citenamefont {Song}, \citenamefont {Nadkarni}, \citenamefont {B{\"o}deker}, \citenamefont {Beta}, \citenamefont {Bae}, \citenamefont {Franck}, \citenamefont {Rappel}, \citenamefont {Loomis},\ and\ \citenamefont {Bodenschatz}}]{song2006dictyostelium}%
  \BibitemOpen
  \bibfield  {author} {\bibinfo {author} {\bibfnamefont {L.}~\bibnamefont {Song}}, \bibinfo {author} {\bibfnamefont {S.~M.}\ \bibnamefont {Nadkarni}}, \bibinfo {author} {\bibfnamefont {H.~U.}\ \bibnamefont {B{\"o}deker}}, \bibinfo {author} {\bibfnamefont {C.}~\bibnamefont {Beta}}, \bibinfo {author} {\bibfnamefont {A.}~\bibnamefont {Bae}}, \bibinfo {author} {\bibfnamefont {C.}~\bibnamefont {Franck}}, \bibinfo {author} {\bibfnamefont {W.-J.}\ \bibnamefont {Rappel}}, \bibinfo {author} {\bibfnamefont {W.~F.}\ \bibnamefont {Loomis}},\ and\ \bibinfo {author} {\bibfnamefont {E.}~\bibnamefont {Bodenschatz}},\ }\bibfield  {title} {\bibinfo {title} {Dictyostelium discoideum chemotaxis: threshold for directed motion},\ }\href@noop {} {\bibfield  {journal} {\bibinfo  {journal} {European journal of cell biology}\ }\textbf {\bibinfo {volume} {85}},\ \bibinfo {pages} {981} (\bibinfo {year} {2006})}\BibitemShut {NoStop}%
\bibitem [{\citenamefont {Van~Haastert}\ and\ \citenamefont {Postma}(2007)}]{van2007biased}%
  \BibitemOpen
  \bibfield  {author} {\bibinfo {author} {\bibfnamefont {P.~J.}\ \bibnamefont {Van~Haastert}}\ and\ \bibinfo {author} {\bibfnamefont {M.}~\bibnamefont {Postma}},\ }\bibfield  {title} {\bibinfo {title} {Biased random walk by stochastic fluctuations of chemoattractant-receptor interactions at the lower limit of detection},\ }\href@noop {} {\bibfield  {journal} {\bibinfo  {journal} {Biophysical journal}\ }\textbf {\bibinfo {volume} {93}},\ \bibinfo {pages} {1787} (\bibinfo {year} {2007})}\BibitemShut {NoStop}%
\bibitem [{\citenamefont {Ueda}\ and\ \citenamefont {Shibata}(2007)}]{ueda2007stochastic}%
  \BibitemOpen
  \bibfield  {author} {\bibinfo {author} {\bibfnamefont {M.}~\bibnamefont {Ueda}}\ and\ \bibinfo {author} {\bibfnamefont {T.}~\bibnamefont {Shibata}},\ }\bibfield  {title} {\bibinfo {title} {Stochastic signal processing and transduction in chemotactic response of eukaryotic cells},\ }\href@noop {} {\bibfield  {journal} {\bibinfo  {journal} {Biophysical journal}\ }\textbf {\bibinfo {volume} {93}},\ \bibinfo {pages} {11} (\bibinfo {year} {2007})}\BibitemShut {NoStop}%
\bibitem [{\citenamefont {Fuller}\ \emph {et~al.}(2010)\citenamefont {Fuller}, \citenamefont {Chen}, \citenamefont {Adler}, \citenamefont {Groisman}, \citenamefont {Levine}, \citenamefont {Rappel},\ and\ \citenamefont {Loomis}}]{fuller2010external}%
  \BibitemOpen
  \bibfield  {author} {\bibinfo {author} {\bibfnamefont {D.}~\bibnamefont {Fuller}}, \bibinfo {author} {\bibfnamefont {W.}~\bibnamefont {Chen}}, \bibinfo {author} {\bibfnamefont {M.}~\bibnamefont {Adler}}, \bibinfo {author} {\bibfnamefont {A.}~\bibnamefont {Groisman}}, \bibinfo {author} {\bibfnamefont {H.}~\bibnamefont {Levine}}, \bibinfo {author} {\bibfnamefont {W.-J.}\ \bibnamefont {Rappel}},\ and\ \bibinfo {author} {\bibfnamefont {W.~F.}\ \bibnamefont {Loomis}},\ }\bibfield  {title} {\bibinfo {title} {External and internal constraints on eukaryotic chemotaxis},\ }\href@noop {} {\bibfield  {journal} {\bibinfo  {journal} {Proceedings of the National Academy of Sciences}\ }\textbf {\bibinfo {volume} {107}},\ \bibinfo {pages} {9656} (\bibinfo {year} {2010})}\BibitemShut {NoStop}%
\bibitem [{\citenamefont {Endres}\ and\ \citenamefont {Wingreen}(2008)}]{endres2008accuracy}%
  \BibitemOpen
  \bibfield  {author} {\bibinfo {author} {\bibfnamefont {R.~G.}\ \bibnamefont {Endres}}\ and\ \bibinfo {author} {\bibfnamefont {N.~S.}\ \bibnamefont {Wingreen}},\ }\bibfield  {title} {\bibinfo {title} {Accuracy of direct gradient sensing by single cells},\ }\href@noop {} {\bibfield  {journal} {\bibinfo  {journal} {Proceedings of the National Academy of Sciences}\ }\textbf {\bibinfo {volume} {105}},\ \bibinfo {pages} {15749} (\bibinfo {year} {2008})}\BibitemShut {NoStop}%
\bibitem [{\citenamefont {Endres}\ and\ \citenamefont {Wingreen}(2009{\natexlab{a}})}]{endres2009accuracy}%
  \BibitemOpen
  \bibfield  {author} {\bibinfo {author} {\bibfnamefont {R.~G.}\ \bibnamefont {Endres}}\ and\ \bibinfo {author} {\bibfnamefont {N.~S.}\ \bibnamefont {Wingreen}},\ }\bibfield  {title} {\bibinfo {title} {Accuracy of direct gradient sensing by cell-surface receptors},\ }\href@noop {} {\bibfield  {journal} {\bibinfo  {journal} {Progress in biophysics and molecular biology}\ }\textbf {\bibinfo {volume} {100}},\ \bibinfo {pages} {33} (\bibinfo {year} {2009}{\natexlab{a}})}\BibitemShut {NoStop}%
\bibitem [{\citenamefont {Mora}\ and\ \citenamefont {Wingreen}(2010)}]{mora2010limits}%
  \BibitemOpen
  \bibfield  {author} {\bibinfo {author} {\bibfnamefont {T.}~\bibnamefont {Mora}}\ and\ \bibinfo {author} {\bibfnamefont {N.~S.}\ \bibnamefont {Wingreen}},\ }\bibfield  {title} {\bibinfo {title} {Limits of sensing temporal concentration changes by single cells},\ }\href@noop {} {\bibfield  {journal} {\bibinfo  {journal} {Physical review letters}\ }\textbf {\bibinfo {volume} {104}},\ \bibinfo {pages} {248101} (\bibinfo {year} {2010})}\BibitemShut {NoStop}%
\bibitem [{\citenamefont {Lalanne}\ and\ \citenamefont {Fran{\c{c}}ois}(2015)}]{lalanne2015chemodetection}%
  \BibitemOpen
  \bibfield  {author} {\bibinfo {author} {\bibfnamefont {J.-B.}\ \bibnamefont {Lalanne}}\ and\ \bibinfo {author} {\bibfnamefont {P.}~\bibnamefont {Fran{\c{c}}ois}},\ }\bibfield  {title} {\bibinfo {title} {Chemodetection in fluctuating environments: Receptor coupling, buffering, and antagonism},\ }\href@noop {} {\bibfield  {journal} {\bibinfo  {journal} {Proceedings of the National Academy of Sciences}\ }\textbf {\bibinfo {volume} {112}},\ \bibinfo {pages} {1898} (\bibinfo {year} {2015})}\BibitemShut {NoStop}%
\bibitem [{\citenamefont {Berg}\ and\ \citenamefont {Purcell}(1977)}]{berg1977physics}%
  \BibitemOpen
  \bibfield  {author} {\bibinfo {author} {\bibfnamefont {H.~C.}\ \bibnamefont {Berg}}\ and\ \bibinfo {author} {\bibfnamefont {E.~M.}\ \bibnamefont {Purcell}},\ }\bibfield  {title} {\bibinfo {title} {Physics of chemoreception},\ }\href@noop {} {\bibfield  {journal} {\bibinfo  {journal} {Biophysical journal}\ }\textbf {\bibinfo {volume} {20}},\ \bibinfo {pages} {193} (\bibinfo {year} {1977})}\BibitemShut {NoStop}%
\bibitem [{\citenamefont {Hu}\ \emph {et~al.}(2010)\citenamefont {Hu}, \citenamefont {Chen}, \citenamefont {Rappel},\ and\ \citenamefont {Levine}}]{hu2010physical}%
  \BibitemOpen
  \bibfield  {author} {\bibinfo {author} {\bibfnamefont {B.}~\bibnamefont {Hu}}, \bibinfo {author} {\bibfnamefont {W.}~\bibnamefont {Chen}}, \bibinfo {author} {\bibfnamefont {W.-J.}\ \bibnamefont {Rappel}},\ and\ \bibinfo {author} {\bibfnamefont {H.}~\bibnamefont {Levine}},\ }\bibfield  {title} {\bibinfo {title} {Physical limits on cellular sensing of spatial gradients},\ }\href@noop {} {\bibfield  {journal} {\bibinfo  {journal} {Physical review letters}\ }\textbf {\bibinfo {volume} {105}},\ \bibinfo {pages} {048104} (\bibinfo {year} {2010})}\BibitemShut {NoStop}%
\bibitem [{\citenamefont {Segota}\ and\ \citenamefont {Franck}(2017)}]{segota2017extracellular}%
  \BibitemOpen
  \bibfield  {author} {\bibinfo {author} {\bibfnamefont {I.}~\bibnamefont {Segota}}\ and\ \bibinfo {author} {\bibfnamefont {C.}~\bibnamefont {Franck}},\ }\bibfield  {title} {\bibinfo {title} {Extracellular processing of molecular gradients by eukaryotic cells can improve gradient detection accuracy},\ }\href@noop {} {\bibfield  {journal} {\bibinfo  {journal} {Physical Review Letters}\ }\textbf {\bibinfo {volume} {119}},\ \bibinfo {pages} {248101} (\bibinfo {year} {2017})}\BibitemShut {NoStop}%
\bibitem [{\citenamefont {Iijima}\ \emph {et~al.}(2002)\citenamefont {Iijima}, \citenamefont {Huang},\ and\ \citenamefont {Devreotes}}]{iijima2002temporal}%
  \BibitemOpen
  \bibfield  {author} {\bibinfo {author} {\bibfnamefont {M.}~\bibnamefont {Iijima}}, \bibinfo {author} {\bibfnamefont {Y.~E.}\ \bibnamefont {Huang}},\ and\ \bibinfo {author} {\bibfnamefont {P.}~\bibnamefont {Devreotes}},\ }\bibfield  {title} {\bibinfo {title} {Temporal and spatial regulation of chemotaxis},\ }\href@noop {} {\bibfield  {journal} {\bibinfo  {journal} {Developmental cell}\ }\textbf {\bibinfo {volume} {3}},\ \bibinfo {pages} {469} (\bibinfo {year} {2002})}\BibitemShut {NoStop}%
\bibitem [{\citenamefont {Rappel}\ and\ \citenamefont {Levine}(2008)}]{rappel2008receptor}%
  \BibitemOpen
  \bibfield  {author} {\bibinfo {author} {\bibfnamefont {W.-J.}\ \bibnamefont {Rappel}}\ and\ \bibinfo {author} {\bibfnamefont {H.}~\bibnamefont {Levine}},\ }\bibfield  {title} {\bibinfo {title} {Receptor noise limitations on chemotactic sensing},\ }\href@noop {} {\bibfield  {journal} {\bibinfo  {journal} {Proceedings of the National Academy of Sciences}\ }\textbf {\bibinfo {volume} {105}},\ \bibinfo {pages} {19270} (\bibinfo {year} {2008})}\BibitemShut {NoStop}%
\bibitem [{\citenamefont {Iglesias}\ and\ \citenamefont {Devreotes}(2008)}]{iglesias2008navigating}%
  \BibitemOpen
  \bibfield  {author} {\bibinfo {author} {\bibfnamefont {P.~A.}\ \bibnamefont {Iglesias}}\ and\ \bibinfo {author} {\bibfnamefont {P.~N.}\ \bibnamefont {Devreotes}},\ }\bibfield  {title} {\bibinfo {title} {Navigating through models of chemotaxis},\ }\href@noop {} {\bibfield  {journal} {\bibinfo  {journal} {Current opinion in cell biology}\ }\textbf {\bibinfo {volume} {20}},\ \bibinfo {pages} {35} (\bibinfo {year} {2008})}\BibitemShut {NoStop}%
\bibitem [{\citenamefont {Andrew}\ and\ \citenamefont {Insall}(2007)}]{andrew2007chemotaxis}%
  \BibitemOpen
  \bibfield  {author} {\bibinfo {author} {\bibfnamefont {N.}~\bibnamefont {Andrew}}\ and\ \bibinfo {author} {\bibfnamefont {R.~H.}\ \bibnamefont {Insall}},\ }\bibfield  {title} {\bibinfo {title} {Chemotaxis in shallow gradients is mediated independently of ptdins 3-kinase by biased choices between random protrusions},\ }\href@noop {} {\bibfield  {journal} {\bibinfo  {journal} {Nature cell biology}\ }\textbf {\bibinfo {volume} {9}},\ \bibinfo {pages} {193} (\bibinfo {year} {2007})}\BibitemShut {NoStop}%
\bibitem [{\citenamefont {Skoge}\ \emph {et~al.}(2010)\citenamefont {Skoge}, \citenamefont {Adler}, \citenamefont {Groisman}, \citenamefont {Levine}, \citenamefont {Loomis},\ and\ \citenamefont {Rappel}}]{skoge2010gradient}%
  \BibitemOpen
  \bibfield  {author} {\bibinfo {author} {\bibfnamefont {M.}~\bibnamefont {Skoge}}, \bibinfo {author} {\bibfnamefont {M.}~\bibnamefont {Adler}}, \bibinfo {author} {\bibfnamefont {A.}~\bibnamefont {Groisman}}, \bibinfo {author} {\bibfnamefont {H.}~\bibnamefont {Levine}}, \bibinfo {author} {\bibfnamefont {W.~F.}\ \bibnamefont {Loomis}},\ and\ \bibinfo {author} {\bibfnamefont {W.-J.}\ \bibnamefont {Rappel}},\ }\bibfield  {title} {\bibinfo {title} {Gradient sensing in defined chemotactic fields},\ }\href@noop {} {\bibfield  {journal} {\bibinfo  {journal} {Integrative Biology}\ }\textbf {\bibinfo {volume} {2}},\ \bibinfo {pages} {659} (\bibinfo {year} {2010})}\BibitemShut {NoStop}%
\bibitem [{\citenamefont {Miao}\ \emph {et~al.}(2017)\citenamefont {Miao}, \citenamefont {Bhattacharya}, \citenamefont {Edwards}, \citenamefont {Cai}, \citenamefont {Inoue}, \citenamefont {Iglesias},\ and\ \citenamefont {Devreotes}}]{miao2017altering}%
  \BibitemOpen
  \bibfield  {author} {\bibinfo {author} {\bibfnamefont {Y.}~\bibnamefont {Miao}}, \bibinfo {author} {\bibfnamefont {S.}~\bibnamefont {Bhattacharya}}, \bibinfo {author} {\bibfnamefont {M.}~\bibnamefont {Edwards}}, \bibinfo {author} {\bibfnamefont {H.}~\bibnamefont {Cai}}, \bibinfo {author} {\bibfnamefont {T.}~\bibnamefont {Inoue}}, \bibinfo {author} {\bibfnamefont {P.~A.}\ \bibnamefont {Iglesias}},\ and\ \bibinfo {author} {\bibfnamefont {P.~N.}\ \bibnamefont {Devreotes}},\ }\bibfield  {title} {\bibinfo {title} {Altering the threshold of an excitable signal transduction network changes cell migratory modes},\ }\href@noop {} {\bibfield  {journal} {\bibinfo  {journal} {Nature cell biology}\ }\textbf {\bibinfo {volume} {19}},\ \bibinfo {pages} {329} (\bibinfo {year} {2017})}\BibitemShut {NoStop}%
\bibitem [{\citenamefont {Camley}\ \emph {et~al.}(2016)\citenamefont {Camley}, \citenamefont {Zimmermann}, \citenamefont {Levine},\ and\ \citenamefont {Rappel}}]{camley2016emergent}%
  \BibitemOpen
  \bibfield  {author} {\bibinfo {author} {\bibfnamefont {B.~A.}\ \bibnamefont {Camley}}, \bibinfo {author} {\bibfnamefont {J.}~\bibnamefont {Zimmermann}}, \bibinfo {author} {\bibfnamefont {H.}~\bibnamefont {Levine}},\ and\ \bibinfo {author} {\bibfnamefont {W.-J.}\ \bibnamefont {Rappel}},\ }\bibfield  {title} {\bibinfo {title} {Emergent collective chemotaxis without single-cell gradient sensing},\ }\href@noop {} {\bibfield  {journal} {\bibinfo  {journal} {Physical review letters}\ }\textbf {\bibinfo {volume} {116}},\ \bibinfo {pages} {098101} (\bibinfo {year} {2016})}\BibitemShut {NoStop}%
\bibitem [{\citenamefont {Ellison}\ \emph {et~al.}(2016)\citenamefont {Ellison}, \citenamefont {Mugler}, \citenamefont {Brennan}, \citenamefont {Lee}, \citenamefont {Huebner}, \citenamefont {Shamir}, \citenamefont {Woo}, \citenamefont {Kim}, \citenamefont {Amar}, \citenamefont {Nemenman} \emph {et~al.}}]{ellison2016cell}%
  \BibitemOpen
  \bibfield  {author} {\bibinfo {author} {\bibfnamefont {D.}~\bibnamefont {Ellison}}, \bibinfo {author} {\bibfnamefont {A.}~\bibnamefont {Mugler}}, \bibinfo {author} {\bibfnamefont {M.~D.}\ \bibnamefont {Brennan}}, \bibinfo {author} {\bibfnamefont {S.~H.}\ \bibnamefont {Lee}}, \bibinfo {author} {\bibfnamefont {R.~J.}\ \bibnamefont {Huebner}}, \bibinfo {author} {\bibfnamefont {E.~R.}\ \bibnamefont {Shamir}}, \bibinfo {author} {\bibfnamefont {L.~A.}\ \bibnamefont {Woo}}, \bibinfo {author} {\bibfnamefont {J.}~\bibnamefont {Kim}}, \bibinfo {author} {\bibfnamefont {P.}~\bibnamefont {Amar}}, \bibinfo {author} {\bibfnamefont {I.}~\bibnamefont {Nemenman}}, \emph {et~al.},\ }\bibfield  {title} {\bibinfo {title} {Cell--cell communication enhances the capacity of cell ensembles to sense shallow gradients during morphogenesis},\ }\href@noop {} {\bibfield  {journal} {\bibinfo  {journal} {Proceedings of the National Academy of Sciences}\ }\textbf {\bibinfo {volume} {113}},\ \bibinfo {pages} {E679} (\bibinfo {year}
  {2016})}\BibitemShut {NoStop}%
\bibitem [{\citenamefont {Bialek}\ and\ \citenamefont {Setayeshgar}(2005)}]{bialek2005physical}%
  \BibitemOpen
  \bibfield  {author} {\bibinfo {author} {\bibfnamefont {W.}~\bibnamefont {Bialek}}\ and\ \bibinfo {author} {\bibfnamefont {S.}~\bibnamefont {Setayeshgar}},\ }\bibfield  {title} {\bibinfo {title} {Physical limits to biochemical signaling},\ }\href@noop {} {\bibfield  {journal} {\bibinfo  {journal} {Proceedings of the National Academy of Sciences}\ }\textbf {\bibinfo {volume} {102}},\ \bibinfo {pages} {10040} (\bibinfo {year} {2005})}\BibitemShut {NoStop}%
\bibitem [{\citenamefont {Hopkins}\ and\ \citenamefont {Camley}(2020)}]{hopkins2020chemotaxis}%
  \BibitemOpen
  \bibfield  {author} {\bibinfo {author} {\bibfnamefont {A.}~\bibnamefont {Hopkins}}\ and\ \bibinfo {author} {\bibfnamefont {B.~A.}\ \bibnamefont {Camley}},\ }\bibfield  {title} {\bibinfo {title} {Chemotaxis in uncertain environments: Hedging bets with multiple receptor types},\ }\href@noop {} {\bibfield  {journal} {\bibinfo  {journal} {Physical Review Research}\ }\textbf {\bibinfo {volume} {2}},\ \bibinfo {pages} {043146} (\bibinfo {year} {2020})}\BibitemShut {NoStop}%
\bibitem [{\citenamefont {Endres}\ and\ \citenamefont {Wingreen}(2009{\natexlab{b}})}]{endres2009maximum}%
  \BibitemOpen
  \bibfield  {author} {\bibinfo {author} {\bibfnamefont {R.~G.}\ \bibnamefont {Endres}}\ and\ \bibinfo {author} {\bibfnamefont {N.~S.}\ \bibnamefont {Wingreen}},\ }\bibfield  {title} {\bibinfo {title} {Maximum likelihood and the single receptor},\ }\href@noop {} {\bibfield  {journal} {\bibinfo  {journal} {Physical review letters}\ }\textbf {\bibinfo {volume} {103}},\ \bibinfo {pages} {158101} (\bibinfo {year} {2009}{\natexlab{b}})}\BibitemShut {NoStop}%
\bibitem [{\citenamefont {Camley}\ and\ \citenamefont {Rappel}(2017)}]{camley2017cell}%
  \BibitemOpen
  \bibfield  {author} {\bibinfo {author} {\bibfnamefont {B.~A.}\ \bibnamefont {Camley}}\ and\ \bibinfo {author} {\bibfnamefont {W.-J.}\ \bibnamefont {Rappel}},\ }\bibfield  {title} {\bibinfo {title} {Cell-to-cell variation sets a tissue-rheology--dependent bound on collective gradient sensing},\ }\href@noop {} {\bibfield  {journal} {\bibinfo  {journal} {Proceedings of the National Academy of Sciences}\ }\textbf {\bibinfo {volume} {114}},\ \bibinfo {pages} {E10074} (\bibinfo {year} {2017})}\BibitemShut {NoStop}%
\bibitem [{\citenamefont {Ipi{\~n}a}\ and\ \citenamefont {Camley}(2022)}]{ipina2022collective}%
  \BibitemOpen
  \bibfield  {author} {\bibinfo {author} {\bibfnamefont {E.~P.}\ \bibnamefont {Ipi{\~n}a}}\ and\ \bibinfo {author} {\bibfnamefont {B.~A.}\ \bibnamefont {Camley}},\ }\bibfield  {title} {\bibinfo {title} {Collective gradient sensing with limited positional information},\ }\href@noop {} {\bibfield  {journal} {\bibinfo  {journal} {Physical Review E}\ }\textbf {\bibinfo {volume} {105}},\ \bibinfo {pages} {044410} (\bibinfo {year} {2022})}\BibitemShut {NoStop}%
\bibitem [{\citenamefont {Govern}\ and\ \citenamefont {ten Wolde}(2012)}]{govern2012fundamental}%
  \BibitemOpen
  \bibfield  {author} {\bibinfo {author} {\bibfnamefont {C.~C.}\ \bibnamefont {Govern}}\ and\ \bibinfo {author} {\bibfnamefont {P.~R.}\ \bibnamefont {ten Wolde}},\ }\bibfield  {title} {\bibinfo {title} {Fundamental limits on sensing chemical concentrations with linear biochemical networks},\ }\href@noop {} {\bibfield  {journal} {\bibinfo  {journal} {Physical review letters}\ }\textbf {\bibinfo {volume} {109}},\ \bibinfo {pages} {218103} (\bibinfo {year} {2012})}\BibitemShut {NoStop}%
\bibitem [{\citenamefont {Tu}(2013)}]{tu2013quantitative}%
  \BibitemOpen
  \bibfield  {author} {\bibinfo {author} {\bibfnamefont {Y.}~\bibnamefont {Tu}},\ }\bibfield  {title} {\bibinfo {title} {Quantitative modeling of bacterial chemotaxis: signal amplification and accurate adaptation},\ }\href@noop {} {\bibfield  {journal} {\bibinfo  {journal} {Annual review of biophysics}\ }\textbf {\bibinfo {volume} {42}},\ \bibinfo {pages} {337} (\bibinfo {year} {2013})}\BibitemShut {NoStop}%
\bibitem [{\citenamefont {Owen}\ and\ \citenamefont {Horowitz}(2023)}]{owen2023size}%
  \BibitemOpen
  \bibfield  {author} {\bibinfo {author} {\bibfnamefont {J.~A.}\ \bibnamefont {Owen}}\ and\ \bibinfo {author} {\bibfnamefont {J.~M.}\ \bibnamefont {Horowitz}},\ }\bibfield  {title} {\bibinfo {title} {Size limits the sensitivity of kinetic schemes},\ }\href@noop {} {\bibfield  {journal} {\bibinfo  {journal} {Nature Communications}\ }\textbf {\bibinfo {volume} {14}},\ \bibinfo {pages} {1280} (\bibinfo {year} {2023})}\BibitemShut {NoStop}%
\bibitem [{\citenamefont {Lang}\ \emph {et~al.}(2014)\citenamefont {Lang}, \citenamefont {Fisher}, \citenamefont {Mora},\ and\ \citenamefont {Mehta}}]{lang2014thermodynamics}%
  \BibitemOpen
  \bibfield  {author} {\bibinfo {author} {\bibfnamefont {A.~H.}\ \bibnamefont {Lang}}, \bibinfo {author} {\bibfnamefont {C.~K.}\ \bibnamefont {Fisher}}, \bibinfo {author} {\bibfnamefont {T.}~\bibnamefont {Mora}},\ and\ \bibinfo {author} {\bibfnamefont {P.}~\bibnamefont {Mehta}},\ }\bibfield  {title} {\bibinfo {title} {Thermodynamics of statistical inference by cells},\ }\href@noop {} {\bibfield  {journal} {\bibinfo  {journal} {Physical review letters}\ }\textbf {\bibinfo {volume} {113}},\ \bibinfo {pages} {148103} (\bibinfo {year} {2014})}\BibitemShut {NoStop}%
\bibitem [{\citenamefont {Govern}\ and\ \citenamefont {ten Wolde}(2014)}]{govern2014energy}%
  \BibitemOpen
  \bibfield  {author} {\bibinfo {author} {\bibfnamefont {C.~C.}\ \bibnamefont {Govern}}\ and\ \bibinfo {author} {\bibfnamefont {P.~R.}\ \bibnamefont {ten Wolde}},\ }\bibfield  {title} {\bibinfo {title} {Energy dissipation and noise correlations in biochemical sensing},\ }\href@noop {} {\bibfield  {journal} {\bibinfo  {journal} {Physical review letters}\ }\textbf {\bibinfo {volume} {113}},\ \bibinfo {pages} {258102} (\bibinfo {year} {2014})}\BibitemShut {NoStop}%
\bibitem [{\citenamefont {Owen}\ \emph {et~al.}(2020)\citenamefont {Owen}, \citenamefont {Gingrich},\ and\ \citenamefont {Horowitz}}]{owen2020universal}%
  \BibitemOpen
  \bibfield  {author} {\bibinfo {author} {\bibfnamefont {J.~A.}\ \bibnamefont {Owen}}, \bibinfo {author} {\bibfnamefont {T.~R.}\ \bibnamefont {Gingrich}},\ and\ \bibinfo {author} {\bibfnamefont {J.~M.}\ \bibnamefont {Horowitz}},\ }\bibfield  {title} {\bibinfo {title} {Universal thermodynamic bounds on nonequilibrium response with biochemical applications},\ }\href@noop {} {\bibfield  {journal} {\bibinfo  {journal} {Physical Review X}\ }\textbf {\bibinfo {volume} {10}},\ \bibinfo {pages} {011066} (\bibinfo {year} {2020})}\BibitemShut {NoStop}%
\bibitem [{sup()}]{supp}%
  \BibitemOpen
  \href@noop {} {}\bibinfo {note} {See Supplemental Material at [URL will be inserted by publisher] for additional details on numerical methods, analytic derivation, and algorithm.}\BibitemShut {Stop}%
\bibitem [{\citenamefont {Cai}\ \emph {et~al.}(2022)\citenamefont {Cai}, \citenamefont {Beppler}, \citenamefont {Eichorst}, \citenamefont {Marchuk}, \citenamefont {Eastman},\ and\ \citenamefont {Krummel}}]{cai2022t}%
  \BibitemOpen
  \bibfield  {author} {\bibinfo {author} {\bibfnamefont {E.}~\bibnamefont {Cai}}, \bibinfo {author} {\bibfnamefont {C.}~\bibnamefont {Beppler}}, \bibinfo {author} {\bibfnamefont {J.}~\bibnamefont {Eichorst}}, \bibinfo {author} {\bibfnamefont {K.}~\bibnamefont {Marchuk}}, \bibinfo {author} {\bibfnamefont {S.~W.}\ \bibnamefont {Eastman}},\ and\ \bibinfo {author} {\bibfnamefont {M.~F.}\ \bibnamefont {Krummel}},\ }\bibfield  {title} {\bibinfo {title} {T cells use distinct topographical and membrane receptor scanning strategies that individually coalesce during receptor recognition},\ }\href@noop {} {\bibfield  {journal} {\bibinfo  {journal} {Proceedings of the National Academy of Sciences}\ }\textbf {\bibinfo {volume} {119}},\ \bibinfo {pages} {e2203247119} (\bibinfo {year} {2022})}\BibitemShut {NoStop}%
\bibitem [{\citenamefont {Yang}\ \emph {et~al.}(2022)\citenamefont {Yang}, \citenamefont {Miao}, \citenamefont {Arnold}, \citenamefont {Li}, \citenamefont {Ly}, \citenamefont {Wang}, \citenamefont {Wang}, \citenamefont {Guo}, \citenamefont {Pathak}, \citenamefont {Zhao} \emph {et~al.}}]{yang2022membrane}%
  \BibitemOpen
  \bibfield  {author} {\bibinfo {author} {\bibfnamefont {S.}~\bibnamefont {Yang}}, \bibinfo {author} {\bibfnamefont {X.}~\bibnamefont {Miao}}, \bibinfo {author} {\bibfnamefont {S.}~\bibnamefont {Arnold}}, \bibinfo {author} {\bibfnamefont {B.}~\bibnamefont {Li}}, \bibinfo {author} {\bibfnamefont {A.~T.}\ \bibnamefont {Ly}}, \bibinfo {author} {\bibfnamefont {H.}~\bibnamefont {Wang}}, \bibinfo {author} {\bibfnamefont {M.}~\bibnamefont {Wang}}, \bibinfo {author} {\bibfnamefont {X.}~\bibnamefont {Guo}}, \bibinfo {author} {\bibfnamefont {M.~M.}\ \bibnamefont {Pathak}}, \bibinfo {author} {\bibfnamefont {W.}~\bibnamefont {Zhao}}, \emph {et~al.},\ }\bibfield  {title} {\bibinfo {title} {Membrane curvature governs the distribution of piezo1 in live cells},\ }\href@noop {} {\bibfield  {journal} {\bibinfo  {journal} {Nature communications}\ }\textbf {\bibinfo {volume} {13}},\ \bibinfo {pages} {7467} (\bibinfo {year} {2022})}\BibitemShut {NoStop}%
\bibitem [{\citenamefont {Xiao}\ \emph {et~al.}(1997)\citenamefont {Xiao}, \citenamefont {Zhang}, \citenamefont {Murphy},\ and\ \citenamefont {Devreotes}}]{xiao1997dynamic}%
  \BibitemOpen
  \bibfield  {author} {\bibinfo {author} {\bibfnamefont {Z.}~\bibnamefont {Xiao}}, \bibinfo {author} {\bibfnamefont {N.}~\bibnamefont {Zhang}}, \bibinfo {author} {\bibfnamefont {D.~B.}\ \bibnamefont {Murphy}},\ and\ \bibinfo {author} {\bibfnamefont {P.~N.}\ \bibnamefont {Devreotes}},\ }\bibfield  {title} {\bibinfo {title} {Dynamic distribution of chemoattractant receptors in living cells during chemotaxis and persistent stimulation},\ }\href@noop {} {\bibfield  {journal} {\bibinfo  {journal} {The Journal of cell biology}\ }\textbf {\bibinfo {volume} {139}},\ \bibinfo {pages} {365} (\bibinfo {year} {1997})}\BibitemShut {NoStop}%
\bibitem [{\citenamefont {Alonso}\ \emph {et~al.}(2024{\natexlab{a}})\citenamefont {Alonso}, \citenamefont {Endres},\ and\ \citenamefont {Kirkegaard}}]{alonso2024receptors}%
  \BibitemOpen
  \bibfield  {author} {\bibinfo {author} {\bibfnamefont {A.}~\bibnamefont {Alonso}}, \bibinfo {author} {\bibfnamefont {R.~G.}\ \bibnamefont {Endres}},\ and\ \bibinfo {author} {\bibfnamefont {J.~B.}\ \bibnamefont {Kirkegaard}},\ }\bibfield  {title} {\bibinfo {title} {Receptors cluster in high-curvature membrane regions for optimal spatial gradient sensing},\ }\href@noop {} {\bibfield  {journal} {\bibinfo  {journal} {arXiv preprint arXiv:2410.03395}\ } (\bibinfo {year} {2024}{\natexlab{a}})}\BibitemShut {NoStop}%
\bibitem [{\citenamefont {Wang}\ \emph {et~al.}(2013)\citenamefont {Wang}, \citenamefont {Senoo}, \citenamefont {Sesaki},\ and\ \citenamefont {Iijima}}]{wang2013rho}%
  \BibitemOpen
  \bibfield  {author} {\bibinfo {author} {\bibfnamefont {Y.}~\bibnamefont {Wang}}, \bibinfo {author} {\bibfnamefont {H.}~\bibnamefont {Senoo}}, \bibinfo {author} {\bibfnamefont {H.}~\bibnamefont {Sesaki}},\ and\ \bibinfo {author} {\bibfnamefont {M.}~\bibnamefont {Iijima}},\ }\bibfield  {title} {\bibinfo {title} {Rho gtpases orient directional sensing in chemotaxis},\ }\href@noop {} {\bibfield  {journal} {\bibinfo  {journal} {Proceedings of the National Academy of Sciences}\ }\textbf {\bibinfo {volume} {110}},\ \bibinfo {pages} {E4723} (\bibinfo {year} {2013})}\BibitemShut {NoStop}%
\bibitem [{\citenamefont {Wu}\ \emph {et~al.}(2014)\citenamefont {Wu}, \citenamefont {Giri}, \citenamefont {Sun},\ and\ \citenamefont {Wirtz}}]{wu2014three}%
  \BibitemOpen
  \bibfield  {author} {\bibinfo {author} {\bibfnamefont {P.-H.}\ \bibnamefont {Wu}}, \bibinfo {author} {\bibfnamefont {A.}~\bibnamefont {Giri}}, \bibinfo {author} {\bibfnamefont {S.~X.}\ \bibnamefont {Sun}},\ and\ \bibinfo {author} {\bibfnamefont {D.}~\bibnamefont {Wirtz}},\ }\bibfield  {title} {\bibinfo {title} {Three-dimensional cell migration does not follow a random walk},\ }\href@noop {} {\bibfield  {journal} {\bibinfo  {journal} {Proceedings of the National Academy of Sciences}\ }\textbf {\bibinfo {volume} {111}},\ \bibinfo {pages} {3949} (\bibinfo {year} {2014})}\BibitemShut {NoStop}%
\bibitem [{\citenamefont {Aquino}\ \emph {et~al.}(2014)\citenamefont {Aquino}, \citenamefont {Tweedy}, \citenamefont {Heinrich},\ and\ \citenamefont {Endres}}]{aquino2014memory}%
  \BibitemOpen
  \bibfield  {author} {\bibinfo {author} {\bibfnamefont {G.}~\bibnamefont {Aquino}}, \bibinfo {author} {\bibfnamefont {L.}~\bibnamefont {Tweedy}}, \bibinfo {author} {\bibfnamefont {D.}~\bibnamefont {Heinrich}},\ and\ \bibinfo {author} {\bibfnamefont {R.~G.}\ \bibnamefont {Endres}},\ }\bibfield  {title} {\bibinfo {title} {Memory improves precision of cell sensing in fluctuating environments},\ }\href@noop {} {\bibfield  {journal} {\bibinfo  {journal} {Scientific reports}\ }\textbf {\bibinfo {volume} {4}},\ \bibinfo {pages} {5688} (\bibinfo {year} {2014})}\BibitemShut {NoStop}%
\bibitem [{\citenamefont {Karmakar}\ \emph {et~al.}(2021)\citenamefont {Karmakar}, \citenamefont {Tang}, \citenamefont {Yue}, \citenamefont {Lombardo}, \citenamefont {Karanam}, \citenamefont {Camley}, \citenamefont {Groisman},\ and\ \citenamefont {Rappel}}]{karmakar2021cellular}%
  \BibitemOpen
  \bibfield  {author} {\bibinfo {author} {\bibfnamefont {R.}~\bibnamefont {Karmakar}}, \bibinfo {author} {\bibfnamefont {M.-H.}\ \bibnamefont {Tang}}, \bibinfo {author} {\bibfnamefont {H.}~\bibnamefont {Yue}}, \bibinfo {author} {\bibfnamefont {D.}~\bibnamefont {Lombardo}}, \bibinfo {author} {\bibfnamefont {A.}~\bibnamefont {Karanam}}, \bibinfo {author} {\bibfnamefont {B.~A.}\ \bibnamefont {Camley}}, \bibinfo {author} {\bibfnamefont {A.}~\bibnamefont {Groisman}},\ and\ \bibinfo {author} {\bibfnamefont {W.-J.}\ \bibnamefont {Rappel}},\ }\bibfield  {title} {\bibinfo {title} {Cellular memory in eukaryotic chemotaxis depends on the background chemoattractant concentration},\ }\href@noop {} {\bibfield  {journal} {\bibinfo  {journal} {Physical Review E}\ }\textbf {\bibinfo {volume} {103}},\ \bibinfo {pages} {012402} (\bibinfo {year} {2021})}\BibitemShut {NoStop}%
\bibitem [{\citenamefont {Skoge}\ \emph {et~al.}(2014)\citenamefont {Skoge}, \citenamefont {Yue}, \citenamefont {Erickstad}, \citenamefont {Bae}, \citenamefont {Levine}, \citenamefont {Groisman}, \citenamefont {Loomis},\ and\ \citenamefont {Rappel}}]{skoge2014cellular}%
  \BibitemOpen
  \bibfield  {author} {\bibinfo {author} {\bibfnamefont {M.}~\bibnamefont {Skoge}}, \bibinfo {author} {\bibfnamefont {H.}~\bibnamefont {Yue}}, \bibinfo {author} {\bibfnamefont {M.}~\bibnamefont {Erickstad}}, \bibinfo {author} {\bibfnamefont {A.}~\bibnamefont {Bae}}, \bibinfo {author} {\bibfnamefont {H.}~\bibnamefont {Levine}}, \bibinfo {author} {\bibfnamefont {A.}~\bibnamefont {Groisman}}, \bibinfo {author} {\bibfnamefont {W.~F.}\ \bibnamefont {Loomis}},\ and\ \bibinfo {author} {\bibfnamefont {W.-J.}\ \bibnamefont {Rappel}},\ }\bibfield  {title} {\bibinfo {title} {Cellular memory in eukaryotic chemotaxis},\ }\href@noop {} {\bibfield  {journal} {\bibinfo  {journal} {Proceedings of the National Academy of Sciences}\ }\textbf {\bibinfo {volume} {111}},\ \bibinfo {pages} {14448} (\bibinfo {year} {2014})}\BibitemShut {NoStop}%
\bibitem [{\citenamefont {Hu}\ \emph {et~al.}(2011)\citenamefont {Hu}, \citenamefont {Chen}, \citenamefont {Rappel},\ and\ \citenamefont {Levine}}]{hu2011geometry}%
  \BibitemOpen
  \bibfield  {author} {\bibinfo {author} {\bibfnamefont {B.}~\bibnamefont {Hu}}, \bibinfo {author} {\bibfnamefont {W.}~\bibnamefont {Chen}}, \bibinfo {author} {\bibfnamefont {W.-J.}\ \bibnamefont {Rappel}},\ and\ \bibinfo {author} {\bibfnamefont {H.}~\bibnamefont {Levine}},\ }\bibfield  {title} {\bibinfo {title} {How geometry and internal bias affect the accuracy of eukaryotic gradient sensing},\ }\href@noop {} {\bibfield  {journal} {\bibinfo  {journal} {Physical Review E}\ }\textbf {\bibinfo {volume} {83}},\ \bibinfo {pages} {021917} (\bibinfo {year} {2011})}\BibitemShut {NoStop}%
\bibitem [{\citenamefont {Hiraiwa}\ \emph {et~al.}(2014)\citenamefont {Hiraiwa}, \citenamefont {Nagamatsu}, \citenamefont {Akuzawa}, \citenamefont {Nishikawa},\ and\ \citenamefont {Shibata}}]{hiraiwa2014relevance}%
  \BibitemOpen
  \bibfield  {author} {\bibinfo {author} {\bibfnamefont {T.}~\bibnamefont {Hiraiwa}}, \bibinfo {author} {\bibfnamefont {A.}~\bibnamefont {Nagamatsu}}, \bibinfo {author} {\bibfnamefont {N.}~\bibnamefont {Akuzawa}}, \bibinfo {author} {\bibfnamefont {M.}~\bibnamefont {Nishikawa}},\ and\ \bibinfo {author} {\bibfnamefont {T.}~\bibnamefont {Shibata}},\ }\bibfield  {title} {\bibinfo {title} {Relevance of intracellular polarity to accuracy of eukaryotic chemotaxis},\ }\href@noop {} {\bibfield  {journal} {\bibinfo  {journal} {Physical biology}\ }\textbf {\bibinfo {volume} {11}},\ \bibinfo {pages} {056002} (\bibinfo {year} {2014})}\BibitemShut {NoStop}%
\bibitem [{\citenamefont {Nakamura}\ and\ \citenamefont {Kobayashi}(2024)}]{nakamura2024gradient}%
  \BibitemOpen
  \bibfield  {author} {\bibinfo {author} {\bibfnamefont {K.}~\bibnamefont {Nakamura}}\ and\ \bibinfo {author} {\bibfnamefont {T.~J.}\ \bibnamefont {Kobayashi}},\ }\bibfield  {title} {\bibinfo {title} {Gradient sensing limit of a cell when controlling the elongating direction},\ }\href@noop {} {\bibfield  {journal} {\bibinfo  {journal} {arXiv preprint arXiv:2405.04810}\ } (\bibinfo {year} {2024})}\BibitemShut {NoStop}%
\bibitem [{\citenamefont {Amselem}\ \emph {et~al.}(2012)\citenamefont {Amselem}, \citenamefont {Theves}, \citenamefont {Bae}, \citenamefont {Beta},\ and\ \citenamefont {Bodenschatz}}]{amselem2012control}%
  \BibitemOpen
  \bibfield  {author} {\bibinfo {author} {\bibfnamefont {G.}~\bibnamefont {Amselem}}, \bibinfo {author} {\bibfnamefont {M.}~\bibnamefont {Theves}}, \bibinfo {author} {\bibfnamefont {A.}~\bibnamefont {Bae}}, \bibinfo {author} {\bibfnamefont {C.}~\bibnamefont {Beta}},\ and\ \bibinfo {author} {\bibfnamefont {E.}~\bibnamefont {Bodenschatz}},\ }\bibfield  {title} {\bibinfo {title} {Control parameter description of eukaryotic chemotaxis},\ }\href@noop {} {\bibfield  {journal} {\bibinfo  {journal} {Physical review letters}\ }\textbf {\bibinfo {volume} {109}},\ \bibinfo {pages} {108103} (\bibinfo {year} {2012})}\BibitemShut {NoStop}%
\bibitem [{\citenamefont {Tweedy}\ \emph {et~al.}(2013)\citenamefont {Tweedy}, \citenamefont {Meier}, \citenamefont {Stephan}, \citenamefont {Heinrich},\ and\ \citenamefont {Endres}}]{tweedy2013distinct}%
  \BibitemOpen
  \bibfield  {author} {\bibinfo {author} {\bibfnamefont {L.}~\bibnamefont {Tweedy}}, \bibinfo {author} {\bibfnamefont {B.}~\bibnamefont {Meier}}, \bibinfo {author} {\bibfnamefont {J.}~\bibnamefont {Stephan}}, \bibinfo {author} {\bibfnamefont {D.}~\bibnamefont {Heinrich}},\ and\ \bibinfo {author} {\bibfnamefont {R.~G.}\ \bibnamefont {Endres}},\ }\bibfield  {title} {\bibinfo {title} {Distinct cell shapes determine accurate chemotaxis},\ }\href@noop {} {\bibfield  {journal} {\bibinfo  {journal} {Scientific reports}\ }\textbf {\bibinfo {volume} {3}},\ \bibinfo {pages} {2606} (\bibinfo {year} {2013})}\BibitemShut {NoStop}%
\bibitem [{\citenamefont {Insall}(2010)}]{insall2010understanding}%
  \BibitemOpen
  \bibfield  {author} {\bibinfo {author} {\bibfnamefont {R.~H.}\ \bibnamefont {Insall}},\ }\bibfield  {title} {\bibinfo {title} {Understanding eukaryotic chemotaxis: a pseudopod-centred view},\ }\href@noop {} {\bibfield  {journal} {\bibinfo  {journal} {Nature reviews molecular cell biology}\ }\textbf {\bibinfo {volume} {11}},\ \bibinfo {pages} {453} (\bibinfo {year} {2010})}\BibitemShut {NoStop}%
\bibitem [{\citenamefont {Bodor}\ \emph {et~al.}(2020)\citenamefont {Bodor}, \citenamefont {P{\"o}nisch}, \citenamefont {Endres},\ and\ \citenamefont {Paluch}}]{bodor2020cell}%
  \BibitemOpen
  \bibfield  {author} {\bibinfo {author} {\bibfnamefont {D.~L.}\ \bibnamefont {Bodor}}, \bibinfo {author} {\bibfnamefont {W.}~\bibnamefont {P{\"o}nisch}}, \bibinfo {author} {\bibfnamefont {R.~G.}\ \bibnamefont {Endres}},\ and\ \bibinfo {author} {\bibfnamefont {E.~K.}\ \bibnamefont {Paluch}},\ }\bibfield  {title} {\bibinfo {title} {Of cell shapes and motion: the physical basis of animal cell migration},\ }\href@noop {} {\bibfield  {journal} {\bibinfo  {journal} {Developmental cell}\ }\textbf {\bibinfo {volume} {52}},\ \bibinfo {pages} {550} (\bibinfo {year} {2020})}\BibitemShut {NoStop}%
\bibitem [{\citenamefont {Alonso}\ \emph {et~al.}(2024{\natexlab{b}})\citenamefont {Alonso}, \citenamefont {Kirkegaard},\ and\ \citenamefont {Endres}}]{alonso2024persistent}%
  \BibitemOpen
  \bibfield  {author} {\bibinfo {author} {\bibfnamefont {A.}~\bibnamefont {Alonso}}, \bibinfo {author} {\bibfnamefont {J.~B.}\ \bibnamefont {Kirkegaard}},\ and\ \bibinfo {author} {\bibfnamefont {R.~G.}\ \bibnamefont {Endres}},\ }\bibfield  {title} {\bibinfo {title} {Persistent pseudopod splitting is an effective chemotaxis strategy in shallow gradients},\ }\href@noop {} {\bibfield  {journal} {\bibinfo  {journal} {arXiv preprint arXiv:2409.09342}\ } (\bibinfo {year} {2024}{\natexlab{b}})}\BibitemShut {NoStop}%
\bibitem [{\citenamefont {Cavanagh}\ \emph {et~al.}(2022)\citenamefont {Cavanagh}, \citenamefont {Kempe}, \citenamefont {Mazalo}, \citenamefont {Biro},\ and\ \citenamefont {Endres}}]{cavanagh2022t}%
  \BibitemOpen
  \bibfield  {author} {\bibinfo {author} {\bibfnamefont {H.}~\bibnamefont {Cavanagh}}, \bibinfo {author} {\bibfnamefont {D.}~\bibnamefont {Kempe}}, \bibinfo {author} {\bibfnamefont {J.~K.}\ \bibnamefont {Mazalo}}, \bibinfo {author} {\bibfnamefont {M.}~\bibnamefont {Biro}},\ and\ \bibinfo {author} {\bibfnamefont {R.~G.}\ \bibnamefont {Endres}},\ }\bibfield  {title} {\bibinfo {title} {T cell morphodynamics reveal periodic shape oscillations in three-dimensional migration},\ }\href@noop {} {\bibfield  {journal} {\bibinfo  {journal} {Journal of the Royal Society Interface}\ }\textbf {\bibinfo {volume} {19}},\ \bibinfo {pages} {20220081} (\bibinfo {year} {2022})}\BibitemShut {NoStop}%
\bibitem [{\citenamefont {Singh}\ \emph {et~al.}(2022)\citenamefont {Singh}, \citenamefont {Leadbetter},\ and\ \citenamefont {Camley}}]{singh2022sensing}%
  \BibitemOpen
  \bibfield  {author} {\bibinfo {author} {\bibfnamefont {A.~R.}\ \bibnamefont {Singh}}, \bibinfo {author} {\bibfnamefont {T.}~\bibnamefont {Leadbetter}},\ and\ \bibinfo {author} {\bibfnamefont {B.~A.}\ \bibnamefont {Camley}},\ }\bibfield  {title} {\bibinfo {title} {Sensing the shape of a cell with reaction diffusion and energy minimization},\ }\href@noop {} {\bibfield  {journal} {\bibinfo  {journal} {Proceedings of the National Academy of Sciences}\ }\textbf {\bibinfo {volume} {119}},\ \bibinfo {pages} {e2121302119} (\bibinfo {year} {2022})}\BibitemShut {NoStop}%
\bibitem [{\citenamefont {Novak}\ and\ \citenamefont {Friedrich}(2021)}]{novak2021bayesian}%
  \BibitemOpen
  \bibfield  {author} {\bibinfo {author} {\bibfnamefont {M.}~\bibnamefont {Novak}}\ and\ \bibinfo {author} {\bibfnamefont {B.~M.}\ \bibnamefont {Friedrich}},\ }\bibfield  {title} {\bibinfo {title} {Bayesian gradient sensing in the presence of rotational diffusion},\ }\href@noop {} {\bibfield  {journal} {\bibinfo  {journal} {New journal of physics}\ }\textbf {\bibinfo {volume} {23}},\ \bibinfo {pages} {043026} (\bibinfo {year} {2021})}\BibitemShut {NoStop}%
\bibitem [{\citenamefont {Levchenko}\ and\ \citenamefont {Iglesias}(2002)}]{levchenko2002models}%
  \BibitemOpen
  \bibfield  {author} {\bibinfo {author} {\bibfnamefont {A.}~\bibnamefont {Levchenko}}\ and\ \bibinfo {author} {\bibfnamefont {P.~A.}\ \bibnamefont {Iglesias}},\ }\bibfield  {title} {\bibinfo {title} {Models of eukaryotic gradient sensing: application to chemotaxis of amoebae and neutrophils},\ }\href@noop {} {\bibfield  {journal} {\bibinfo  {journal} {Biophysical journal}\ }\textbf {\bibinfo {volume} {82}},\ \bibinfo {pages} {50} (\bibinfo {year} {2002})}\BibitemShut {NoStop}%
\bibitem [{\citenamefont {Xiong}\ \emph {et~al.}(2010)\citenamefont {Xiong}, \citenamefont {Huang}, \citenamefont {Iglesias},\ and\ \citenamefont {Devreotes}}]{xiong2010cells}%
  \BibitemOpen
  \bibfield  {author} {\bibinfo {author} {\bibfnamefont {Y.}~\bibnamefont {Xiong}}, \bibinfo {author} {\bibfnamefont {C.-H.}\ \bibnamefont {Huang}}, \bibinfo {author} {\bibfnamefont {P.~A.}\ \bibnamefont {Iglesias}},\ and\ \bibinfo {author} {\bibfnamefont {P.~N.}\ \bibnamefont {Devreotes}},\ }\bibfield  {title} {\bibinfo {title} {Cells navigate with a local-excitation, global-inhibition-biased excitable network},\ }\href@noop {} {\bibfield  {journal} {\bibinfo  {journal} {Proceedings of the National Academy of Sciences}\ }\textbf {\bibinfo {volume} {107}},\ \bibinfo {pages} {17079} (\bibinfo {year} {2010})}\BibitemShut {NoStop}%
\bibitem [{\citenamefont {Janetopoulos}\ and\ \citenamefont {Firtel}(2008)}]{janetopoulos2008directional}%
  \BibitemOpen
  \bibfield  {author} {\bibinfo {author} {\bibfnamefont {C.}~\bibnamefont {Janetopoulos}}\ and\ \bibinfo {author} {\bibfnamefont {R.~A.}\ \bibnamefont {Firtel}},\ }\bibfield  {title} {\bibinfo {title} {Directional sensing during chemotaxis},\ }\href@noop {} {\bibfield  {journal} {\bibinfo  {journal} {FEBS letters}\ }\textbf {\bibinfo {volume} {582}},\ \bibinfo {pages} {2075} (\bibinfo {year} {2008})}\BibitemShut {NoStop}%
\bibitem [{\citenamefont {Hadjitheodorou}\ \emph {et~al.}(2021)\citenamefont {Hadjitheodorou}, \citenamefont {Bell}, \citenamefont {Ellett}, \citenamefont {Shastry}, \citenamefont {Irimia}, \citenamefont {Collins},\ and\ \citenamefont {Theriot}}]{hadjitheodorou2021directional}%
  \BibitemOpen
  \bibfield  {author} {\bibinfo {author} {\bibfnamefont {A.}~\bibnamefont {Hadjitheodorou}}, \bibinfo {author} {\bibfnamefont {G.~R.}\ \bibnamefont {Bell}}, \bibinfo {author} {\bibfnamefont {F.}~\bibnamefont {Ellett}}, \bibinfo {author} {\bibfnamefont {S.}~\bibnamefont {Shastry}}, \bibinfo {author} {\bibfnamefont {D.}~\bibnamefont {Irimia}}, \bibinfo {author} {\bibfnamefont {S.~R.}\ \bibnamefont {Collins}},\ and\ \bibinfo {author} {\bibfnamefont {J.~A.}\ \bibnamefont {Theriot}},\ }\bibfield  {title} {\bibinfo {title} {Directional reorientation of migrating neutrophils is limited by suppression of receptor input signaling at the cell rear through myosin ii activity},\ }\href@noop {} {\bibfield  {journal} {\bibinfo  {journal} {Nature Communications}\ }\textbf {\bibinfo {volume} {12}},\ \bibinfo {pages} {6619} (\bibinfo {year} {2021})}\BibitemShut {NoStop}%
\bibitem [{\citenamefont {Fancher}\ and\ \citenamefont {Mugler}(2017)}]{fancher2017fundamental}%
  \BibitemOpen
  \bibfield  {author} {\bibinfo {author} {\bibfnamefont {S.}~\bibnamefont {Fancher}}\ and\ \bibinfo {author} {\bibfnamefont {A.}~\bibnamefont {Mugler}},\ }\bibfield  {title} {\bibinfo {title} {Fundamental limits to collective concentration sensing in cell populations},\ }\href@noop {} {\bibfield  {journal} {\bibinfo  {journal} {Physical review letters}\ }\textbf {\bibinfo {volume} {118}},\ \bibinfo {pages} {078101} (\bibinfo {year} {2017})}\BibitemShut {NoStop}%
\bibitem [{\citenamefont {Bi}\ \emph {et~al.}(2015)\citenamefont {Bi}, \citenamefont {Lopez}, \citenamefont {Schwarz},\ and\ \citenamefont {Manning}}]{bi2015density}%
  \BibitemOpen
  \bibfield  {author} {\bibinfo {author} {\bibfnamefont {D.}~\bibnamefont {Bi}}, \bibinfo {author} {\bibfnamefont {J.}~\bibnamefont {Lopez}}, \bibinfo {author} {\bibfnamefont {J.~M.}\ \bibnamefont {Schwarz}},\ and\ \bibinfo {author} {\bibfnamefont {M.~L.}\ \bibnamefont {Manning}},\ }\bibfield  {title} {\bibinfo {title} {A density-independent rigidity transition in biological tissues},\ }\href@noop {} {\bibfield  {journal} {\bibinfo  {journal} {Nature Physics}\ }\textbf {\bibinfo {volume} {11}},\ \bibinfo {pages} {1074} (\bibinfo {year} {2015})}\BibitemShut {NoStop}%
\bibitem [{\citenamefont {Bi}\ \emph {et~al.}(2016)\citenamefont {Bi}, \citenamefont {Yang}, \citenamefont {Marchetti},\ and\ \citenamefont {Manning}}]{bi2016motility}%
  \BibitemOpen
  \bibfield  {author} {\bibinfo {author} {\bibfnamefont {D.}~\bibnamefont {Bi}}, \bibinfo {author} {\bibfnamefont {X.}~\bibnamefont {Yang}}, \bibinfo {author} {\bibfnamefont {M.~C.}\ \bibnamefont {Marchetti}},\ and\ \bibinfo {author} {\bibfnamefont {M.~L.}\ \bibnamefont {Manning}},\ }\bibfield  {title} {\bibinfo {title} {Motility-driven glass and jamming transitions in biological tissues},\ }\href@noop {} {\bibfield  {journal} {\bibinfo  {journal} {Physical Review X}\ }\textbf {\bibinfo {volume} {6}},\ \bibinfo {pages} {021011} (\bibinfo {year} {2016})}\BibitemShut {NoStop}%
\bibitem [{\citenamefont {Aitken}(1936)}]{aitken1936iv}%
  \BibitemOpen
  \bibfield  {author} {\bibinfo {author} {\bibfnamefont {A.~C.}\ \bibnamefont {Aitken}},\ }\bibfield  {title} {\bibinfo {title} {Iv.—on least squares and linear combination of observations},\ }\href@noop {} {\bibfield  {journal} {\bibinfo  {journal} {Proceedings of the Royal Society of Edinburgh}\ }\textbf {\bibinfo {volume} {55}},\ \bibinfo {pages} {42} (\bibinfo {year} {1936})}\BibitemShut {NoStop}%
\end{thebibliography}%


\begin{thebibliography}{3}%
\makeatletter
\providecommand \@ifxundefined [1]{%
 \@ifx{#1\undefined}
}%
\providecommand \@ifnum [1]{%
 \ifnum #1\expandafter \@firstoftwo
 \else \expandafter \@secondoftwo
 \fi
}%
\providecommand \@ifx [1]{%
 \ifx #1\expandafter \@firstoftwo
 \else \expandafter \@secondoftwo
 \fi
}%
\providecommand \natexlab [1]{#1}%
\providecommand \enquote  [1]{``#1''}%
\providecommand \bibnamefont  [1]{#1}%
\providecommand \bibfnamefont [1]{#1}%
\providecommand \citenamefont [1]{#1}%
\providecommand \href@noop [0]{\@secondoftwo}%
\providecommand \href [0]{\begingroup \@sanitize@url \@href}%
\providecommand \@href[1]{\@@startlink{#1}\@@href}%
\providecommand \@@href[1]{\endgroup#1\@@endlink}%
\providecommand \@sanitize@url [0]{\catcode `\\12\catcode `\$12\catcode `\&12\catcode `\#12\catcode `\^12\catcode `\_12\catcode `\%12\relax}%
\providecommand \@@startlink[1]{}%
\providecommand \@@endlink[0]{}%
\providecommand \url  [0]{\begingroup\@sanitize@url \@url }%
\providecommand \@url [1]{\endgroup\@href {#1}{\urlprefix }}%
\providecommand \urlprefix  [0]{URL }%
\providecommand \Eprint [0]{\href }%
\providecommand \doibase [0]{http://dx.doi.org/}%
\providecommand \selectlanguage [0]{\@gobble}%
\providecommand \bibinfo  [0]{\@secondoftwo}%
\providecommand \bibfield  [0]{\@secondoftwo}%
\providecommand \translation [1]{[#1]}%
\providecommand \BibitemOpen [0]{}%
\providecommand \bibitemStop [0]{}%
\providecommand \bibitemNoStop [0]{.\EOS\space}%
\providecommand \EOS [0]{\spacefactor3000\relax}%
\providecommand \BibitemShut  [1]{\csname bibitem#1\endcsname}%
\let\auto@bib@innerbib\@empty
%</preamble>
\bibitem [{\citenamefont {Wu}\ \emph {et~al.}(2014)\citenamefont {Wu}, \citenamefont {Giri}, \citenamefont {Sun},\ and\ \citenamefont {Wirtz}}]{wu2014three}%
  \BibitemOpen
  \bibfield  {author} {\bibinfo {author} {\bibfnamefont {P.-H.}\ \bibnamefont {Wu}}, \bibinfo {author} {\bibfnamefont {A.}~\bibnamefont {Giri}}, \bibinfo {author} {\bibfnamefont {S.~X.}\ \bibnamefont {Sun}}, \ and\ \bibinfo {author} {\bibfnamefont {D.}~\bibnamefont {Wirtz}},\ }\href@noop {} {\bibfield  {journal} {\bibinfo  {journal} {Proceedings of the National Academy of Sciences}\ }\textbf {\bibinfo {volume} {111}},\ \bibinfo {pages} {3949} (\bibinfo {year} {2014})}\BibitemShut {NoStop}%
\bibitem [{\citenamefont {Nakamura}\ and\ \citenamefont {Kobayashi}(2024)}]{nakamura2024gradient}%
  \BibitemOpen
  \bibfield  {author} {\bibinfo {author} {\bibfnamefont {K.}~\bibnamefont {Nakamura}}\ and\ \bibinfo {author} {\bibfnamefont {T.~J.}\ \bibnamefont {Kobayashi}},\ }\href@noop {} {\bibfield  {journal} {\bibinfo  {journal} {arXiv preprint arXiv:2405.04810}\ } (\bibinfo {year} {2024})}\BibitemShut {NoStop}%
\bibitem [{\citenamefont {Novak}\ and\ \citenamefont {Friedrich}(2021)}]{novak2021bayesian}%
  \BibitemOpen
  \bibfield  {author} {\bibinfo {author} {\bibfnamefont {M.}~\bibnamefont {Novak}}\ and\ \bibinfo {author} {\bibfnamefont {B.~M.}\ \bibnamefont {Friedrich}},\ }\href@noop {} {\bibfield  {journal} {\bibinfo  {journal} {New journal of physics}\ }\textbf {\bibinfo {volume} {23}},\ \bibinfo {pages} {043026} (\bibinfo {year} {2021})}\BibitemShut {NoStop}%
\end{thebibliography}%

\end{document}

\title{Supplemental Material for ``Optimal cell shape for accurate chemical gradient sensing in eukaryote chemotaxis"}
\maketitle

\section{non-uniform distribution of receptors}

\begin{figure}[b]
    \centering

    \begin{minipage}[b]{0.25\textwidth}
    \subfigure[]{
        \includegraphics[width=0.9\textwidth]{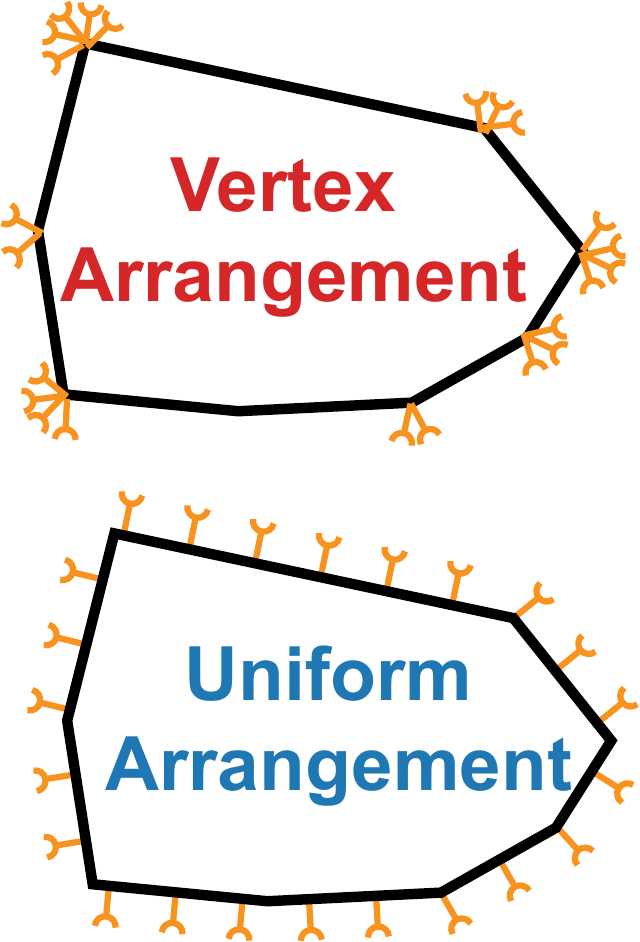}
        \label{fig:convex_sampling:sketch}
        }
    \vspace*{3em}
    \end{minipage}
    \subfigure[]{
        \begin{minipage}[b]{0.6\textwidth}
            \includegraphics[width=0.99\linewidth]{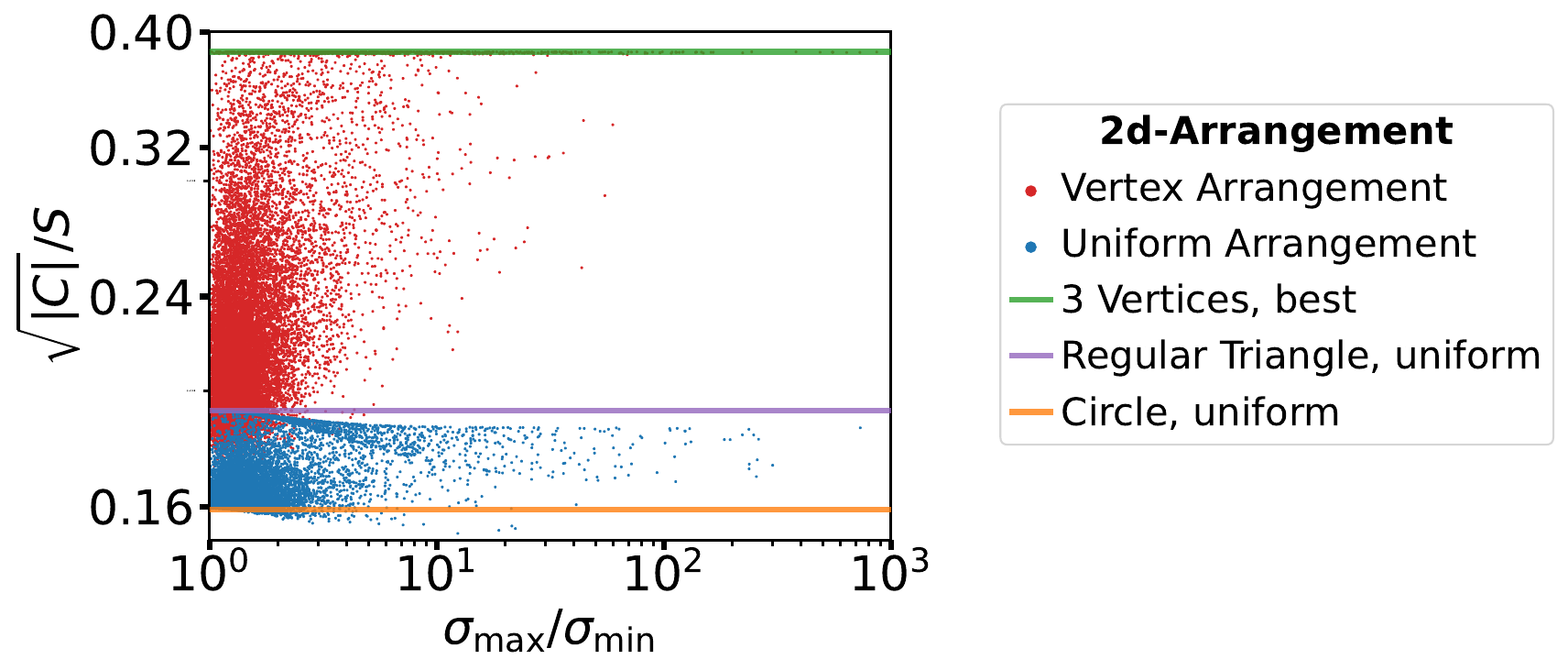}\\
            %\hspace*{1em}
            \includegraphics[width=0.99\linewidth]{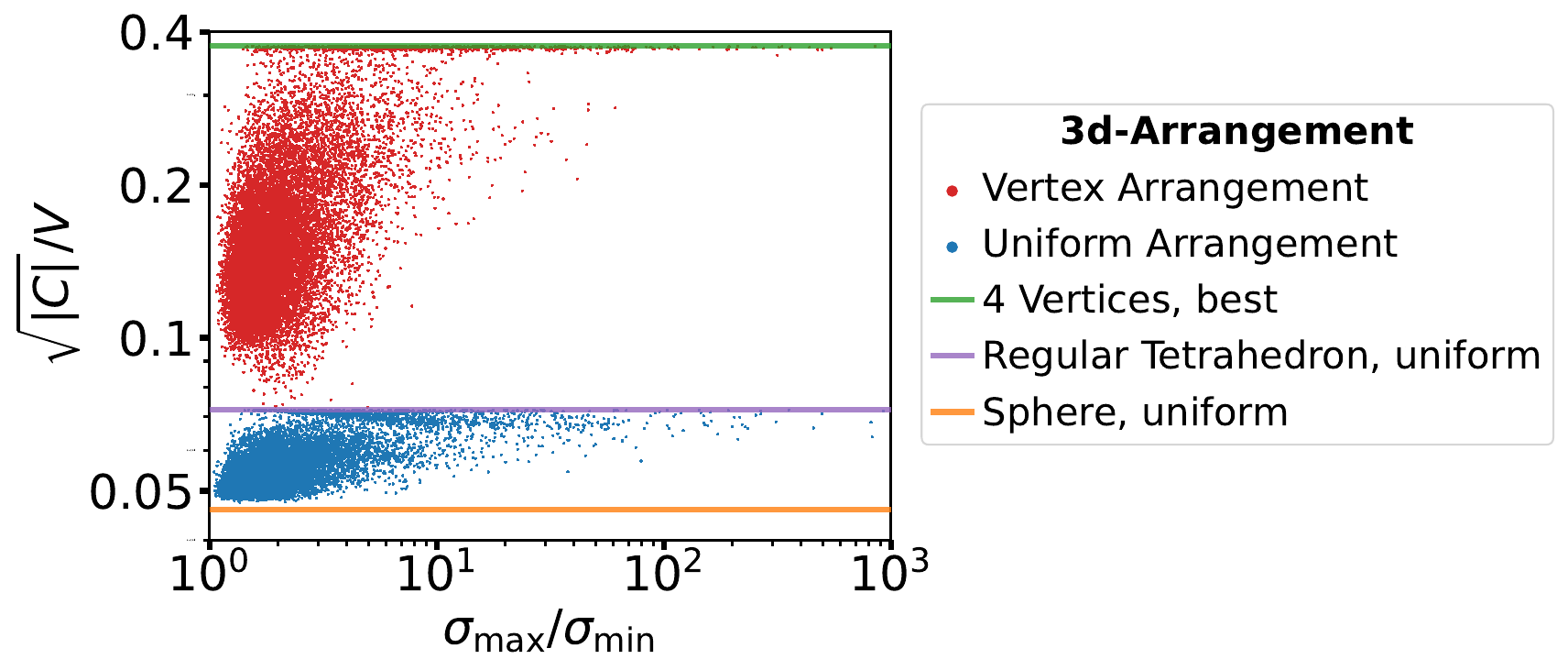}
            \label{fig:convex_sampling:result}
        \end{minipage}
    }

    \caption{Bound of $\det |\bm C|$ verified by random convex shape sampling.
    \subref{fig:convex_sampling:sketch} 
        After generating convex hull of a cell (black lines),
            we tried two different ways to maximize $|\bm C|$.
            1) we assume that all of the sensing units are
                located on its vertices,
                and we try to optimize $\alpha_i$ on each vertex $i$.
            2) we assume that sensing units 
                can only be arranged uniformly on the cell's surface,
                and therefore matrix $\bm C$ is fixed for a certain shape.
    \subref{fig:convex_sampling:result}
        Simulation results for 2D and 3D cases.
        Lines are calculated analytically on the special cases of sensing units on vertices and surface of triangles, tetrahedrons, circles and spheres, respectively.
    }
    \label{fig:convex_sampling}
\end{figure}

It is worth pointing out that the upper bound on $\det |\bm C|$ established in Appendix C 
    is relatively loose. 
We further verify its validity 
    by simulations.
Through randomly sampling $n$-vertex convex graph,
    we optimize $\{\alpha_i\}$ associated with each vertex $\bm r_i$
    to maximize $\det \vert\bm C\vert$, 
    as illustrated in Fig.~\ref{fig:convex_sampling:sketch}.
With each 1000 samples for 17 different $n$ 
    ($3\sim19$ for 2d cases and $4\sim 20$ for 3d cases),
    we present the results 
    in Fig.~\ref{fig:convex_sampling:result}.
Our simulation suggests a tighter bound (green lines) compared to the bound in previous section. The specific values of the bound coefficients are given in Table.~\ref{tab:bound}.
Interestingly, one of the optimal shape in 2D corresponds to an equal partition of receptors on three vertices in a triangle, i.e., $\alpha_i=1/3$. In 3D, one of the optimal shape is a tetrahedron with $\alpha_i=1/4$. 

\begin{table}[h]
    \centering 
    \caption{Upper Bound of $\bm C$ found by Random Sampling in Convex Cells}
    \label{tab:bound}
    \begin{tabular}{ccccc}
        \hline 
        $\quad$Dimension$\quad$ & $\quad$Index$\quad$ 
            & $\quad$Theory Bound$\quad$ & $\quad$Simulation Bound$\quad$ & $\quad$Simulation Bound (uniform)$\quad$\\
        \hline
        2 & $\sqrt{|\bm C|}/S_h$ &$1/2$ & $\sqrt{4/27}$ & $\sqrt{1/27}$ \\
        3 & $\sqrt{|\bm C|}/V_h$ &$3/2$ & $3/8$ & $\sqrt{1/192}$\\
        \hline
    \end{tabular}
\end{table}
    
When the receptors are uniformly distributed on the cell surface,
    the cell can only indirectly adjust the receptor distributions by modifying its shape, 
    but cannot directly manipulate the values $\{\alpha_i\}$.
As shown by blue dots in Fig.~\ref{fig:convex_sampling:result},
    the performance of cells with uniform receptor arrangement falls
         far below the simulation upper bound (green lines).
This observation suggests that a uniform distribution of receptors may not fully allow the cell to exploit its potential for accurate gradient sensing.

To quantify the non-uniformity of receptor distribution on the cell surface, we employ the negative Shannon entropy, defined as:
\begin{equation}
    -H = \int_0^{2\pi} f(u) \log f(u) \, du,
\end{equation}
where $u$ is a periodic coordinate on the one-dimensional cell boundary with a period of $2\pi$. 
As the receptor distribution transitions from a uniform distribution to a vertex distribution, the negative entropy $-H$ increases from $-\ln(2\pi)$ to $+\infty$ correspondingly.

\iffalse
By utilizing the von Mises distribution ($I_0(\kappa)$ is the modified Bessel function of the first kind of order 0)
\begin{equation}
    f(u | \mu, \kappa) = \frac{e^{\kappa \cos(u - \mu)}}{2\pi I_0(\kappa)},
\end{equation}
the receptor distribution transitions from a uniform distribution to a vertex distribution as the parameter $\kappa$ increases from 0 to $\infty$. Correspondingly, the negative entropy $-H$ increases from $-\ln(2\pi)$ to $+\infty$, reflecting the growing non-uniformity of the receptor arrangement.
\fi

\section{Chemotactic Index and Alignment}\label{sec:CI}

\newcommand*{\g}{\mathop{}\!\mathcal{G}}
\newcommand*{\R}{\mathop{}\!\mathcal{R}}
    
In 2D space, the distribution of 
    measurement result $\tilde{\bm g} / |\tilde{\bm g}|$ is
    related with 3 factors: 
    the relative signal strength on 2 spindles $g/\sigma_p, g/\sigma_q$,
    and the angle $\phi_p$ between main axis $p$ and true gradient direction $\bm g/|\bm g|$.
Set $\bm g/|\bm g|$ as the $x+$ direction, 
    the probability density function (PDF) of $\tilde{\bm g}$ can be written as
\begin{equation*}
    f(\tilde{\bm g}) = 
    \frac{1}{2\mathrm{\pi}\sigma_p\sigma_q}
    \exp\left[-\frac{1}{2}(\frac{a^2}{\sigma_p^2} + \frac{b^2}{\sigma_q^2})\right]
    ,\quad
    \text{where }
    \tilde{\bm g} = \begin{bmatrix}
        g + a\cos\phi-b\sin\phi\\
        a\sin\phi+b\cos\phi
    \end{bmatrix}.
\end{equation*}
Let $\g = \sqrt{\mathcal{S}} = g/\!\sqrt{\sigma_p\sigma_q}$ represents the relative signal strength,
  and $k^2$ for $\sigma_p=k\sigma,\, \sigma_q=k^{-1}\sigma$ is 
    the aspect ratio of the uncertainty from two spindles.
Then the above equation can be rewritten as 
\begin{equation*}
    f(\alpha, \beta) \dif\alpha \dif\beta = 
    \frac{1}{2\mathrm{\pi}}
    \exp\left(-\frac{\alpha^2 + \beta^2}{2}\right)
    \dif\alpha \dif\beta
    ,\quad
    \text{where }
    \tilde{\bm g}/\sigma = \begin{bmatrix}
        \g + \alpha k\cos\phi - \beta k^{-1}\sin\phi\\
        \alpha k\sin\phi + \beta k^{-1}\cos\phi
    \end{bmatrix}.
\end{equation*}
Define $\R = |\tilde{\bm g}|/\sigma$ as the measured steepness
    normalized by $\sigma$, and
\begin{equation*}
    \tilde{\bm g}/\sigma 
    = \begin{bmatrix}
        \g + \alpha k\cos\phi - \beta k^{-1}\sin\phi\\
        \alpha k\sin\phi + \beta k^{-1}\cos\phi
        \end{bmatrix}
    = \begin{bmatrix}
        \R \cos\theta\\
        \R \sin\theta
        \end{bmatrix},
\end{equation*}
which leads to 
\begin{equation*}
    \begin{bmatrix}
        \alpha \\ \beta
        \end{bmatrix}
    = \begin{bmatrix}
        k^{-1}\cdot \left(\R \cos(\theta-\phi) - \g \cos\phi \right)\\
        k\cdot \left(\R \sin(\theta-\phi) + \g \sin\phi \right)
        \end{bmatrix}
    ,\quad
    \text{therefore }
    |\bm J| = 
    \begin{vmatrix}
        k^{-1}\cos(\theta-\phi) &-k^{-1}\R\sin(\theta-\phi)\\
        k\sin(\theta-\phi)  &k\R\cos(\theta-\phi)
    \end{vmatrix}
    =\R.
\end{equation*}
We can do a coordinate transformation
\begin{equation}
    f(\theta|\phi) \dif\theta
    = \frac{1}{2\mathrm{\pi}}
        \int_0^\infty \dif \R \R 
        \exp\left[ - \frac{
            k^{-2} \left(\R \cos(\theta-\phi) - \g \cos\phi \right)^2
            + 
            k^2 \left(\R \sin(\theta-\phi) + \g \sin\phi \right)^2
        }{2}\right]
        \dif \theta.
\end{equation}
The distribution of $\phi$ is denoted as $p(\phi)$.
    Finally, we have the distribution of angle error $\theta$ as
\begin{equation}
    f(\theta) = 
    \int \dif\phi\, p(\phi) \cdot
    \frac{1}{2\mathrm{\pi}} H(\theta, \phi, k, \g)
    \label{eq:ap:dist_q_byphi}
\end{equation}
where 
\begin{equation}
    \begin{split}
        &H(\theta, \phi, k, \g) \\
        = &\int_0^\infty \dif \R \R 
        \exp\left[ - \frac{
            k^{-2} \left(\R \cos(\theta-\phi) - \g \cos\phi \right)^2
            + 
            k^2 \left(\R \sin(\theta-\phi) + \g \sin\phi \right)^2
        }{2}\right]\\
        = 
        &\frac{k}{\sqrt{2} (\cos^2 \Delta + k^4 \sin^2\Delta)^{3/2}} \cdot 
        \Bigg\{ 
            \sqrt{2} k
                \exp\left(-\frac{\g^2}{2} \frac{\cos^2\phi+k^4\sin^2\phi}{k^2}\right)
                \sqrt{\cos^2\Delta + k^4 \sin^2\Delta}
                \ +\\
            &\qquad \g \sqrt{\mathrm{\pi}} 
            \exp\left(- \frac{\g^2}{2} \frac{k^2 \sin^2\theta}{\cos^2 \Delta + k^4 \sin^2\Delta}\right)
            \left( \cos\Delta\cos\phi-k^4\sin\Delta\sin\phi \right)
            \left[ 1 + \mathrm{Erf}\left(
                \frac{\g}{\sqrt{2}} 
                \frac{\cos\Delta\cos\phi-k^4\sin\Delta\sin\phi}{k\sqrt{\cos^2 \Delta + k^4 \sin^2\Delta}}
            \right) \right]
        \Bigg\}\\
    \end{split}
    \label{eq:ap:H}
\end{equation}
Here, we use $\Delta=\theta-\phi$ to simplify the expression,
and error function 
    $\mathrm{Erf}(z) = 
        \frac{2}{\sqrt{\mathrm{\pi}}} 
        \int_{0}^z \dif t\, e^{-t^2} $.
The chemotactic index (CI) is
\begin{equation}
    \begin{split}
    \langle\cos\theta\rangle
    = &\frac{1}{2\mathrm{\pi}}
        \int \dif\phi\, p(\phi) \cdot
        \int \dif\theta\, 
        \cos\theta\, 
        H(\theta, \phi, k, \g) \\
    = &\frac{k\g}{2\sqrt{2\mathrm{\pi}}}
        \iint \dif\phi \dif\theta \,
        p(\phi)\,
        \frac{\cos\theta (\cos\Delta \cos\phi -k^4 \sin\Delta\sin\phi)}
                {(\cos^2\Delta+k^4\sin^2\Delta)^{3/2}}
        \exp\left(- \frac{\g^2}{2} \frac{k^2 \sin^2\theta}{\cos^2 \Delta + k^4 \sin^2\Delta}\right).
    \end{split}
    \label{eq:ap:cos}
\end{equation}
    We need to point out that both $\phi$ and $\theta$ 
        are integral variables.

For the prototypical circular model, 
    the allocation is spatially symmetric 
    so that the noise on each direction 
    will be the same, i.e. $\sigma_p=\sigma_q,\ k=1$.
In this special case, the distribution of 
    the direction measurement error $\theta$ 
    can be calculated analytically as 
\begin{equation}
    f(\theta) = 
    \frac{\exp(-\g^2/2)}{4\mathrm{\pi}}
    \left[
        2 + 
        \exp\left(\frac{\g^2\cos^2\theta}{2}\right) 
        \g \sqrt{2\mathrm{\pi}} 
        \cos\theta\left( 1+\mathrm{Erf}\frac{\g\cos\theta}{\sqrt{2}} \right)
    \right]. 
\end{equation}
The alignment angle $\phi$ is no longer important 
    as all directions can be regarded as the main axis.
And therefore, 
    we get CI
\begin{equation}
    \langle \cos\theta\rangle = 
    \frac{\sqrt{2 \mathrm{\pi}} \g \cdot\mathrm{e}^{- \g^2/4}}{4}
    \left[I_0\left(\frac{\g^2}{4}\right)+I_1\left(\frac{\g^2}{4}\right)\right]
\end{equation}
where $I_n (x)$ denotes modified Bessel functions 
    of the first kind.

\begin{figure}[h]
    \centering
    \includegraphics[width=0.7\textwidth]{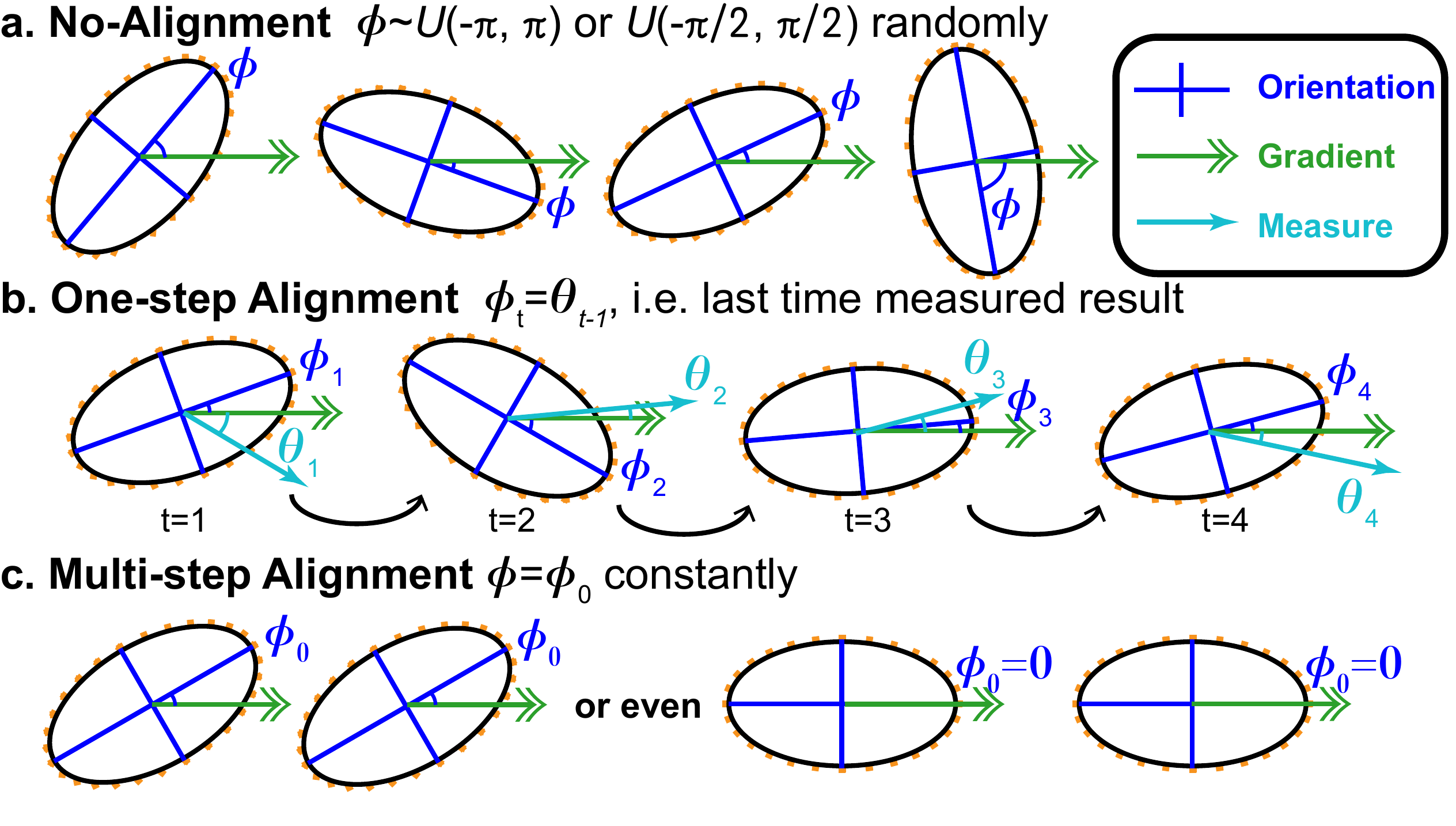}
    \caption{
    An additional parameter $\phi$ is used to describe the polarity of cell's orientation, and its distribution depends on the different alignment strategies.
    3 types of alignment schemes without memory is provided here: 
    (a) no-alignment, the orientation of cells in space is completely random;
    (b) one-step alignment, align to the direction of last-time-measured gradient $\tilde{\bm g}$;
    (c) multi-step alignment, the cells would strictly align to the direction of true gradient $\bm g$ with a fixed angular difference $\phi_0$.
    }
    \label{fig:alignment}
    \end{figure}
    
\begin{figure}[h]
    \centering
    \subfigure[]{
        \includegraphics[width=.225\textwidth]{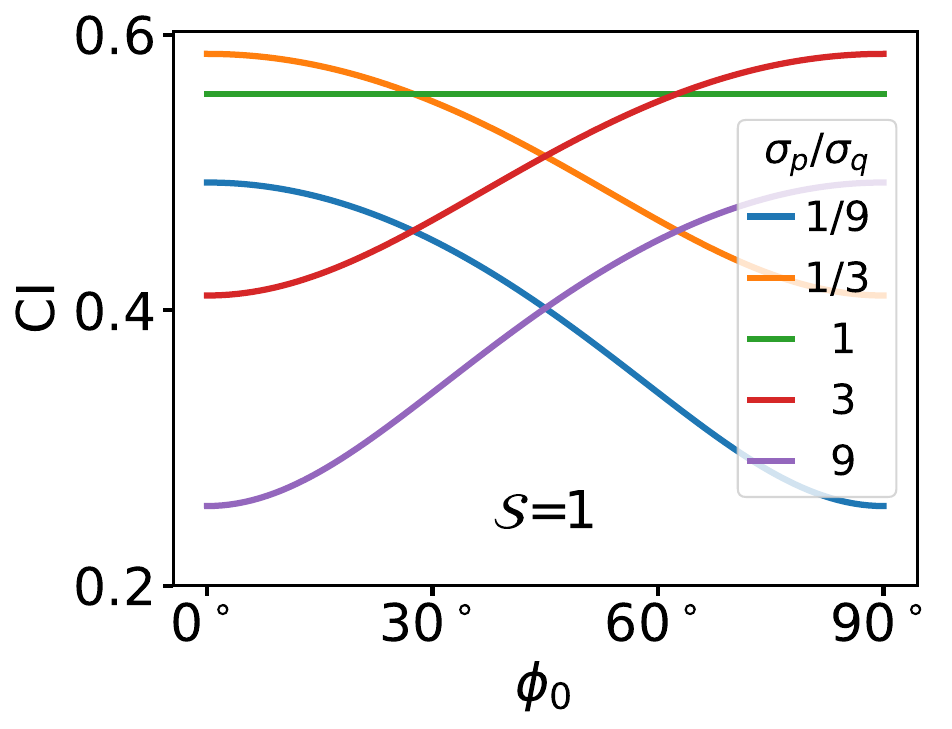}
        \label{fig:preAlign:phi}
    }
    \subfigure[]{
        \includegraphics[width=.235\textwidth]{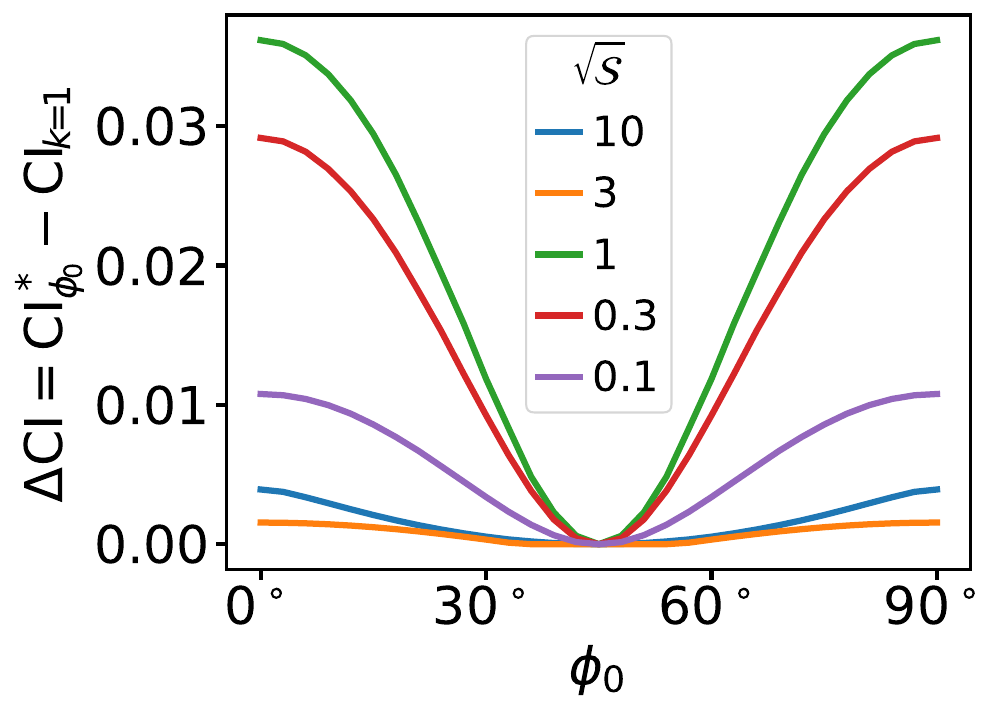}
        \label{fig:preAlign:improvement}
    }
    \subfigure[]{
        \includegraphics[width=.47\textwidth]{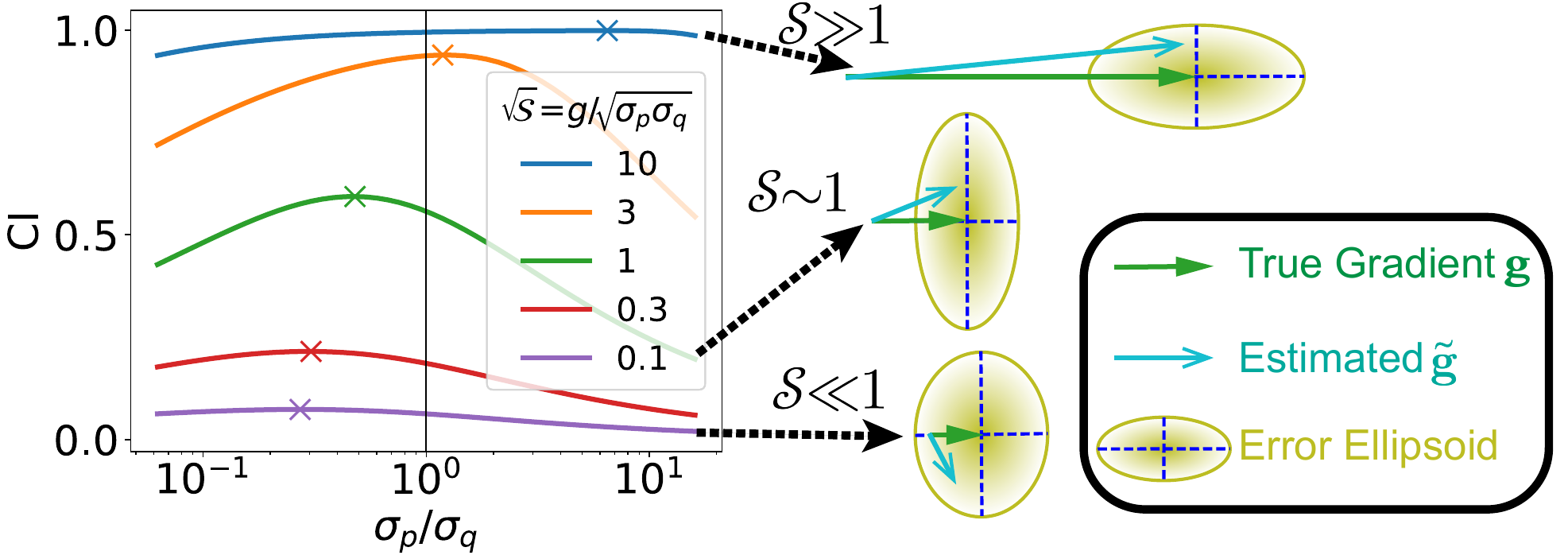}
        \label{fig:preAlign:sketch}
    }\\     
    \caption{Improvement of $\mathrm{CI}$ with multi-step alignment strategy.
    \subref{fig:preAlign:phi}
        $\mathrm{CI}$ under 
            different alignment angles $\phi_0$.
        Each line corresponds to a fixed aspect-ratio $\sigma_p/\sigma_q$.
    \subref{fig:preAlign:improvement}
        Improvement of $\mathrm{CI}$ under fixed signal-noise-ratio $\mathcal{S}$.
        The data points are calculated by 
            $\mathrm{CI}^*(\phi_0|\mathcal{S}) = \max_{k} \mathrm{CI}(\mathcal{S}, \sigma_p/\sigma_q=k^2, \phi_0)$,
        and then compared with $\mathrm{CI}(k=1\,|\,\mathcal{S})$.
    \subref{fig:preAlign:sketch}
        Chemotactic index $\mathrm{CI}$ under 
            different aspect-ratio $\sigma_p/\sigma_q$ with $\phi_0\equiv0$.
        We use cross to mark the best point on each line.
    }
    \label{fig:preAlign}
    \end{figure}

However, there might be asymmetries in the directions of the two spindles,
    i.e. $\sigma_p \neq \sigma_q$. 
    The axis-alignment parameter $\phi$ starts 
    to influence the measurement accuracy.
Different models give different $p(\phi)$.
    In the following, we will focus on three specific alignment methods:
     no-alignment, one-step alignment and multi-step alignment (Fig.~\ref{fig:alignment}).

\textit{No-Alignment}, $p(\phi) = 1/2\mathrm{\pi}$.
If a cell do not have any prior information
    about the direction for the true gradient $\bm g$,
    all it can do is to choose a random direction to align its spindle and then measure.
Take the distribution $p(\phi) = 1 / 2\mathrm{\pi}$ for 
    $-\mathrm{\pi} \le\ \phi \le \mathrm{\pi}$.
Eq.~\ref{eq:ap:cos} can be simplified as 
\begin{equation}
    \langle \cos\theta\rangle 
    = \frac{1}{\sqrt{2\mathrm{\pi}}} 
        \int_{0}^{\mathrm{\pi}} \dif w \,
        \sqrt{ u_w }
        \mathrm{e}^{- u_w}
        \left[I_0 (u_w) + I_1(u_w)\right]
    \quad\text{where }
    u_w = \frac{\g^2/4}{\cosh \kappa + \sinh\kappa\cos w} 
\end{equation}
$\kappa = -\log k^2$ represents the level of asymmetry between the two spindle directions.
As expected, for random alignment case, optimal CI is achieved at $k=1$.

\textit{One-step Alignment}.
In the absence of any prior knowledge about external gradient $\bm g$, 
    one potential strategy is to align a cell's spindle with the gradient direction measured in the previous step. This approach leverages the most recent information to guide alignment in the subsequent measurement.
Over a prolonged period, the sensing process might reach a steady state. In this steady state, the distribution of the current measurement error, $f(\theta)$ becomes equivalent to the distribution of alignment angle, $p(\theta)$ in Eq.~\ref{eq:ap:dist_q_byphi}.

\textit{Multi-step Alignment}. 
If a cell has a long memory, it can take the average of multiple steps to infer $\phi$. In this case, variance in $\phi$ will approach zero as the number of memorized steps becomes infinite, allowing the cell to always align with the true gradient for subsequent instantaneous gradient inference.
In this scenario, $p(\phi)=\delta(\phi-\phi_0)$, with $\phi_0$ being the true gradient direction. As shown in Fig.~\ref{fig:preAlign:sketch}, at high SNR, CI is optimal when the error space's long axis (corresponding to cell's short axis) is aligned with the gradient. At low SNR, CI is optimal when the error space's short axis (corresponding to cell's long axis) is aligned with the gradient. Compared to the isotropic error space, the best improvement with this multi-step alignment happens near SNR$\sim 1$, but the improvement is less than $15\%$ (Fig.~\ref{fig:preAlign:improvement}), which is marginal.

To determine the stable distribution of measurement error $\theta$
    under the one-step alignment strategy,
    we used a grid-based numerical method on the angular space.
We calculated the numerical solution of $p(\theta_i | \phi_j)$
    and constructed the transition matrix $\bm W_{N\times N}$ where $i,j=0,1,2,...,N-1$.
The eigenvector with the largest eigenvalue is just the stable distribution 
    of $\theta$.

There are some other possible alignment strategies,
    e.g., dynamically adjusting aspect ratio $k$ 
        based on the last measured steepness $|\tilde{\bm g}|$,
        or using shape orientation $\phi$ as a memory 
        of previous estimation.
However, it's reasonable to expect that these alternative schemes wouldn't outperform the multi-step alignment strategy. Even multi-step alignment, despite its benefits, can only provide limited improvement in scenarios where the mean signal-to-noise ratio is relatively low $\mathcal{S}\sim1$.

\section{An infinite expanded shape}\label{sec:nStar}

\begin{figure}[b]
\centering
\subfigure[]{
    \includegraphics[width=.3\textwidth]{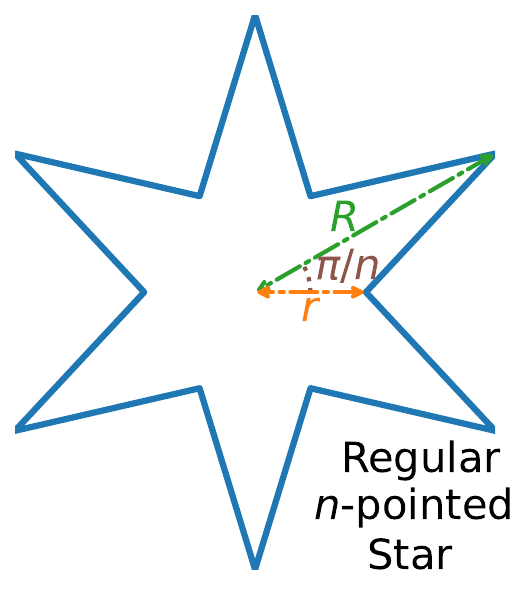}
    \label{fig:nStar_shape}
}
\subfigure[]{
    \includegraphics[width=.5\textwidth]{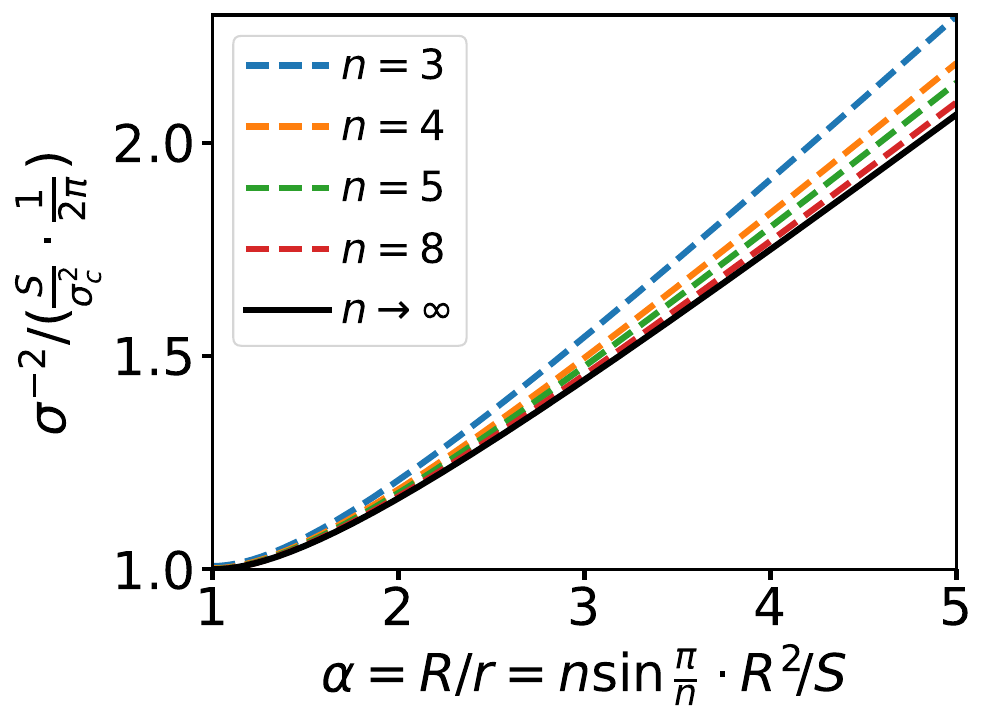}
    \label{fig:nStar_solution}
}\\        
\caption{Gradient sensing in the regular $n$-pointed star shape cell.
\subref{fig:nStar_shape}
    The shape of the regular $n$-pointed star with $n=6$,
        where concentration sensing units 
        are evenly allocated on its perimeter.
\subref{fig:nStar_solution}
    Improvements introduced 
        by the deformation factor $\alpha=R/r$ 
        with fixed 2D volume $S$.
    The vertical axis is normalized 
        by solution of the circle with uniformly distributed receptors.
    Dashed line for small $n$ 
        and black line for limit $n\rightarrow\infty$.
}
\label{fig:nStar}
\end{figure}

Here, we first consider a particular class of non-convex shapes:
    regular $n$-pointed star.
As shown in Fig.~\ref{fig:nStar_shape},
    it is a rotationally symmetric polygon 
    with of 2n vertices.
    Each vertex can be located at a distance of either $R$ (outer radius) or $r$ (inner radius) to the center.
For simplicity, 
    we model this shape as a 2D cell 
    with sensing units uniformly distributed 
    along its perimeter (i.e. the blue line in Fig.~\ref{fig:nStar_shape}). 
We define $\alpha = R/r$ as the ratio of outer and inner radius.  
     For $n \ge 3$ (so that $\mathrm{var}(x)=\mathrm{var}(y)$)
\begin{equation*}
    \sqrt{\vert \bm C \vert}/S = 
        \frac{1}{6}
        \frac{\alpha^2 + 1 + \alpha \cos\pi/n}{\alpha\, n \sin \pi/n},
\end{equation*}
where $S$ is the area of the cell.
For large $n$, we can further obtain 
    its asymptotic behavior
\begin{equation}
    \sigma^{-2} = \frac{ \sqrt{\vert \bm C \vert} }{\sigma_{\hat c}^2}
    \approx \frac{S}{\sigma_{\hat c}^2}
        \cdot \frac{1}{2\mathrm{\pi}}
        \cdot \frac{1}{3}
            \left(\alpha + \frac{1}{\alpha} + 1\right).
    \label{eq:nStar_solution_asymptotic}
\end{equation}
In fact, for $n\ge 3$,
    all regular $n$-pointed stars 
    performed close to Eq.~\ref{eq:nStar_solution_asymptotic},
    as in Fig.~\ref{fig:nStar_solution}.
Therefore, the elongation of a non-convex cell 
    can lead to an improvement 
    proportional to $\alpha$ (for $n$-star).
There is no inherent upper bound for this improvement,
    suggesting that for cells with high deformability,
    it is a very efficient strategy for gradient sensing, but at a cost of diverging perimeter.

\section{Algorithm for searching optimal cell shape}
In order to search for the optimal shape, 
    we use a simplified model 
    assuming that there exists a special point within the shape that can be directly connected to any other point 
    without venturing outside the cell's boundary.
In simple terms, 
    the cell's surface can be represented using 
    a polar coordinate system centering at this special point, 
    with angular coordinate $\theta$.
Consequently, the entire shape can be described 
    by a series of points $(R_i, \theta_i)$.
    To ensuring $R_i > 0$, 
    we perform the optimization process on $z_i = \log R_i$ instead.
In fact, the model can be extended to account for a minimum circular area required for the cell's nucleus. In such cases, we can define $R_i = R_0 + \exp z_i$ , where $R_0$ represents the minimum nuclear radius (non-negative).

With the shape model above, 
    we can calculate the loss function 
    $\mathcal{Z} = -\mathrm{CI} + \lambda \mathcal{L}$
    and optimize it by gradient descent method.
Here we use no-align cases in the function of $\mathrm{CI}$.

\begin{algorithm}[h]
    \SetInd{0.5em}{0.5em}
    \SetNlSkip{0.5em}
    \SetAlgoLined
    \caption{Optimize the Projected 2d-Shapes}
    \label{alg:optim}
    \KwIn{penalty factor $\lambda$, 
        normalized exogenous gradient $g_0$,
        number of grids $N$}
    \KwOut{$\{R_i\}$ for direction $\theta_i = 2\mathrm{\pi} \cdot i / N$}
    
    init direction $\theta_i = 2\mathrm{\pi} \cdot i / N$ for $i=0,1,2,\dots,N-1$\;
    init $z_i = \log R_i$ by random Gaussian distribution\;
    \While{target $\mathcal{Z}$ not converged}
    {
        \Indp
        \tcc{Obtain coordinates $(x_i, y_i)$ from the rescaled shape of area $S_0=1$.}
        $R_i \leftarrow \exp z_i$\;

        Area $S \leftarrow \sum\nolimits_i 1/2\cdot R_i R_{i+1} \sin 2\mathrm{\pi}/N$, rescale $R_i \leftarrow R_i / \sqrt{S}$\; 
        Coordinates $x_i\leftarrow R_i \cos\theta_i, y_i\leftarrow R_i \sin\theta_i$

        \tcc{Calculate penalty factor.}
        calculate circumference $\mathcal{L}$\;

        \tcc{Calculate main target $\mathrm{CI}$}
        calculate $\mathrm{var}(x), \mathrm{var}(y), \mathrm{cov}(x,y)$ so get $\bm C$\;
        calculate eigenvalues of matrix $\bm C$ to get $k$ and $|\bm C|$\;
        Using integration to calculate $\mathrm{CI}(|\bm C|, k)$ and its derivatives $\nabla \mathrm{CI}$

        \tcc{Gradient descent.}
        calculate loss $\mathcal{Z} \leftarrow -\mathrm{CI} + \lambda \mathcal{L} + 
        \varepsilon (S - S_0)^2$
        \tcp*{$\varepsilon$ used to push the no-rescaled shape towards $S_0=1$.}
        
        update $z_i \leftarrow z_i - \eta \,\nabla \mathcal{Z}$\;
    }
    return $R_i = \mathrm{e}^{z_i} / \sqrt{S}$\;
\end{algorithm}

\begin{figure}[h]
    \centering
    \subfigure[]{
        \includegraphics[width=.39\textwidth]{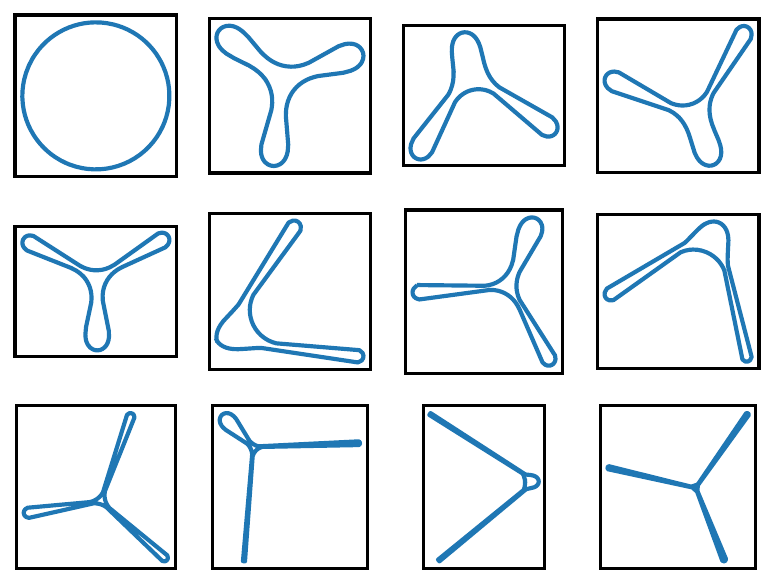}
        \label{fig:alg_colk_notgood:shapes}
    }
    \subfigure[]{
        \includegraphics[width=.45\textwidth]{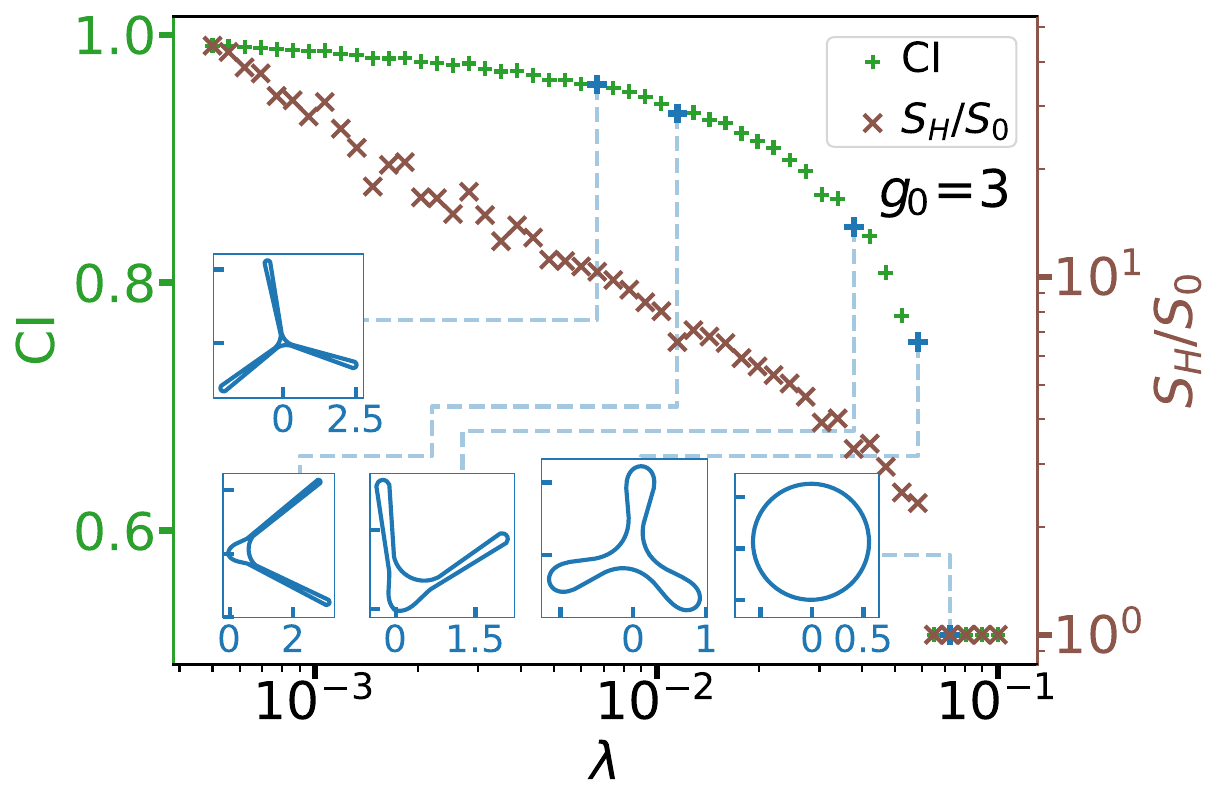}
        \label{fig:alg_colk_notgood:CI}
    }\\     
    \caption{Optimal cell shape for barrel-shaped-cells (3D) under $g_0=3$.
    \subref{fig:alg_colk_notgood:shapes}
        Some of the optimized shapes.
        Each shape has the same area $S_0=1$ and 
            corresponds to a data point in \subref{fig:alg_colk_notgood:CI}.
    \subref{fig:alg_colk_notgood:CI}
        Relationship between optimized shapes and 
            their chemotactic index.
    }
    \label{fig:alg_colk_notgood}
    \end{figure}

The randomness introduced by the initial values of $z_i$ (logarithm of radii) necessitates running the optimization multiple times with different starting points. This helps mitigate the risk of getting stuck in local minima or saddle points, which can prevent the algorithm from finding the optimal solution.

Our simulations revealed that the optimization process can become particularly slow for shapes with two tentacles (2-tentacle-shapes). Interestingly, even though these 2-tentacle-shapes might not be the absolute optimal solutions, they still exhibit good performance.
Fig.~\ref{fig:alg_colk_notgood},
    showcases some results obtained during the optimization of disk-shaped cells under a specific gradient strength ($g_0 = 3$) with varying penalty factor $\lambda$.
As the figure illustrates, the algorithm successfully converged for these disk-shape optimizations, resulting in a range of shapes. Interestingly, some of the optimized shapes appear visually quite different, resembling either 2 or 3 tentacles.

Despite the seemingly significant visual differences between the optimized shapes, their impact on the loss function ($\mathcal{Z}$) and CI is surprisingly minimal.
Fig.~\ref{fig:alg_colk_notgood:CI} demonstrates that even the incompletely optimized 2-tentacle-shapes only lead to a slight decrease in CI.

This observation suggests that the optimization landscape between 2-tentacle-shapes and 3-tentacle-shapes might be relatively flat, making it challenging for the algorithm to achieve clear differentiation. However, this also implies that 2-tentacle-shapes can be quite effective for gradient sensing. Consequently, both 2-tentacle and 3-tentacle shapes could be considered desirable configurations for gradient perception.

\section{Receptors located on 2D cell contour}

\begin{figure}[h]
    \centering

    \subfigure[]{
        \includegraphics[width=.4\textwidth]{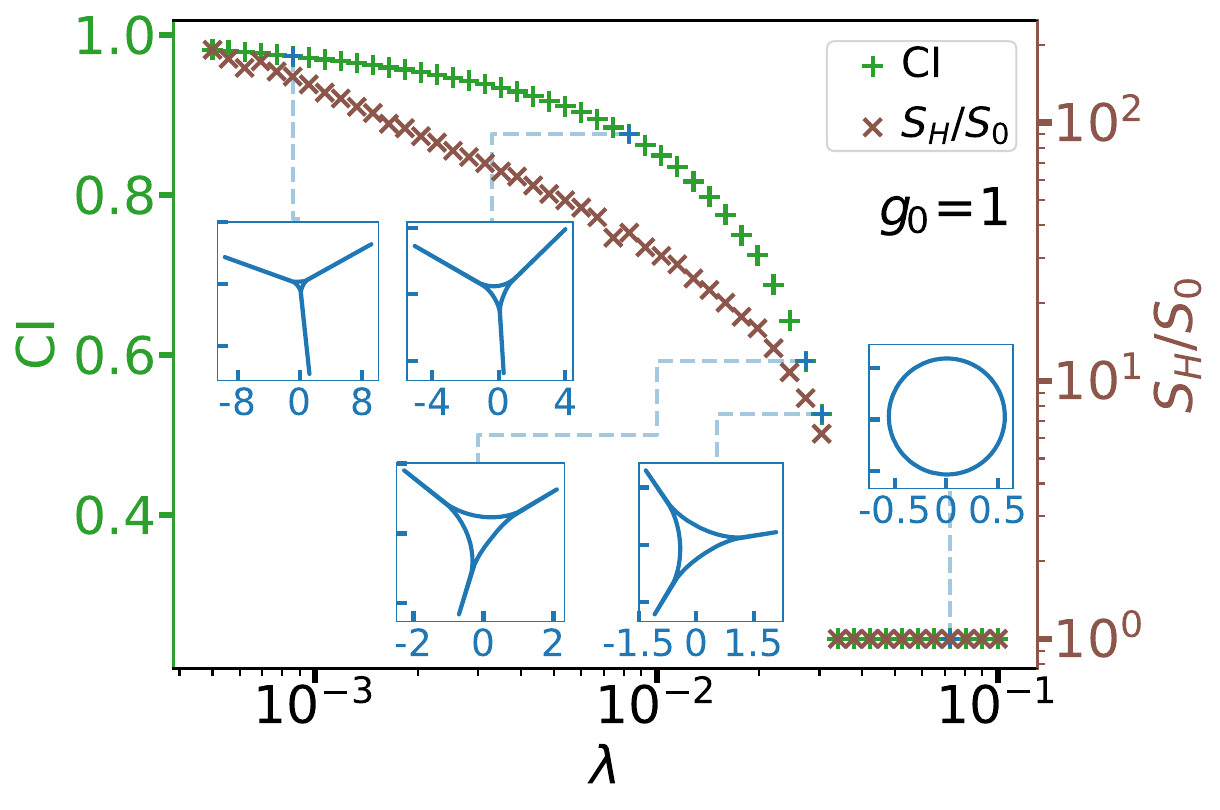}
        \label{fig:optim2d_shape:simul}
    }
    \subfigure[]{
        \includegraphics[width=.25\textwidth]{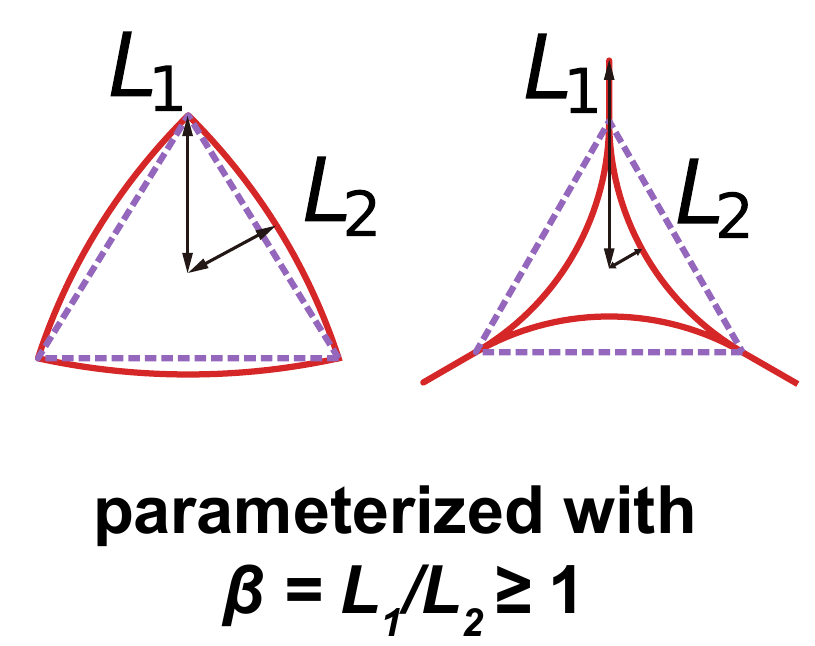}
        \label{fig:optim2d_shape:sketch}
    }
    
    \caption{\subref{fig:optim2d_shape:simul} Optimal shape for 2D cells under external gradient $g= \sigma_c/\sqrt{S_0}$,
        where all receptors are evenly distributed on the edges of the graph.
    \subref{fig:optim2d_shape:sketch} A 2D cell shape parameterized with its two axis $L_1$ and $L_2$.
    }
    \label{fig:optim2d_shape}
    
    \end{figure}

We apply the optimization program to 2D cells. In this scenario, the sensing units are uniformly distributed along the perimeter of the shape (as discussed in section~\ref{sec:nStar}), 
as opposed to being located within an inner region or the surface of a 3D disk-shape cell. The optimal shapes and CI are given in Fig.~\ref{fig:optim2d_shape}. Similar to the disk-shape cell, phase transitions and three-branched cell-shape are observed. 

To understand the transition near $\lambda_c$, we introduce a simplified representation using a single parameter $\beta$, which is the ratio of the longest side ($L_1$) to the shortest side ($L_2$) from the center (Fig.~\ref{fig:optim2d_shape:sketch}). As $\beta$ changes from $1$ to $\infty$, the shape smoothly transitions from a perfect circle ($\beta=1$) to infinitely long, three equally spaced tentacles.

\section{3D optimization}
\begin{figure}[h]
    \centering
    \subfigure[]{
        \includegraphics[width=.44\textwidth]{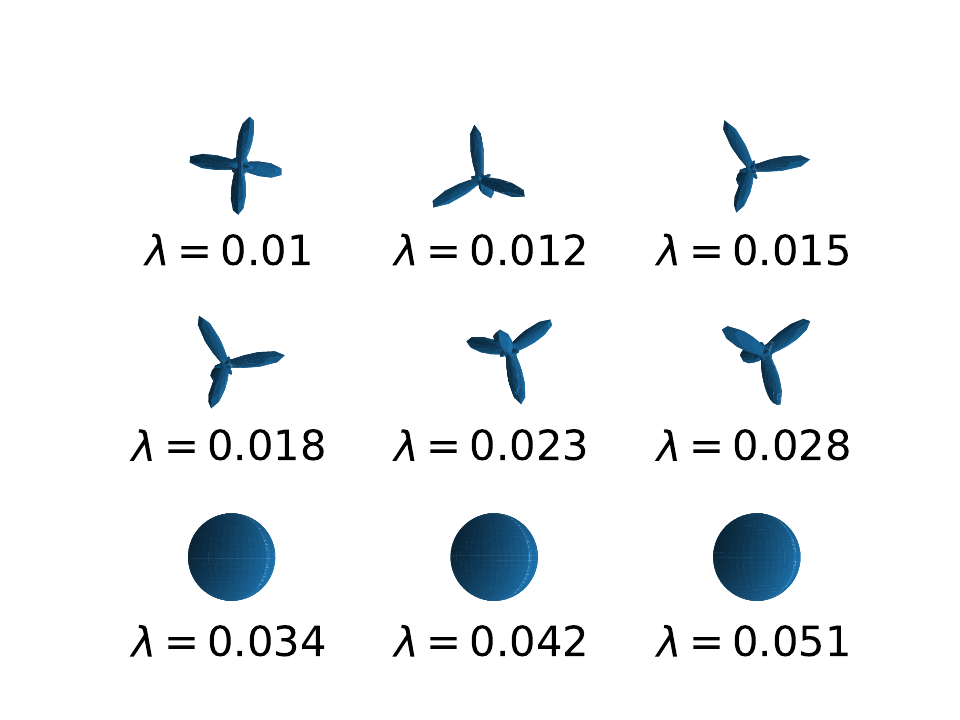}
        \label{fig:3d_simul:shape}
    }
    \subfigure[]{
        \includegraphics[width=.4\textwidth]{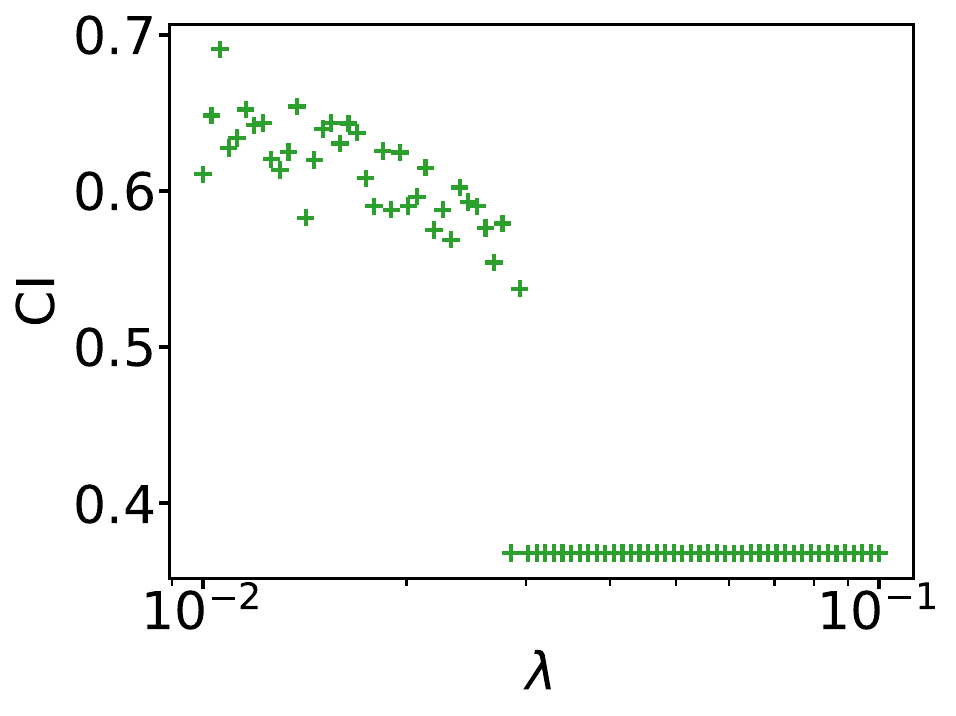}
        \label{fig:3d_simul:CI}
    }

    \caption{Optimal shape for 3D chemotaxis cells under $g_0=2$.
    \subref{fig:3d_simul:shape}
        Some of the optimized shapes.
        Each shape corresponds to a data point in \subref{fig:3d_simul:CI}.
    \subref{fig:3d_simul:CI}
        CI decreases with $\lambda$ with a sharp transition.
    }
    \label{fig:3d_simul}
    \end{figure}

Using the spherical basis function $Y_l^m(\theta, \phi)$, we define the cell surface under spherical coordinates $R(\theta, \phi)=R_0 + \exp \left[\sum\nolimits_{lm} c_l^m \cdot Y_l^m(\theta,\phi) \right]$.
Similar to optimization protocol above, we calculate the loss function $\mathcal{Z}=-\mathrm{CI} + \lambda \mathcal{A}$, where $\mathcal{A}$ is cell surface area with volume 1 and $\mathrm{CI}$ is computed without alignment. As shown in Fig.~\ref{fig:3d_simul}, the optimal shape also exhibits phase transition behavior.
For highly deformable cells, 4 tentacles are preferred corresponding to isotropic error space in 3D space.

\section{A dynamic model of the run-and-expansion strategy}
During an infinitesimal time interval $\delta t$, the inferred gradient follows $\hat{\bm g}\sim \mathcal{N}(\vec{\bm g},  \bm S\cdot 2\Delta/\delta t)$, where $\Delta$ is the characteristic timescale of concentration measurement. Here, $\bm S=\sigma_c^2\bm C^{-1}$ represents the noise covariance matrix. This information is written into the memory vector $\bm m$, which evolves according to an over-damped Langevin equation\cite{wu2014three}:
\[
\partial_t\bm m=\frac{\bm g-\bm m}{\tau}+\bm \xi(t).
\]
Here $\tau^{-1}$ is the memory update rate, and the noise vector satisfies $\langle\xi_i(t)\rangle=0$, and $\langle\xi_i(t)\xi_j(t')\rangle=2\nu\tau\sigma_c^2\delta(t-t')\delta_{ij}$, with $\nu$ being the white noise strength (scaled by $\sigma_c^2$). This memory dynamics ensures that $\langle\bm m\rangle=\langle\bm g\rangle$, maintaining an unbiased estimate of the gradient.

The corresponding covariance matrix of $\bm m$ evolves as 
\begin{equation}
    \partial_t\bm \Sigma=-\frac{2}{\tau}\bm \Sigma + \frac{2\Delta}{\tau^2}\bm S + \nu \bm I.
\label{Sigma_evol}
\end{equation}
for cells with isotropic error space, the matrix equation simplifies to a scalar form using $\bm S=\sigma_c^2/l^2\cdot \bm I$, $\bm \Sigma = \zeta(t) \sigma_c^2\cdot \bm I$, where $l$ is the length of the error space's principle axes; $\zeta(t)$ is a the memory noise strength (scaled by $\sigma_c^2$).

The key feature of our model is that different migration modes correspond to different parameters in the above equation. In particular, the expansion-phase and run-phase have different error spaces, with $l_R<l_E$. For an efficient chemotaxis, the cell should update its memory when sensing is most accurate, i.e., during the expansion-phase. This memory should persist through the run-phase. Therefore, we adopt a dynamic memory update rule that minimizes memory noise by minimizing the average of the right hand side of Eq.\ref{Sigma_evol}, leading to:
\begin{equation}
    \tau(t)=\frac{2\Delta}{\zeta(t) l^2}.
\end{equation}
This dynamic rule is similar to the approaches in \cite{nakamura2024gradient, novak2021bayesian}. Intuitively, a cell adapts its memory update rate based on sensing accuracy.

Cell motion is guided by its memory. CI describes the alignment between cell's moving direction and the gradient direction, which is computed from Eq. 7 in the main text. However, in this case, the SNR is based on the memory $\mathcal{S}=g^2/(\zeta\sigma_c^2)$ instead of the instantaneous estimate $\hat{\bm g}$. The cell's velocity is $v_R$ and $v_E$ in the run and expansion phase, respectively. The effective speed along the gradient direction is given by:
\begin{equation}
v_\mathrm{eff} = \frac{1}{T} \int_{0}^{T} v(\omega(t)) \cdot \mathrm{CI}(\mathcal{S}) \, \mathrm{d}t.
\end{equation}
Here, $\omega=0,1$ represents the run and expansion modes. $v=v_R$ if $\omega=0$ and $v=v_E$ if $\omega=1$.

\begin{figure}[h]
    \centering
    \includegraphics[width=0.47\textwidth]{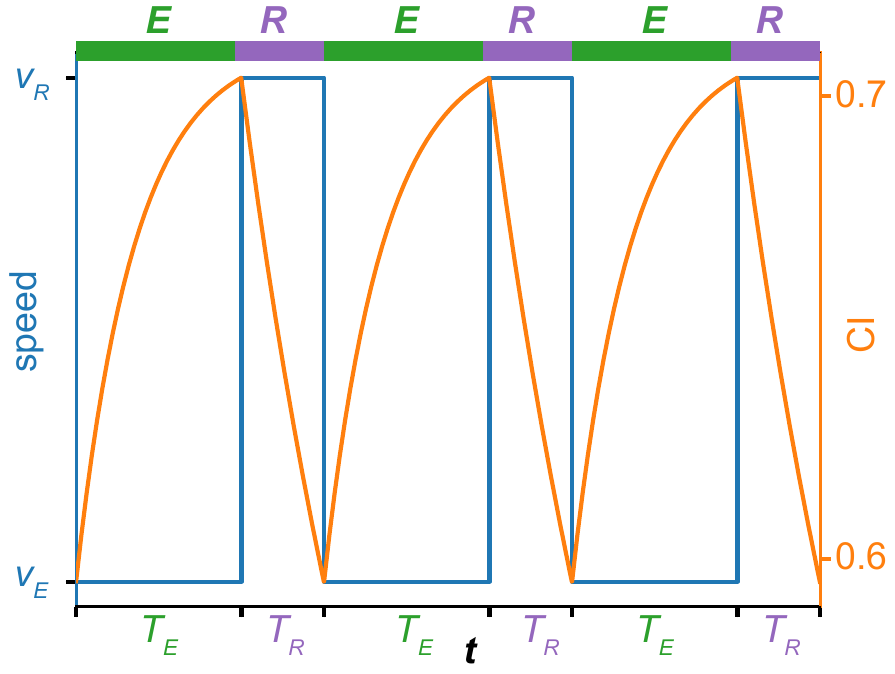}
    
    \caption{A typical trajectory of cell switching between run and expansion phase, with different cell speed and dynamic CI.
    }
    \label{fig:RE_sketch}
    
    \end{figure}

We seek the optimal switching dynamics that maximize $v_\text{eff}$. Given the above dynamics, there are two free parameters: the duration of the expansion phase $T_E$ and the run phase $T_R$, as shown in fig.~\ref{fig:RE_sketch}. 
The proportion of the expansion phase is defined as $r_E=T_E/(T_R+T_E)$.
The dimensionless parameters used in the main text Fig.5 are $v_E=0.5, v_R=1,l_E=10, l_R=1, \Delta=1, \nu=0.1$.

\bibliography{ref}
\vspace{-5em}